\documentclass[12pt]{article}

\usepackage[font=scriptsize,labelfont=bf]{caption}
\usepackage[compact]{titlesec}
\titlespacing{\section}{%
  0pt}{
  0em}{
  0pt}%
\titlespacing{\subsection}{%
  0pt}{
  0em}{
  0pt}%
\titlespacing{\subsubsection}{%
  0pt}{
  0em}{
  0pt}%
\titlespacing{\paragraph}{%
  0pt}{
  0em}{
  0.5em}%

\usepackage{amsmath,amssymb}
\usepackage{graphicx,psfrag,epsf}
\usepackage{subfig}

\usepackage{enumerate}

\usepackage{enumitem}

\usepackage{natbib}
\usepackage{soul}
\usepackage[normalem]{ulem}
\usepackage{lscape}

\usepackage{float,multirow,booktabs}
\usepackage[table,xcdraw]{xcolor}
\usepackage{url}

\usepackage{multicol}
\usepackage[section]{placeins}
\usepackage{rotating,titlesec}
\usepackage[english]{babel}

\usepackage[utf8]{inputenc}

\usepackage{algorithm,amsmath,tabularx}
\usepackage[noend]{algpseudocode}
\usepackage{hyperref}

\hypersetup{
  colorlinks=true,
  allcolors=blue
}

\usepackage{tcolorbox}
\usepackage{cleveref}

\newcounter{algsubstate}

\makeatletter
\newcommand{\multiline}[1]{%
	\begin{tabularx}{\dimexpr\linewidth-\ALG@thistlm}[t]{@{}X@{}}
		#1
	\end{tabularx}
}
\makeatother

\newcommand{\blind}{1}

\newcommand{\Supp}[1]{#1}

\begin{document}
	\setstcolor{red}

	\def\spacingset#1{\renewcommand{\baselinestretch}%
		{#1}\small\normalsize} \spacingset{1}
	\def\0{\mbox{\boldmath{$\mathbf{0}$}}}
\def\1{\mbox{\boldmath{$\mathbf{1}$}}}
\def\bzeta{\mbox{\boldmath$\zeta$}}
\def\bnu{\mbox{\boldmath$\nu$}}
\def\btheta{\mbox{\boldmath$\theta$}}
\def\bTheta{\mbox{\boldmath$\Theta$}}
\def\bmu{\mbox{\boldmath$\mu$}}
\def\bbeta{\mbox{\boldmath$\beta$}}
\def\bchi{\mbox{\boldmath$\chi$}}
\def\boldeta{\mbox{\boldmath$\eta$}}
\def\bzeta{\mbox{\boldmath$\zeta$}}
\def\bepsilon{\mbox{\boldmath$\epsilon$}}
\def\bomega{\mbox{\boldmath$\omega$}}
\def\bOmega{\mbox{\boldmath$\Omega$}}
\def\bgamma{\mbox{\boldmath$\gamma$}}
\def\bGamma{\mbox{\boldmath$\Gamma$}}
\def\bsigma{\mbox{\boldmath$\sigma$}}
\def\bSigma{\mbox{\boldmath$\Sigma$}}
\def\bdelta{\mbox{\boldmath$\delta$}}
\def\bDelta{\mbox{\boldmath$\Delta$}}
\def\blambda{\mbox{\boldmath$\lambda$}}
\def\bLambda{\mbox{\boldmath$\Lambda$}}
\def\calF{\mbox{$\mathcal{F}$}}
\def\calH{\mbox{$\mathcal{H}$}}
\def\calI{\mbox{$\mathcal{I}$}}
\def\calJ{\mbox{$\mathcal{J}$}}
\def\calX{\mbox{$\mathcal{X}$}}
\def\A{\mbox{\boldmath{$\mathbf{A}$}}}
\def\B{\mbox{\boldmath{$\mathbf{B}$}}}
\def\C{\mbox{\boldmath{$\mathbf{C}$}}}
\def\D{\mbox{\boldmath{$\mathbf{D}$}}}
\def\G{\mbox{\boldmath{$\mathbf{G}$}}}
\def\I{\mbox{\boldmath{$\mathbf{I}$}}}
\def\J{\mbox{\boldmath{$\mathbf{J}$}}}
\def\M{\mbox{\boldmath{$\mathbf{M}$}}}
\def\N{\mbox{\boldmath{$\mathbf{N}$}}}
\def\R{\mbox{\boldmath{$\mathbf{R}$}}}
\def\S{\mbox{\boldmath{$\mathbf{S}$}}}
\def\U{\mbox{\boldmath{$\mathbf{U}$}}}
\def\X{\mbox{\boldmath{$\mathbf{X}$}}}
\def\W{\mbox{\boldmath{$\mathbf{W}$}}}
\def\Y{\mbox{\boldmath{$\mathbf{Y}$}}}
\def\Z{\mbox{\boldmath{$\mathbf{Z}$}}}
\def\a{\mbox{\boldmath{$\mathbf{a}$}}}
\def\c{\mbox{\boldmath{$\mathbf{c}$}}}
\def\e{\mbox{\boldmath{$\mathbf{e}$}}}
\def\f{\mbox{\boldmath{$\mathbf{f}$}}}
\def\h{\mbox{\boldmath{$\mathbf{h}$}}}
\def\j{\mbox{\boldmath{$\mathbf{j}$}}}
\def\p{\mbox{\boldmath{$\mathbf{p}$}}}
\def\x{\mbox{\boldmath{$\mathbf{x}$}}}
\def\y{\mbox{\boldmath{$\mathbf{y}$}}}
\def\z{\mbox{\boldmath{$\mathbf{z}$}}}
	
	
	
	
	\if1\blind
	{   \title{\Large \bf Functional Integrative Bayesian Analysis of High-dimensional Multiplatform Genomic Data}
            
		\author{Rupam Bhattacharyya$^{a}$, Nicholas C. Henderson$^{a}$\\ and Veerabhadran Baladandayuthapani$^{a, *}$\\~\\
			$^{a}${\small Department of Biostatistics, University of Michigan, Ann Arbor, MI 48105}\\
			$^{*}${\small Corresponding author. E-mail: \href{mailto:veerab@umich.edu}{veerab@umich.edu}}
			}

            \maketitle
	} \fi
	

	
	\if0\blind
	{
		\bigskip
		\begin{center}
			{\Large \bf Functional Integrative Bayesian Analysis of High-dimensional Multiplatform Genomic Data}
		\end{center}
	} \fi
	
	\bigskip
	
	\begin{abstract}

    Rapid advancements in collection and dissemination of multi-platform molecular and genomics data has resulted in enormous opportunities to aggregate such data in order to understand, prevent, and treat human diseases. While significant improvements have been made in multi-omic data integration methods to discover biological markers and mechanisms underlying both prognosis and treatment, the precise cellular functions governing these complex mechanisms still need detailed and data-driven de-novo evaluations. We propose a framework called Functional Integrative Bayesian Analysis of High-dimensional Multiplatform Genomic Data (fiBAG), that allows simultaneous identification of upstream functional evidence of proteogenomic biomarkers and the incorporation of such knowledge in Bayesian variable selection models to improve signal detection. fiBAG employs a conflation of Gaussian process models to quantify (possibly non-linear) functional evidence via Bayes factors, which are then mapped to a novel calibrated spike-and-slab prior, thus guiding selection and providing functional relevance to the associations with patient outcomes. Using simulations, we illustrate how integrative methods with functional calibration have higher power to detect disease related markers than non-integrative approaches. We demonstrate the profitability of fiBAG via a pan-cancer analysis of 14 cancer types to identify and assess the cellular mechanisms of proteogenomic markers associated with cancer stemness and patient survival.

	\end{abstract}
	
	\noindent%
	
	\vfill
	
	\newpage
	
	\spacingset{1.8}
	
	\section{Introduction}\label{sec: intro}
	
	Rapid advancements in collection, processing, and dissemination of multi-platform molecular and genomics ({\it multi-omics}, in short) data has resulted in enormous opportunities to aggregate such data in order to understand, prevent, and treat diseases. This has catalyzed development of integrative methods that can collectively mine multiple types and scales of multi-omics data, in order to provide a more holistic view of human disease evolution and progression (\citealt{subramanian2020multi}). Specifically, in the context of cancer, a disease driven predominantly by agglomerations of several molecular changes (\citealt{sun2021role}), the importance of synthesizing information from multi-platform omics and clinical sources to understand the cellular basis of the disease is even further underscored. Cellular oncological mechanisms, triggered at different molecular levels of the DNA $\rightarrow$ RNA $\rightarrow$ Protein path, can confer profound phenotypic advantages/disadvantages. While significant improvements have been made in multi-omics data integration methods to unveil such mechanisms, focused on both prognosis (\citealt{duan2021evaluation}) and treatment (\citealt{finotello2020multi}), the precise functions governing them need detailed and data-driven de-novo evaluations.  Our work, in the same vein, aims at two different but inter-related scientific axes: (i) selection of biomarkers associated with cancer prognosis and clinical outcomes, and (ii) learning the mechanism of these biomarkers' effects upon such outcomes via integrating upstream molecular information - we provide some additional scientific context below. 
	
	\paragraph{Classes of Integrative Omics Models} First, we briefly discuss existing integrative omics approaches in order to contextualize the need for our framework. Broadly, most of the existing integrative statistical methods can be classified into two categories - horizontal (meta-analysis type) and vertical (multi-omics) integration procedures (\citealt{tseng2015integrating}). Horizontal meta-analysis methods focus on integrating data on similar omics features from different sources such as laboratories, cohorts, sites, etc; examples include works by \cite{tu2015network} and \cite{angel2020simple}.
    Vertical integrative methods, on the other hand, are focused on integrating data sets on the same cohort of samples obtained from different omics experiments, wherein the data sets can be vertically aligned; examples include works by \cite{cheng2015transcription} and \cite{kaplan2017prediction};
    see \cite{richardson2016statistical} and \cite{morris2017statistical} for a comprehensive review of integrative methods. Most, if not all such studies perform the integration in an agnostic manner -- they neither take into account known biological structures nor utilize data illustrating functional roles of the markers of interest. Incorporating such structures and molecular regulatory information into integrative models can improve both the power to detect true biomarkers of a disease and the understanding of their cellular roles in the progression of it, as we discuss next.
	
    \paragraph{Importance of Functional Information} Broadly, by \ul{functional information}, we mean the knowledge of the specific molecular functions of the cellular genomic, epigenomic, and transcriptomic elements, leading to disease outcomes. Incorporation of such information in biomarker association models is important due to several reasons. First, different omic components of the molecular configuration of a disease, while interconnected and hierarchical, can provide complementary information. A recent review by \cite{buccitelli2020mrnas} indicates how in the DNA $\rightarrow$ RNA $\rightarrow$ Protein path, termed the {`gene expression pathway'}, lower or higher correlations of the expressions of proteins and their coding genes may be observed due to changes in the functional regulatory elements. Second, specifically for cancer, recent literature indicates that recurrent regulatory structures drive tumor progression through aberrations at different omics levels via common {\it master regulatory mechanisms} (\citealt{califano2017recurrent}). Finally, experimental validation and characterization of functional information on a biomarker-by-biomarker basis is resource-intensive - especially with many plausible candidates. Thus, computational models that can identify and incorporate functional information about genes/proteins (referred to as {\it proteogenomic} data henceforth) into the models rather than post-hoc analyses in a natural, inherent way can facilitate this understanding and can lead to \ul{de-novo data-driven prioritization} of the relevant biomarkers, especially for translational and clinical utility. To this end, we propose a framework called \ul{F}unctional \ul{I}ntegrative \ul{B}ayesian \ul{A}nalysis of High-dimensional Multiplatform \ul{G}enomic Data (\ul{fiBAG}, in short), that allows simultaneous identification of upstream functional evidence  of proteogenomic biomarkers and the incorporation of such knowledge in Bayesian (biomarker) selection models to improve signal detection.
    
\begin{figure}[hbt!]
	    \centering
	    \includegraphics[scale = 0.24]{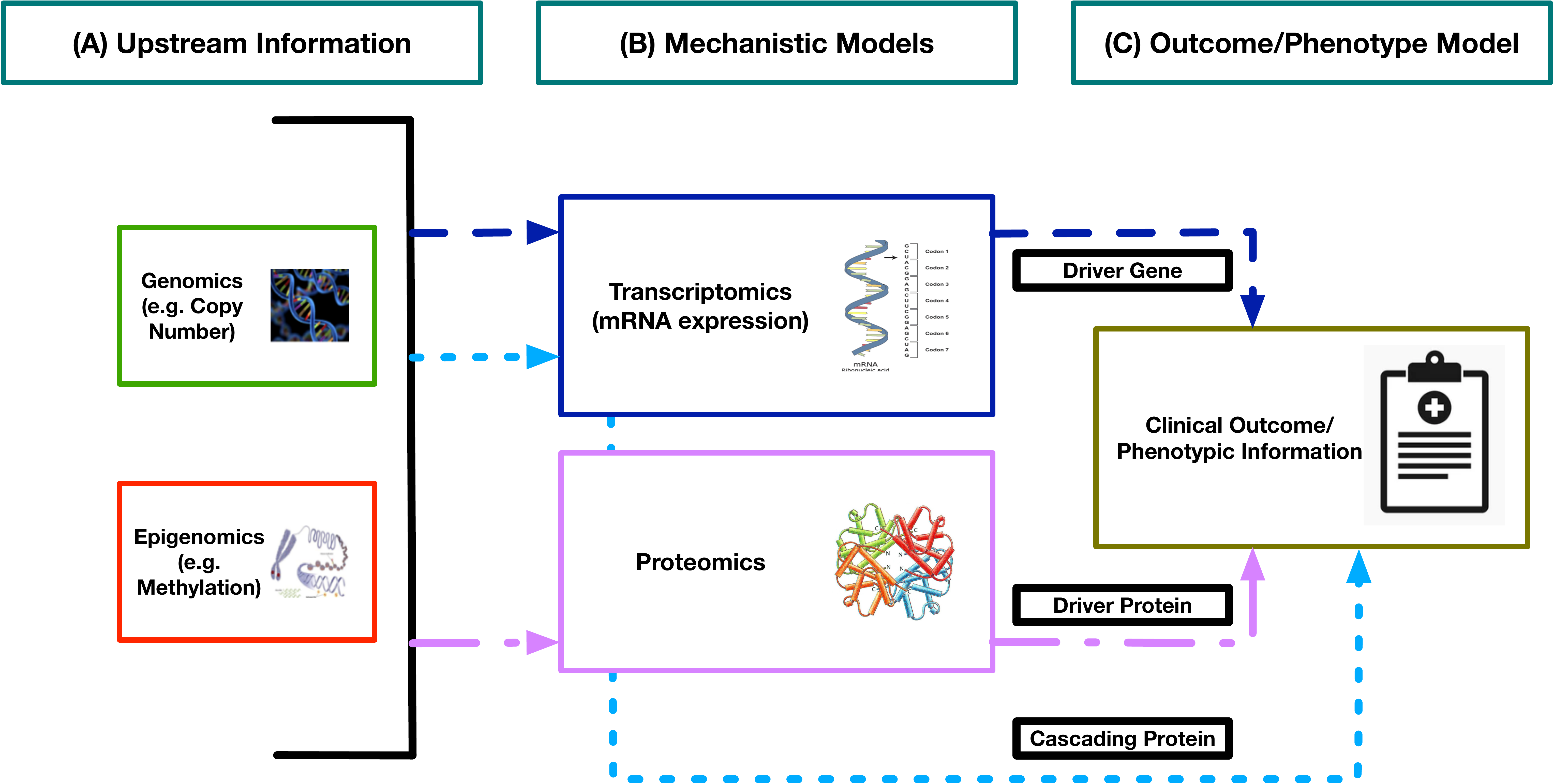}
	    \caption{\textbf{Conceptual schematic summarizing the fiBAG procedure.} Panels (A), (B), and (C) respectively describe the upstream platforms, the mechanistic models, and the outcome model. The dashed, dotted-dashed, and dotted arrows indicate the regulatory paths for the driver gene, driver protein and cascading protein functional axes, respectively.}
	    \label{fig:pipeline}
	\end{figure}

	
	\paragraph{Goals and Utility of fiBAG} Our scientific goals are multifold. We focus on integrating high-dimensional multi-omics data and clinical responses in an approach similar to that of \cite{wang2013ibag, jennings2013bayesian}, deciphering and delineating functional roles of proteogenomic markers using mechanism-driven (\ul{mechanistic}) models, and incorporating this functional information into \ul{outcome} models, thus providing functional relevance to the findings. In particular, we want to contextualize the available functional information such as the type of proteogenomic activity following existing literature such as \cite{gevaert2013identification} and \cite{song2019insights}. Using functional evidence from these mechanistic models, we then guide selection and hence the degree of penalization (thus prioritization) of covariates in the final outcome models. \Cref{fig:pipeline} provides a broad-scale summary of how the fiBAG procedure achieves these goals. The first column describes the upstream data utilized to infer functional information. For the purpose of this work, we only use copy number and DNA methylation; other data platforms having potential functional relevance (such as microRNA) may also be utilized. The middle column describes the three axes of mechanistic information that we intend to infer on using gene and protein expression data, namely, \ul{driver gene} (dashed line), \ul{driver protein} (dashed-dotted line), and \ul{cascading protein} (dotted line). We describe the construction of these mechanistic models in more detail in the following {Section(s)}. The third and final column describes the ``calibrated" outcome model, where summary information from the mechanistic models are incorporated alongside outcome data to improve selection of genes and proteins. Briefly, our study offers both methodological novelty via proposing a calibrated Bayesian variable selection procedure for the outcome model, and scientific innovation via performing integration of both patient outcomes and tumor features with multiomic data.
	

    \paragraph{Methodological Novelty} We employ a mapping function to {calibrate numerical evidences of significance} obtained from the mechanistic models to a prior inclusion probability scale, that subsequently inform the outcome model of prior functional evidence in favor of specific proteogenomic candidates. Our method is flexible -- the calibration function can be adapted according to the choice of the mechanistic model and the resulting quantification of significance. This hierarchical evidence sharing procedure allows our method to integrate data across any number of genomic, epigenomic, and other relevant platforms of choice. Using Gaussian processes to identify the mechanistic evidence, our model is better equipped to detect nonlinear cellular associations than a standard linear model, as used in \cite{wang2013ibag}, and is computationally simpler, eliminating the needs of choosing the number of knots and incorporating penalization in a spline-type setting, such as the approach taken by \cite{jennings2013bayesian}. We calibrate the evidence summarized from these models to the Bayesian variable selection setting by proposing a generalized version of the spike-and-slab prior originally proposed by \cite{george1997approaches}, termed the \ul{calibrated spike-and-slab prior}. This calibrated prior structure improves the selection of the covariates by borrowing strength across multi-platform data, choosing to continuously up-weight prior inclusion probabilities of biomarkers in a data-driven manner. While we take the spike-and-slab route to build an adaptive and flexible mechanism to incorporate external knowledge in this work, other existing methods perform the incorporation of such {\it a priori} information in an outcome model via adaptive shrinkage, penalization, or some different prior structure. Using simulation studies under multiple synthetic and real data-based scenarios, we compare both the selection and estimation performances of our method against standard penalized regression (\citealt{tibshirani1996regression}), grouped penalized regression (\citealt{boulesteix2017ipf}), and prior-informed selection (\citealt{velten2021adaptive, zeng2021incorporating}) methods. Our method exhibits comparable selection and estimation performances with state-of-the-art methods for higher sample size to number of covariates ratios but accords significantly better false discovery rate controls, and exhibits substantial improvements in performance for low-sample high-dimensional settings. We also offer computational flexibility using both a Markov chain Monte Carlo (MCMC) implementation and a computationally efficient expectation-maximization based variable selection procedure (\citealt{rovckova2014emvs}). Our calibrated Bayesian variable selection (\ul{cBVS}) procedure has broad applicability to general regression problems even outside the multi-omic integration domain where some type of covariate importance quantification is available, allowing the user to incorporate prior evidence quantities via other calibration functions tuned to the scale of such evidence.

	\paragraph{Scientific Innovation} Multiple works from recent biostatistical literature have focused on incorporating existing evidence or external information into final models of interest via various approaches, such as data-adaptive shrinkage (\citealt{boonstra2015data}), adaptive Bayesian updates (\citealt{boonstra2020incorporating}), or calibrated maximum-likelihood type procedures (\citealt{chatterjee2016constrained}).
	Our method is different from such approaches in the sense that it offers a framework to both learn {de novo} evidence within the pipeline and incorporate the said evidence into the final outcome models. As discussed before, omics elements at different hierarchical levels of the \textit{gene expression pathway} may provide partly independent and complementary information (\citealt{buccitelli2020mrnas}). Our integrative approach allows learning such information across interconnected axes of functional activity such as DNA, RNA, and protein level quantifications. Additionally, our integrative analysis of pan-cancer proteogenomic data from The Cancer Genome Atlas utilizes both traditional \ul{prognostic outcomes} (survival data) and recently developed \ul{cellular descriptors} of cancer growth (stemness indices) to identify the proteogenomic signatures driving such features, along with the molecular basis of these signatures. Cancer stem-like cells lead to sustained proliferation via resisting apoptosis, evading growth suppression, and exhibiting increased invasive and metastatic potential (\citealt{fulda2013regulation,adorno2015cancer}). We look at the challenge of identifying the cellular molecular basis of the behavior of such stem-like cells, by using the mRNA-based stemness index (SI) proposed by \cite{malta2018machine} as an outcome variable. From the survival and stemness outcome analyses across four common groups of cancers: Pan-gynecological, Pan-gastrointestinal, Pan-squamous and Pan-kidney, we identify both known and novel markers associated with the outcomes alongside insights on their functional roles. In particular, the genes RPS6KA1 (protein p90RSK) and YAP1 (proteins YAP and YAPPS127) are identified as top driver genes in the pan-gyne cancers, along with significant associations with the stemness outcome across multiple cancers in the group; both have been known to be crucial agents impacting gynecological cancers. Similarly, our analyses found the gene ERBB2 (protein HER2) to be positively associated with stemness for gastrointestinal cancers, supported by existing literature.
	
	The rest of the paper is organized as follows. Sections \ref{subsec:concept}-\ref{subsec:cBVS} describe fiBAG both conceptually and with the mathematical details of the mechanistic and outcome models and the evidence calibration function. \Cref{subsec: outcome_model} describes the computational steps behind the fitting, estimation, and selection procedures for the two models. \Cref{sec: simulation} summarizes simulation studies in both synthetic and real data-based settings comparing our method to existing benchmarks. \Cref{sec: pancancer} summarizes results from our pan-cancer integrative proteogenomic analyses. We conclude the paper with a discussion on the methodological and biological aspects of our work, along with some potential future directions in \Cref{sec: discussion}. All our results are available in an interactive R-Shiny dashboard (\Supp{Supplementary Materials Section 5}, denoted by SM henceforth). All the software codes used to perform the analyses, along with the processed datasets are available as a zip file along with the submission.
	
	
	\section{\lowercase{fi}BAG Model}\label{sec: funcibag}
	
	\subsection{Conceptual Factorization of the Multi-omics Model}\label{subsec:concept}
	
	We begin with the data structure and some notations. Let $n$ denote the number of samples, $q_g$ be the number of genes, and $q_p$ be the number of proteins in the dataset of interest. The gene (mRNA) expression data matrix $\mathbf{G}$ and the protein expression matrix $\mathbf{P}$ respectively have dimensions ${n \times q_g}$ and ${n \times q_p}$. Further, let the data matrices corresponding to the upstream covariates copy number alteration and DNA methylation be denoted as $\mathbf{C}$ and $\mathbf{M}$, respectively, each having $n$ rows and are matched to their respective genes and proteins (i.e. {\it cis}-level summarizations). Let $\mathbf{Y}$ denote the $n \times 1$ outcome data vector. Thus, the proteogenomic, upstream, and outcome data available for a cohort of samples can be aligned vertically (i.e. matched across samples). We write the joint model of the outcome and the proteogenomic data conditional on the upstream data as
	\vspace{-8pt}
    \small
	\begin{equation}
	    \mathcal{P}[\mathbf{Y}, \mathbf{G}, \mathbf{P} | \mathbf{C}, \mathbf{M}, \boldsymbol{\theta}] = \underbrace{\mathcal{P}[\mathbf{Y} | \mathbf{G}, \mathbf{P}, \boldsymbol{\theta}_{Y}]}_{Outcome} \underbrace{\mathcal{P}[\mathbf{G}, \mathbf{P} | \mathbf{C}, \mathbf{M}, \boldsymbol{\theta}_{F}]}_{Mechanistic}.
	    \label{eq: factorization}
	\end{equation}
	\normalsize
	Here, $\boldsymbol{\theta} = (\boldsymbol{\theta}_{Y}, \boldsymbol{\theta}_{F})$ denotes a conceptual parameter (possibly multi-dimensional) that connects the two layers of models. The omics-only part ($\mathcal{P}[\mathbf{G}, \mathbf{P} | \mathbf{C}, \mathbf{M}, \boldsymbol{\theta}_{F}]$) represents the \ul{mechanistic model}, which concerns the functional mechanisms of proteogenomic expression as driven by DNA-level cellular activities. Via the parameter $\boldsymbol{\theta}_{F}$, this mechanistic information is then learned and incorporated into the \ul{outcome model} ($\mathcal{P}[\mathbf{Y} | \mathbf{G}, \mathbf{P}, \boldsymbol{\theta}_{Y}]$). The parameter vector $\boldsymbol{\theta}$ enables the mechanistic layer to inform the outcome layer, in line with our scientific aims of integrating functional information.
	
	
	\paragraph{Biological Rationale for the Factorization} The connection between the two layers drives potentially improved identification of proteogenomic features in the final outcome model. This conceptual framework aligns with the idea that different tumors, although potentially driven by changes in different agents or genomic locations, are controlled by a recurrent regulatory architecture where genomic alterations cluster upstream of functional proteins (master regulators) (\citealt{califano2017recurrent}). The interconnections between these regulators form the tumor checkpoints that can potentially be useful as biomarkers and therapeutic targets. Further, epigenetic and genetic changes determining a cancer cell state can be intertwined, and the mutual dependencies between such traits can contribute to tumor progression via sequential layers of cellular activity (\citealt{alizadeh2015toward}). Thus, it makes sense to model multi-platform omics data in a way where functional contributions to the variability in expressions of genes and proteins can be inferred separately and can be utilized to calibrate the identification of the roles of those genes and proteins in tumor progression. Over the next few subsections, we describe the specific approaches undertaken in this work to formulate these two model layers to utilize this biological framework.
	
	\subsection{Mechanistic Models}\label{subsec:mechmodels}
	
	We focus on \ul{three axes of mechanistic information} (one for each gene and two for each protein) based on the available upstream data (\Cref{fig:pipeline}). We first describe the mathematical settings and the interpretations of the three axes. Let $j$ denote the index for the specific proteogenomic biomarker of interest in a mechanistic model, with the understanding that it is a gene if $j \in \{1,\ldots,q_g\}$, and a protein if $j \in \{q_g+1,\ldots,q_g+q_p\}$. For biomarker $j$ and sample $i$, let the corresponding sub-vectors of $\mathbf{C}$ and $\mathbf{M}$ be denoted by $\mathbf{C}_{ij}$ and $\mathbf{M}_{ij}$, respectively. We also assume that all proteogenomic expression data are mean-centered. The general forms of the  models corresponding to the three mechanistic axes are as follows: 
	\vspace{-27pt}
	\small
	\begin{eqnarray}
	    \textbf{Driver gene model:} \; \; \; G_{ij} &=& f_{1j}\big((\mathbf{C}_{ij}^T, \mathbf{M}_{ij}^T)^T\big) + e_{1ij}, \nonumber \\
	    \textbf{Driver protein model:} \; \; \; P_{ij} &=& f_{2j}\big((\mathbf{C}_{ij}^T, \mathbf{M}_{ij}^T)^T\big) + e_{2ij}, \nonumber \\
	    \textbf{Cascading Protein model:} \; \; \; P_{ij} &=& f_{3j}\big((G_{ij}, \mathbf{C}_{ij}^T, \mathbf{M}_{ij}^T)^T\big) + e_{3ij}, \nonumber
	\end{eqnarray}
	\normalsize
	\vspace{-3pt}
\noindent where $f_{\bullet}$ are nonlinear functionals and $e_{\bullet}$ denote the (normal) error components. The upstream to gene correspondences are defined by the physical co-location of the coding segment of the gene in the genome, where-in a window of $\pm 500$ kb are taken into account. The gene to protein correspondences are defined using which gene codes for which protein. The upstream to protein correspondences are defined analogously. The three models can be biologically interpreted in the following manner:
	
    \setlist{nosep}
    \begin{enumerate}[noitemsep]
	    
	    \item \textbf{Driver gene:} Whether the regulations corresponding to the gene are unique at transcript levels, and whether there is a significant relationship between the genomic/epigenomic events and the resulting gene expression which, in turn, {drives} cancer progression and outcomes (\citealt{gevaert2013identification}).
	    
	    \item \textbf{Driver protein:} Whether the regulations corresponding to the protein are unique at transcript levels, and whether there is a significant relationship between the genomic/epigenomic events and the resulting protein expression which, in turn, {drives} cancer progression and outcomes.
	    
	    \item \textbf{Cascading protein:} Whether the protein-specific regulations transit through multiple prior omics levels, i.e., whether there is a cascade of effects via the DNA $\rightarrow$ RNA $\rightarrow$ protein path (\citealt{song2019insights}).
	    
	\end{enumerate}
	
	The middle column of \Cref{fig:pipeline}  describes the data structures as presented in the above equations. For each model, we are interested in a null hypothesis of the type $f_{\bullet} = 0$, indicating that the corresponding proteogenomic expression does not depend on the covariates i.e. $\mathbf{C}$ and $\mathbf{M}$. Compelling evidence against such hypotheses would provide evidence for the corresponding mechanistic activity being present for the gene/protein in question. For example, a strong level of evidence for the driver gene model would mean that the upstream DNA-level events impact the expression of the gene of interest significantly; the other models can be interpreted similarly. We now describe the characterization of these relationships via suitable modeling choices for the $f_{\bullet}$s, and the hypothesis testing setting to quantify the strength of evidence in the data.

	
	
    \paragraph{Characterization of $f$ via Gaussian Processes}	The flexibility of the mechanistic model setting lies in the freedom in choosing the specific form of $f$. The simplest choice would be a linear function - which would lead to multiple linear regression models. These models, explored by \cite{wang2013ibag}, are easy to handle computationally and easy to interpret, since the regression parameters are explicitly available for inference (estimation/testing). However, they could miss potentially nonlinear associations prevalent across the multi-omics levels of cellular activity, as has been shown previously (\citealt{solvang2011linear,litovkin2014methylation}). A more sophisticated choice of $f$, allowing such nonlinear associations would be to use a set of basis functions to describe $f$ (such as using a spline model, as explored by \cite{jennings2013bayesian}). Such models, however, require specifications of the knots based on a priori knowledge and demand additional penalization to obtain a stable fit, rendering the procedure to be more computationally intensive and less interpretable. To allow computationally tractable and interpretable identification of nonlinear associations while avoiding the need to specify knot locations over a multivariate domain, we use Gaussian process (GP) models. GPs have been utilized in the context of genomic data in past literature including modeling gene expression dynamics (\citealt{rattray2019modelling}), transcriptional regulations (\citealt{lawrence2007modelling}), and pathway analyses (\citealt{liu2007semiparametric}). Our simulation studies indicate that in scenarios with a high degree of non-linearity among the covariates in the generating model, GPs are better equipped in capturing significant associations than linear models (\Cref{sec: simulation}). 
    
    We now describe the GP specifics for the driver gene mechanistic model here; the other models can be expressed similarly. To recall, the $j^{\textrm{th}}$ driver gene model is written as $G_{ij} = f_{1j}((\mathbf{C}_{ij}^T, \mathbf{M}_{ij}^T)^T) + e_{1ij}$, where we assume $e_{1ij} \overset{\textrm{iid}}{\sim} \textrm{N}(0, \tau_{1j}^2)$. Let us also denote $f_{1j}^{(i)} = f_{1j}((\mathbf{C}_{ij}^T, \mathbf{M}_{ij}^T)^T)$ for all $i$. Then, the GP prior on $f_{1j}$ is placed as follows:
    \vspace{-8pt}
	\small
    \begin{eqnarray}
	   \textbf{GP prior:} &&(f_{1j}^{(1)}, \ldots, f_{1j}^{(n)})^T \sim \mathbf{N}(\mathbf{0}, \mathbf{K}_{1j}), \nonumber \\
    \textbf{Covariance matrix:} &&\mathbf{K}_{1j(i,k)} = K_{1j}((\mathbf{C}_{ij}^T, \mathbf{M}_{ij}^T)^T, (\mathbf{C}_{kj}^T, \mathbf{M}_{kj}^T)^T), \nonumber \\
	    \textbf{Kernel function:} &&K_{1j}(\mathbf{u}, \mathbf{v}) = g\tau_{1j}^2\textrm{exp}\big( -\frac{||\mathbf{u} - \mathbf{v}||^2}{\lambda_{1j}^2} \big). \nonumber
	\end{eqnarray}
	\normalsize
	\vspace{-6pt}
    The hyperpriors specify $\tau_{1j}^2 \sim \textrm{Inverse-Gamma}(\frac{\nu_{01j}}{2}, \frac{\nu_{01j}\tau_{01j}^2}{2})$ and $\lambda_{1j} \sim \textrm{exp}(\lambda_{01j})$. For all our analyses, we set $g = n$. Although we use the  standard squared exponential kernel, a common default choice (\citealt{micchelli2006universal}), other kernels can be adopted as well. As stated before, the gene expression data is mean-centered.
    
    \paragraph{Hypothesis Tests for Drivers and Cascades via Bayes Factors} We describe our Bayesian hypothesis testing procedure for the mechanistic models using our driver gene models; the other models can be tested similarly. For the $j^{\textrm{th}}$ driver gene model, we test $H_{0j}: f_{1j} = 0 \textrm{ vs } H_{1j}: H_{0j} \textrm{ is false (equivalent to } H_{0j}: K_{1j}(\bullet, \bullet) \equiv 0)$. We compute a log of the Bayes factor (lBF) corresponding to the comparison of the full model vs the null (zero-mean, error-only) model to perform the test. Bayes factors (BFs) are particularly useful in this setup from both a statistical and a scientific point of view. From a methodological perspective, elegant almost sure convergence results for Bayes factors are available for rather general settings under standard assumptions (\citealt{chatterjee2020short}). Further, Bayes factors have been successfully utilized to quantify significance and compare model performances in omics models in past literature (\citealt{stephens2009bayesian}). For our case, the final expression of the lBF for the driver gene model $j$, ignoring the constants $a_n, b_n, c_n$, and $a$ (dependent on data dimensions and hyperparameters), is given below (detailed derivations in \Supp{SM 2.1}). The integral in \Cref{equation: BFexpression} is computed numerically. For any matrix $\mathbf{A}$, $\mathbf{A}_{\bullet j}$ denotes the $j^{\textrm{th}}$ column of $\mathbf{A}$, and $\mathbf{A}_{i \bullet}$ denotes the $i^{\textrm{th}}$ row of $\mathbf{A}$.
    \vspace{-8pt}
	\small
    \begin{eqnarray}
	    && \textrm{lBF}_{1j} = \big[a_n + b_n\ln\Big( a+\sum_{i=1}^{n}G_{ij}^2-\frac{(\sum_{i=1}^{n}G_{ij})^2}{c_n} \Big) \nonumber\\
	    &+& \ln\int_0^{\infty} \exp(-\lambda_{01j}\lambda_{1j})\frac{2^{b_n}|\mathbf{K}_{1j}/\tau_{1j}^2+I|^{-\frac{1}{2}}}{\big\{\mathbf{G}_{\bullet j}^{T}(\mathbf{K}_{1j}/\tau_{1j}^2+I)\mathbf{G}_{\bullet j}+a)\big\}^{b_n}} d\lambda_{1j}\big] / \ln(10).
	    \addtocounter{equation}{-1}\refstepcounter{equation}\label{equation: BFexpression}
	\end{eqnarray}
	\normalsize
	\vspace{-3pt}
    We decide the strength of the evidence posed by the lBF from a mechanistic model using the following standard significance ranges: $<0.5$ (no evidence), $0.5-1$ (substantial), $1-2$ (strong), and $>2$ (decisive) (\citealt{kass1995bayes}). For each gene, we have one lBF from the driver gene model, whereas for each protein, we have a maximum of two such lBFs from the driver and cascading protein models. We now describe our approach to calibrate these evidence quantities into the variable selection models for the outcomes of interest.
    
	
	\subsection{Calibrated Bayesian Variable Selection}\label{subsec:cBVS}
	
	Functional information is captured by the lBFs from the mechanistic models - each lBF summarizes the strength of evidence for the functional role of the corresponding gene/protein. The lBF metric is particularly useful for two reasons - first, it provides us with an interpretable, unidirectional and continuous scale of evidence strength, and second, across a large proteogenomic panel, it provides a highly parallelizable procedure to gather scalar evidence of functional relevance which are useful numerical quantities in their own terms and pertinent prior information for future models, such as our outcome model.
	
	Thus, following \Cref{eq: factorization}, we want to incorporate this mechanistic information into our final outcome model. We pose this problem in context of a general regression framework where some quantitative summaries of covariate importance are available beforehand, and such information is to be combined with the mechanics of a typical variable selection setting. Generally, we denote such prior information as $\boldsymbol{\mathcal{E}}$, with $\boldsymbol{\mathcal{E}_j}$ denoting the possibly multi-dimensional evidence summary for covariate $j$ (i.e., lBF for gene/protein $j$ in our case). We intend to inform the final outcome model using these evidences in terms of selection/non-selection of the predictors. Specifically, if there is sufficiently \ul{strong evidence} in favor of a covariate, we want to \ul{up-weight its probability of inclusion}. Otherwise, we want to put a uniform probability on selection/non-selection for that particular covariate. To achieve this, we utilize a hierarchical Bayesian framework with spike-and-slab priors for each effect, with the spike probabilities calibrated using the evidence available. The rest of this subsection describes the components of our fiBAG outcome model, called calibrated Bayesian variable selection, or \ul{cBVS} in short.

	\paragraph{Notations} Following notations introduced at the beginning of this section, let $Y_i$ denote the outcome for individual $i$.  For the purpose of exposition, we assume that the outcome of interest is continuous, and each $Y_i$ is assumed to be observed. Generalizations to censored or categorical outcomes are straightforward. Let us denote the design matrix corresponding to any additional covariates (other than genes and proteins, such as clinical information or demographics) by $\mathbf{B}_{n \times q_b}$. The combined design matrix will have a dimension $p = 1 + q_{b} + q_{g} + q_{p}$. We then propose a hierarchical Bayesian outcome model, as described below.
	\vspace{-8pt}
	\small
	\begin{eqnarray}
	    Y_{i} &=& \beta_{0} + \mathbf{B}_{i\bullet}^{T}\bbeta_{B} + \mathbf{G}_{i\bullet}^{T}\bbeta_{G} + \mathbf{P}_{i\bullet}^{T}\bbeta_{P} + \epsilon_{i}, \quad \forall i \in \{1,...,n\}, \nonumber \\
        \epsilon_{i} &\overset{\textrm{iid}}{\sim}& \textrm{Normal}(0, \sigma^{2}), \quad \forall i \in \{1,...,n\}. \nonumber
    \end{eqnarray}
	\normalsize
	\vspace{-8pt}
	Let $\bbeta_{p \times 1}$ denote the complete regression coefficient vector. We set up a hierarchical \ul{calibrated spike-and-slab} prior on its last $q_g + q_p$ components, as described below. A standard conjugate prior is put on the residual variance parameter as $\sigma^{2} \sim \textrm{Inverse-Gamma}(\tfrac{\nu}{2}, \tfrac{\nu\lambda}{2})$.
	\vspace{-8pt}
	\small
	\begin{eqnarray}
        (\beta_{0}, \bbeta_{B}^{T}, \bbeta_{G}^{T}, \bbeta_{P}^{T})^{T} = \bbeta | \bgamma_{q_g + q_p \times 1}, \sigma &\sim& \textrm{Normal}( \mathbf{0}, \mathbf{D}_{\bgamma, \sigma}), \nonumber \\
        \gamma_{j}|\omega_{j} &\sim& \textrm{Bernoulli}(\omega_{j}), \quad \forall j \in \{1,..., q_{g} + q_{p}\}, \nonumber \\
        \omega_{j} &\sim& \textrm{Beta}\Big( \mathcal{F}(\boldsymbol{\mathcal{E}}_j), \frac{1}{\mathcal{F}(\boldsymbol{\mathcal{E}}_j)} \Big), \quad \forall j \in \{1,..., q_{g} + q_{p}\}, \nonumber
    \end{eqnarray}
    \normalsize
	\vspace{-3pt}
    where $\mathbf{D}_{\bgamma, \sigma} = \sigma^{2}\mathbf{A}_{\bgamma}$, $\mathbf{A}_{\bgamma \; p \times p} = \textrm{diag}\{v_{1}\mathbf{1}_{1+q_b}, \gamma_{1}v_{1} + (1 - \gamma_{1})v_{0}, \ldots, \gamma_{q_{g} + q_{p}}v_{1} + (1 - \gamma_{q_{g} + q_{p}})v_{0}\}$ where $v_{1} \geq v_{0} > 0$ are respectively the slab and spike variances,
    and where $\mathcal{F}$ is a calibration function mapping the evidence $\boldsymbol{\mathcal{E}}_{j}$ to the prior inclusion probability $\omega_{j}$.
    
    
    \paragraph{Calibration Functions incorporating External Information} The function $\mathcal{F}$ in the outcome model maps the functional evidence from the mechanistic models to a parameter scale to inform the Beta hyperprior for selection of each covariate. The specific mathematical properties and form of this function depends on the user-specific context of what type, range, and scale of functional evidence is being used. Further, it is possible to garner multiple quantities of evidence from different sources for each covariate of interest, and depending on the scenario, the calibration function may take inputs in $\mathbb{R}^K$ for some $K>1$.
    
\begin{figure}[hbt!]
	    \centering
	    \includegraphics[scale = 0.30]{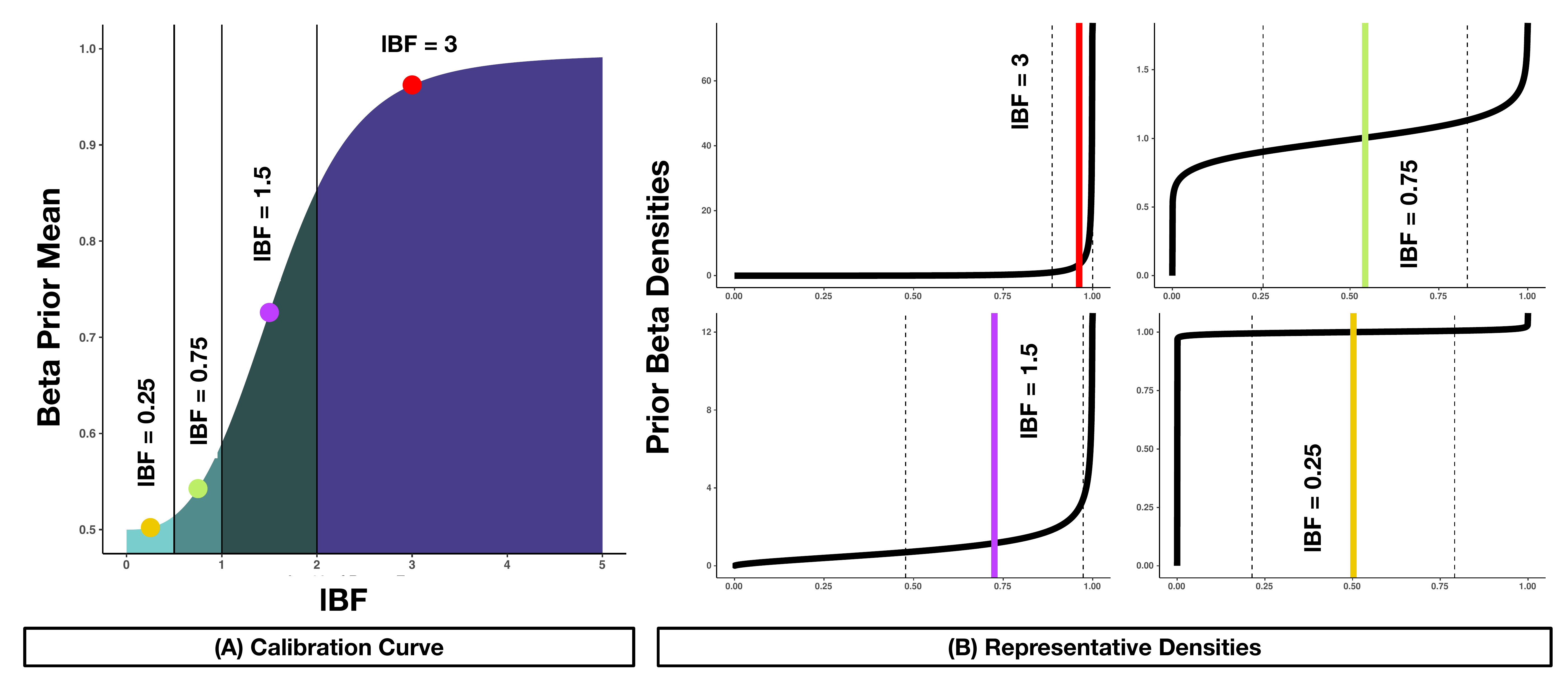}
	    \caption{\textbf{Calibration function for the outcome model}. Panel (A) plots the calibrated prior Beta means against the lBFs. Panel (B) presents the densities (means indicated by solid vertical lines, broken vertical lines indicate the $\pm1$ sd ranges around the means) for four representative values from the four ranges of interest in the x-axis of panel (A).}
	    \label{fig:calibration_fun}
    \end{figure}

    For the suitability of exposition, we describe a specific form of the mapping function used for our analyses where it is assumed that the evidence measures $\mathcal{E}_{j}$ are in the form of lBF's from the mechanistic models for genes and proteins.
    Our calibration function then needs to be a map from $\mathbb{R} \rightarrow \mathbb{R}$ that can reflect the changes in the strength of evidence in the lBFs in the desired ranges. For lBFs, the strength of evidence is unidirectional (i.e., higher evidence quantity implies higher evidence strength). In such a scenario, a mapping function is required to be non-decreasing. Such  a mapping function can either be discrete or continuous, and the jumps of the discrete curve or the slope of the continuous curve can be tuned depending on the importance to be put on different scalar values of the evidence. In particular, we map the standard lBF significance ranges as described before to a parameter describing a Beta prior distribution for the covariate-inclusion probabilities $\omega_j$. The shape of the lBF to Beta prior mean curve and the representative densities for some lBF values of interest in each of the standard lBF significance ranges ($<0.5$ (no evidence), $0.5-1$ (substantial), $1-2$ (strong), and $>2$ (decisive)) for this calibration function are summarized in \Cref{fig:calibration_fun}. Effectively, we build a function having the following properties.
    
    \begin{itemize}
    
        \item If the mechanistic models \ul{do not provide} us with \ul{enough evidence} regarding the functional utility of a gene/protein, we put a \ul{uniform prior} (mean prior probability $= 0.5$) for the corresponding selection probability parameter in the outcome model. This is illustrated by the leftmost point on the X axis of \Cref{fig:calibration_fun}A.
        
        \item If the mechanistic models provide us with \ul{strong evidence}, on the other hand, we put a \ul{strong prior} with a large prior mean on the selection probability parameter. This is illustrated by the increment in the Y axis of \Cref{fig:calibration_fun}A.
        
    \end{itemize}

    Specifically, we use a four-parameter logistic function (as shown in \Cref{fig:calibration_fun}A) for our applications and the details on choosing the calibration function are summarized in \Supp{SM 2.2}. For this specification, the prior mean probability of inclusion for a covariate ranges from $0.5$ (when the lBF is $0$ or negative) to approximately $1$ (when the lBF is large, say $> 5$). The calibrated prior mean probabilities for some representative points from these intervals are as follows, with respect to \Cref{fig:calibration_fun}B: 0.502 (lBF $= 0.25$, bottom-right), 0.543 (lBF $= 0.75$, top-right), 0.726 (lBF $= 1.5$, bottom-left), 0.962 (lBF $= 3$, top-left). This construction enables the functional evidence to inform selection/estimation in the outcome model, while retaining the flexibility to still allow the model to ignore prior evidence, if necessary, for the proteogenomic markers without any observed association with the outcome of interest, as are illustrated by our pan-cancer applications presented in \Cref{sec: pancancer}.
    
    \paragraph{Benefits and Utilities of Calibration} The benefits of this calibrated model formulation are multifold. First, by modifying the calibration function $\mathcal{F}$, our framework allows the user to incorporate any form of model summary information (such as z-scores, p-values, etc.) into the outcome model. Unlike existing grouped shrinkage-based procedures, this eliminates the restriction of only using external information through categorical covariates. Second, unlike the shrinkage-based or Bayesian approaches where the external covariates directly inform the final model, our model only relies on the supply of the summary statistics $s_{j}$ - rendering the computations less challenging and allowing the use of any upstream dataset, however large, in learning the functional information from multiple categorical/continuous sources. In the same vein, any such calibration function can easily be adapted to a scenario where multiple sources of evidence are available for a single covariate rather than just a single summary statistic $s_{j}$. For example, if there are $K$ lBFs ($\boldsymbol{\mathcal{E}}_j = (\textrm{lBF}_{1j},\ldots,\textrm{lBF}_{Kj})$) available from as many mechanistic models for the $j$th biomarker, then we can think of the calibration function as a composition: $\mathcal{F}(\boldsymbol{\mathcal{E}}_j) = \mathcal{F}_1(\mathcal{F}_2(\boldsymbol{\mathcal{E}}_j))$, where $\mathcal{F}_1$ can be similar to the $\mathcal{F}$ we use, and $\mathcal{F}_2: \mathbb{R}^E \rightarrow \mathbb{R}$ aggregates the multiple lines of evidence to a single scalar value $s_j$. One possible choice for this is a linear map, as $s_j = \mathcal{F}_2(\boldsymbol{\mathcal{E}}_j) = \sum_{k=1}^{K} \alpha_{kj} \textrm{lBF}_{kj}$. Here $\alpha_{kj}$s are convex weights specific to biomarker $j$, interpreted as quantifications of the importance of each source of evidence. Several choices of the $\alpha_{kj}$s are possible, as described below.

    \begin{enumerate}
    
    \item \textbf{Average evidence:} $\alpha_{kj} = 1/K$ (takes a simple average of all available evidences).
    
    \item \textbf{Maximal evidence:} $\alpha_{kj} = I\big(\textrm{lBF}_{kj} = \underset{k^\prime\in\{1,\ldots,K\}}{\max}\textrm{lBF}_{k^\prime j}\big)$ (only takes into account the strongest evidence available from any source).
    
    
    \item \textbf{Precision-weighted evidence:} $\alpha_{kj} = \rho_{kj}/\sum_{k=1}^{K}\rho_{kj}$ (weights the evidences by some metric of reliability of the evidences, such as $\rho_{kj} = \hat{\tau}^{-2}_{kj}$ where $\hat{\tau}^{2}_{kj}$ is the estimated noise variance for the source mechanistic model of $\textrm{lBF}_{kj}$).
    
    \end{enumerate}
	
	This provides an additional layer of flexibility that allows the user to choose how many sources of evidence to use and how best to combine them.
	
	\paragraph{Parameters of Interest and Model Dependence Structure} Having described the mechanistic and outcome layers of the modeling framework and established the connection between them via the evidence calibration function, we now revisit \Cref{eq: factorization} to interpret the conceptual parameters in context of our modeling scheme. For the mechanistic models, the quantities $\boldsymbol{\theta}_{F}$ represent the parameter vector for the Gaussian process models, namely, $\{(\tau_{lj}, \lambda_{lj}): l$ varies across each mechanistic model available for each gene/protein $j\}$. The quantification of evidence is not performed via direct estimation of these parameters but via computing and calibrating the Bayes factors corresponding to each model. Therefore, one component of $\boldsymbol{\theta}_{Y}$ is driven by $\boldsymbol{\theta}_{F}$ via the evidence calibration. The rest of $\boldsymbol{\theta}_{Y}$ is specified by $(\bbeta^T, \bgamma^T, \boldsymbol{\omega}^T, \sigma)^T$, i.e. the parameters of interest from the outcome model. The inferential task then is to fit the outcome model, estimate the parameters of interest $(\bbeta^T, \bgamma^T, \boldsymbol{\omega}^T, \sigma)^T$, and perform covariate selection based on the posterior distribution of $(\bgamma^T, \boldsymbol{\omega}^T)^T$. The overall dependence structures in the mechanistic and outcome model settings are summarized in Supplementary Figure 1 (denoted by SF henceforth). In the next section, we describe our model-fitting procedures.
	
	
	\paragraph{Generalizations to Non-Gaussian Outcomes} Such generalizations typically involve minimal changes to the outcome model. As an example, we discuss an extension to survival outcomes briefly. Let us denote $Y_{i} = \log T_{i}$ as the possibly unobserved log-survival time and $C_{i}$ as the possibly unobserved censoring time for individual $i$. The observed response for individual $i$ is $(Z_{i}, \delta_{i})$, where $Z_{i} = \min\{ \log T_{i}, \log C_{i} \}$ and $\delta_{i} = I( T_{i} < C_{i} )$. Under the assumptions that the censoring distribution is free of $\bbeta$ and $\bgamma$ and that the censoring times are independent of the true outcomes conditional on $\bbeta$ and $\bgamma$, changing the outcome model from Gaussian to an accelerated failure time model with log-Normal outcomes only introduces truncations of the censored (unobserved) outcomes in the log-posterior. Therefore, the only change in the model fitting procedure occurs in the expression of the full log-posterior, which now includes the terms corresponding to the truncations.
	
	\subsection{Model Fitting and Parameter Estimation} \label{subsec: outcome_model}
	
	\paragraph{Mechanistic Model Fitting} As described in \Cref{subsec:mechmodels}, we are interested in quantifying the functional evidence for each gene/protein via the GP-based mechanistic model using the lBF - therefore, we do not require a full Bayesian exploration of the model. To ensure computational efficiency, we directly compute the lBF following expressions as in \Cref{equation: BFexpression}. This requires the evaluation of an integral, which we perform numerically.
	
	\paragraph{Outcome Model Fitting} As described in the previous subsection, the complete set of parameters to be estimated by cBVS is $(\bbeta^T, \bgamma^T, \boldsymbol{\omega}^T, \sigma)^T$. $\bbeta^T$ provides estimates of the effect sizes of the proteogenomic covariates on the outcome, and the $(\bgamma^T, \boldsymbol{\omega}^T)^T$ guides their selection. The parameter estimation is focused on the posterior of  $(\bbeta^T, \bgamma^T, \boldsymbol{\omega}^T, \sigma)^T$. For large proteogenomic panels, this posterior will be computationally resource-intensive to directly sample from. The question, therefore, is of a trade-off between computational simplicity and estimation accuracy. Due to this reason, we offer three implementations of cBVS in increasing order of computational efficiency -- a standard Markov chain Monte Carlo (MCMC) using Gibbs sampler to sample from the complete posterior, a selection-only MCMC to sample from the marginal posterior of $\bgamma$, and an expectation-maximization based variable selection (EMVS) procedure to approximate the posterior modes of the parameters. Briefly, the selection-only MCMC focuses on estimating $\bgamma$ first and then estimates $\bbeta$ using a Bayesian model averaging-type procedure (\citealt{hinne2020conceptual}), and the EMVS sacrifices the full posterior along with error estimates to achieve fast point estimation (\citealt{rovckova2014emvs}). The exact details of each are in \Supp{SM 2.3}.

	\paragraph{Model Summaries} The MCMC/EMVS procedures provide us with posterior inclusion probabilities (PIPs) $\hat{\omega_j}$ and regression coefficient estimates $\hat{\beta_j}$ for each covariate in the model. A cut-off on the PIPs is computed using a false discovery rate adjustment procedure at a specified level of significance, treating the $1 - \hat{\omega_j}$ as p-value type quantities (\Supp{SM 2.4}).
	
	
	\section{Simulation Studies}\label{sec: simulation}
	
	To illustrate the utilities of cBVS, we performed two sets of simulation studies. \ul{Simulation 1} deals with continuous outcomes in synthetic datasets, comparing metrics of selection and estimation from the cBVS procedure with existing benchmarks across a class of variable selection methods. We include both penalized/grouped penalized selection procedures and Bayesian prior-based selection procedures as benchmarks. {Simulation 1} is expected to assess cBVS for both low and high sample size to dimension ratios and quantify the improvement and/or preservation of performance along this spectrum compared to the benchmarks. In \ul{Simulation 2}, the data generation procedure is informed by the patient datasets for breast invasive carcinoma from our application. The next two subsections describe the design, settings, and results from these two simulation studies.

	\subsection{Simulation 1: Synthetic data-based Simulations}
	
	\paragraph{Data Generation and Choices of True Effects} To compare performances across a grid of varying sample size/number of covariates ($n/p$) ratios, we fix the number of proteogenomic biomarkers generated in the simulated datasets at $p = 200$ and vary the sample size across $n = 50, 100, 200, 400, 800$, covering a range of $n/p = 1/4$ to $4$. For simplicity, we assume that there is only one upstream covariate for each biomarker.
    100 replicates are generated for each n. Briefly, we generate the upstream covariates (akin to copy number/methylation data) from the standard normal distribution. The proteogenomic covariates are generated such that there are 12 groups of size 5 each -- all possible combinations of effect size (low, medium, high) and prior functional evidence (no evidence, substantial, strong, decisive). The rest all have zero true effect sizes. The continuous outcome is generated from a normal distribution. Further details for each step are available in \Supp{SM 3.1}.

	\paragraph{Brief Overview of Benchmark Methods} We compare the performance of the cBVS model against three classes of methods that perform variable selection based on different approaches. We use one Bayesian variable selection method without any external evidence, namely, the expectation-maximization based variable selection (EMVS) (\citealt{rovckova2014emvs}) - as an uncalibrated counterpart to cBVS. We also include two penalized selection procedures - additional to LASSO (\citealt{tibshirani1996regression}), we include a grouped penalized selection procedure termed Integrative LASSO with Penalty Factors (IPF-LASSO) (\citealt{boulesteix2017ipf}). The reason behind including a grouped penalized regression procedure is that the prior evidence levels provide a natural grouping for the covariates to be used in a variable selection setting. Finally, we implement two recently proposed Bayesian variable selection methods incorporating external information, namely, graper (\citealt{velten2021adaptive}), which incorporates categorical external covariates only, and xtune (\citealt{zeng2021incorporating}), which can handle continuous covariates as well. Our simulation scheme (i) allows a comprehensive comparison of calibrated vs non-calibrated methods and (ii) provided a benchmark for calibration to evaluate cBVS.
	
	\paragraph{Summary of Metrics Used} Each method provides a coefficient estimate $\hat{\bbeta}$. For LASSO, we get a  $\hat{\bgamma}$ corresponding to the best-fit $\lambda$. For each of the other methods, we compute $\hat{\bomega}$, on which we use the FDR-based adjustment method as described in \Supp{SM 2.4} to infer selection. We compute several standard metrics of selection performance, namely, area under the receiver operating characteristic curve (AUC), scaled AUC between 0.8 to 1 specificity (AUC20), true positive rate (TPR), false positive rate (FPR), and Matthew's correlation coefficient (MCC). The use of MCC is particularly useful since at a specified selection threshold it aggregates performance summary from both the TPR and FPR metrics. Using AUC and AUC20 allows threshold-free evaluation of selection performance.
	
\begin{figure}[hbt!]
	    \centering
	    \includegraphics[scale = 0.45]{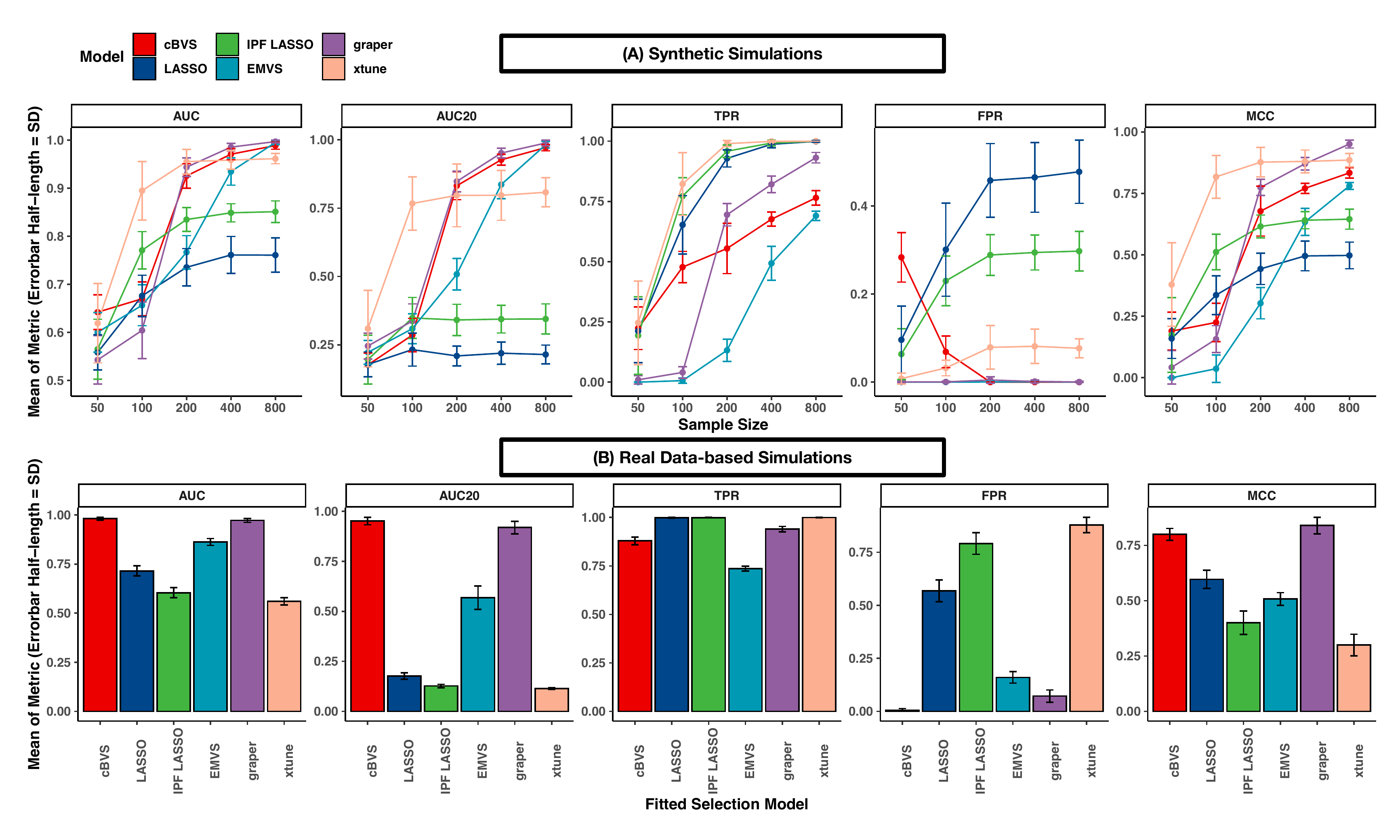}
	    \caption{\textbf{Summary of results from (A) synthetic and (B) real data-based simulations with continuous response.} For (A), 100 replicates were generated for each sample size. The number of covariates was 200, with 60 of them having non-zero true effect sizes, mixed across a grid of low, medium, and high values, each with equal proportions of lBFs at the levels no evidence, substantial, strong, and decisive. For (B), 100 replicates were generated as well. Covariate matrix, effect sizes, and calibrated evidence were picked from a calibrated model for BRCA (breast invasive carcinoma).}
	    \label{fig: sims}
	\end{figure}

	\paragraph{Results and Discussion} The results from {Simulation 1} are summarized in \Cref{fig: sims}A. In terms of AUC, our calibrated outcome regression model performs the best for the smallest $n/p$ ratio $= 1/4$, and is only second to graper by a very small margin for the $n/p$ ratios $= 2, 4$. In terms of MCC, our calibrated outcome regression model is only second to xtune for $n/p$ ratios $= 1/4, 1/2$, and is only behind graper and xtune for the $n/p$ ratios $= 1, 2$, and $4$. cBVS shows promise particularly in terms of preventing false positives, by exhibiting the lowest FPR among the method compared for all $n/p$ ratios $\geq 1$. In particular, cBVS has lower FPR than IPF-LASSO for all simulation scenarios other than $n/p = 1/4$, and comparable or lower FPR than the calibrated procedure xtune for the same scenarios as well. These summaries indicate several utilities of the calibrated outcome regression procedure. First of all, based on the metrics summarized, calibration based on prior evidence seems to have an evident benefit, as all the methods utilizing prior evidence (cBVS, graper, xtune) outperform those not incorporating any prior evidence (EMVS, LASSO, IPF-LASSO) across all the $n/p$ ratios used and across all metrics. Further, as evident from the AUC summaries, cBVS offers an improved preservation of selection performance for $n/p$ ratios $<1$ over the methods incorporating external covariates, while maintaining a comparable and satisfactory level of performance in the more favorable scenarios, i.e., $n/p >1$, as well. This is particularly reassuring, since unlike MCC or other standard metrics of selection, AUC is a threshold-free summary which enables us to sum up the performance of the methods across the whole spectrum of lenient to stringent selection criteria.
	
	\subsection{Simulation 2: Real Data-based Simulations}
	
	\paragraph{Data Generation and Choices of True Effects} Before the pan-cancer multi-platform proteogenomic analysis using data from the Cancer Genome Atlas, we perform a modified version of {Simulation 1}, based on breast invasive carcinoma (BRCA) patient data. We aggregate and annotate DNA methylation, copy number alteration, proteogenomic expression, and censored survival data for BRCA following \Cref{sec: pancancer}. Since the rest of the simulation scheme is analogous to {Simulation 1}, we only summarize the changes briefly. The upstream covariate data, proteogenomic covariate data, gene/protein evidence quantities (lBFs) are all taken from the BRCA patient data ($n = 790$ and $p = 365$). These lBFs are then grouped as before ($< 0.5$ (no evidence), $0.5 - 1$ (substantial), $1 - 2$ (strong), $> 2$ (decisive)). The true effect sizes are taken to be estimates from a survival cBVS model using the real data. These effect sizes are again clustered into four groups based on their quartiles, meaning that we now have $4 \times 4 = 16$ possible combinations. 
	The outcome data generation procedure, benchmark methods, summary metrics, and number of simulations remain the same as before. 
	
	\paragraph{Results and Discussion} The resulting metrics are summarized in \Cref{fig: sims}B. Notably, cBVS performs the best among all the methods compared in terms of AUC, AUC20 and FPR, and is at the second place, only behind graper, for MCC. Again, two calibrated selection methods (cBVS, graper) outperform uncalibrated methods (EMVS, LASSO, IPF-LASSO) in terms of all metrics but TPR. cBVS provides the lowest FPR among all, even lower than the FPRs for both the calibrated competitors - graper and xtune. Thus, cBVS is well-equipped for the real data scenarios to be encountered in the pan-cancer setting.
	
	\paragraph{Additional Simulations} We perform a simulation study to {compare the performances of GPs and linear models} in capturing {non-linear associations}.
    The generating models, simulation procedure and results are described in \Supp{SM 3.2}.
    Our results show that \ul{GPs are better equipped in identifying nonlinear evidence than linear models}. We also perform comparisons between model outputs from the MCMC and EMVS implementations to ensure EMVS provides estimates comparable to those from the MCMC. These details are available in \Supp{SM 3.3}. Our results reassure that although EMVS sacrifices quantifying uncertainty in lieu of faster computations, it \ul{does not compromise in terms of estimation accuracy}.
    
	\section{Integrative Pan-cancer Proteogenomic Analyses}\label{sec: pancancer}
	
	\subsection{Data Description, Preprocessing and Analysis}
	
	\paragraph{Pan-cancer Multi-omics Data} We analyze pan-cancer multi-platform proteogenomic data from the Cancer Genome Atlas (\href{https://www.cancer.gov/tcga}{TCGA}). We include 14 cancers across four cancer groups, classified by commonalities in the sites/tissues of occurrence. For each cancer, DNA methylation, copy number alteration, transcriptomic and proteomic expression data were obtained. The details on the cancer groups and types are available in SM 4.1. The codes used for data collation and cleaning along with the processed datasets are available in the code archive included in the submission. The sample size summaries for each cancer are available in the interactive shiny dashboard. The largest and the smallest sample sizes were available for breast ovarian carcinoma ($n = 1188$) and uterine carcinosarcoma ($n = 57$). We only used the genes corresponding to the proteins available in the TCGA reverse-phase protein array (RPPA) panel, resulting in a number of covariates generally between $350-400$ across cancers, depending on data availability and quality.

	\paragraph{Outcome Data} We investigate two different outcomes. We use the censored survival data available from TCGA and implement the cBVS model to identify cancer-specific proteogenomic biomarkers associated with survival. However, survival outcome alone does not provide biologically interpretable insights into the molecular progression of the disease. To this end, we use another outcome called mRNA-based stemness index (SI, in short) - a metric of cancer growth in terms of its cellular features. Briefly, SIs for TCGA cancers are computed using a one-class logistic regression model trained on pluripotent stem cell samples from the Progenitor Cell Biology Consortium dataset (\citealt{daily2017molecular, malta2018machine}). The SIs quantify the stem-cell-like behavior of the tumor of interest. We build cBVS models selecting proteogenomic biomarkers associated with SIs.
	
	\paragraph{Scientific Questions} Our analyses are driven by three broad scientific aims. First, we intend to \ul{identify cancer-specific and pan-cancer functional drivers and cascades} from the proteogenomic candidates using the functional evidence learned via the mechanistic models. Second, using the cBVS models, we \ul{select cancer-specific biomarkers associated with changes in the survival and stemness outcomes}. Finally, we assess the mechanistic and outcome model results in conjunction to \ul{evaluate the utility of calibrating outcome models using mechanistic evidence}. We present the numerical results in the following two subsections, followed by the biological interpretations and discussions of the results in \Cref{subsec: biological}.

    \paragraph{Modeling} For each cancer, the mechanistic model analyses for each gene and protein are performed using GPs as described in \Cref{subsec:mechmodels}, and lBFs are computed. As explained in the previous section, we have a single evidence quantity (the lBF from the driver gene model) for each gene and a maximum of two evidence quantities (lBFs from the driver and cascading protein models) for each protein.
    We use the maximal approach for evidence aggregation as described in \Cref{subsec:cBVS}.
    Two cBVS models - one using stemness and one using survival data - are built for each cancer, using the selection-only MCMC procedure as described in \Cref{subsec: outcome_model} to estimate $(\bbeta^T, \boldsymbol{\omega}^T, \sigma)^T$ in each. For each gene and protein, we thus obtain five quantities of interest: lBF, and $(\hat{\beta}_j, \hat{\omega}_j)^T$ - one each from each outcome model. An overview of the analysis pipeline is presented in \Supp{SF 4}.
    
	
	\subsection{Mechanistic Model Results for Pan-Cancer Groups}
	
\begin{sidewaysfigure}
	    \centering
	    \includegraphics[scale = 0.345]{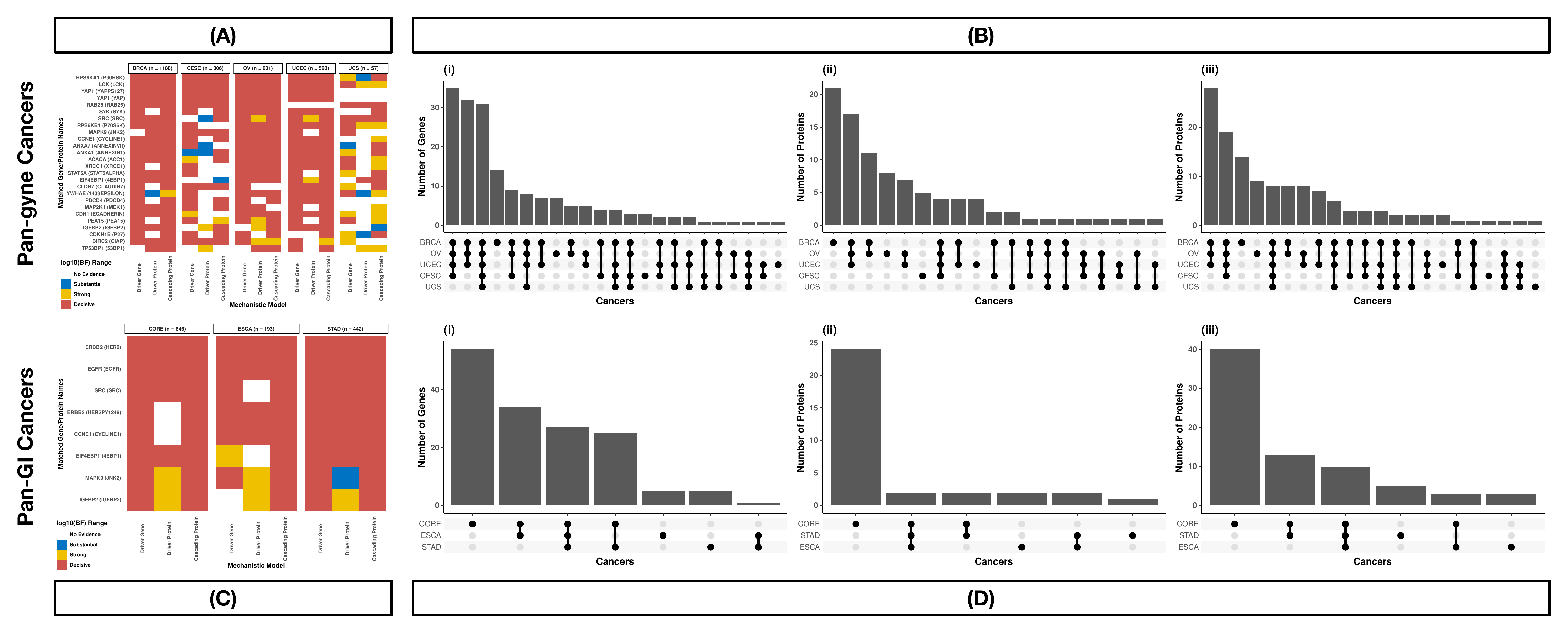}
	    \caption{\textbf{Summary of mechanistic model results} for the pan-gyne group cancers: BRCA (breast invasive carcinoma), CESC (cervical squamous cell carcinoma and endocervical adenocarcinoma), OV (ovarian serous cystadenocarcinoma), UCEC (uterine corpus endometrial carcinoma), UCS (uterine carcinosarcoma) (panels A and B), and the pan-GI group cancers: CORE (colon and rectum adenocarcinoma), ESCA (esophageal carcinoma), STAD (stomach adenocarcinoma) (panels C and D). (A/C) The lBF ranges are defined as: $< 0.5$ (no evidence), $0.5 - 1$ (substantial), $1 - 2$ (strong), $> 2$ (decisive). Only the gene-protein pairs with some evidence across at least 75\% of the cancers in at least two out of the three mechanistic model types are shown here. (B/D) Upset plots exhibiting the number of genes/proteins at the decisive level of significance for the (i) driver gene, (ii) driver protein, and (iii) cascading protein mechanistic models.}
	    \label{fig:mm_gyne_gi}
	\end{sidewaysfigure}

	We summarize the mechanistic model outputs across cancers for the pan-gyne and pan-GI cancer groups in \Cref{fig:mm_gyne_gi}; the rest of the groups are presented in \Supp{SF 5-10}. \Cref{fig:mm_gyne_gi}A, C exhibit heatmaps summarizing lBF classes for the gene-protein pairs which have some evidence across at least three-fourths of the cancers in at least two out of the three mechanistic model types, along with corresponding upset plots in \Cref{fig:mm_gyne_gi}B, D describing the number of genes/proteins that are at the decisive level of significance for the three mechanistic models across the intersections of the different cancers.
	
	For the pan-gyne cancers, 26 gene-protein pairs are at strong/decisive level of evidence across at least four cancers in at least two out of the three mechanistic model types, including genes such as RPS6KA1 (protein p90RSK), YAP1 (proteins YAP and YAPPS127), and DIABLO (protein SMAC) (\Cref{fig:mm_gyne_gi}A). The largest sharing of decisive driver gene signatures is observed across BRCA, CESC, OV, and UCEC, and that for cascading proteins is observed across BRCA, OV, and UCEC (\Cref{fig:mm_gyne_gi}B).
	
	For the pan-GI cancers, eight gene-protein pairs are at strong/decisive level of evidence across all three cancers in at least two out of the three mechanistic model types, including genes such as ERBB2 (proteins HER2 and HER2PY1248), CCNE1 (protein CYCLINE1), and MAPK9 (protein JNK2) (\Cref{fig:mm_gyne_gi}C). The largest number of decisive driver gene, driver protein, and cascading protein signatures is observed from CORE, and the largest numbers of pan-cancer signatures for the driver gene and the cascading protein models are observed between CORE and ESCA (\Cref{fig:mm_gyne_gi}D).
	
	\subsection{Stemness cBVS Results for TCGA Cancers}
	
\begin{figure}[hbt!]
	    \centering
	    \includegraphics[scale = 0.33]{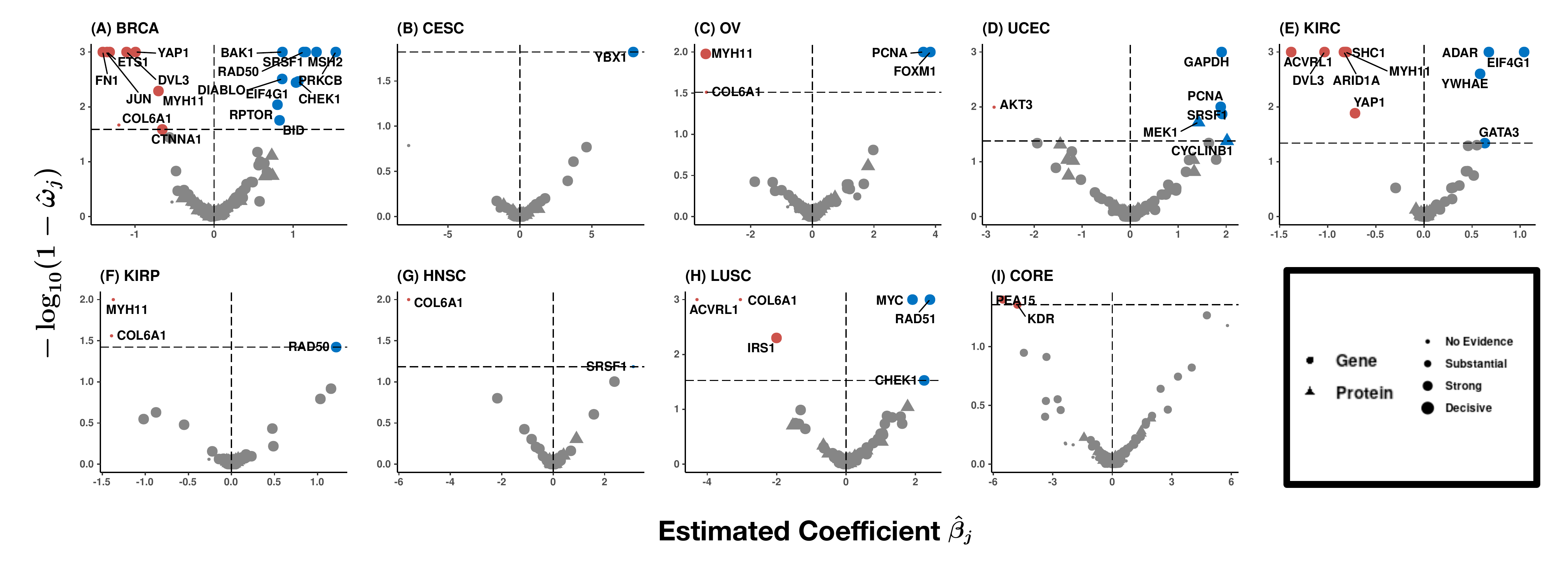}
        \caption{\textbf{Plots summarizing outcome model results based on stemness indices (SI) for TCGA cancers.} Proteins are represented by triangles and genes by circles. The shapes are colored red if the estimated $\hat{\beta}_j$ from fiBAG is negative, and green if positive. The x-axis shows the $\hat{\beta}_j$s, and the y-axis shows the $-\log_{10}(1 - \hat{\omega}_j)$s. An FDR check to adjust for multiple comparisons is performed treating $1 - \hat{\omega}_j$ as a p-value type quantity at the 10\% FDR level. Only the selected biomarkers are marked in non-gray colors and labeled. The sizes of the points are in the increasing order of evidence from the mechanistic models: lBF ranges are defined as: $< 0.5$ (no evidence), $0.5 - 1$ (substantial), $1 - 2$ (strong), $> 2$ (decisive).}
	    \label{fig:om_si}
	\end{figure}

	We summarize the stemness outcome model outputs for each TCGA cancer using plots showing $-\log_{10}(1 - \hat{\omega})$ on the y-axis and $\hat{\beta}$ on the x-axis for each cancer in \Cref{fig:om_si} and \Supp{SF 11-15}. Further, we present bar diagrams of the lBFs for the selected genes/proteins and histograms with density plots of lBFs for the predictors not selected in our shiny app. Here, we discuss the results from the BRCA and CORE analyses since they are the cancers with the largest sample sizes in the pan-gyne and pan-GI groups, respectively.
	
	For BRCA, several genes are selected, such as YAP1, DIABLO, and RAPTOR (\Cref{fig:om_si}A); however, no proteins are selected using the 10\% FDR cut-off on $\hat{\bomega}$ from the stemness cBVS model. The genes selected cover a large span of evidence range from the mechanistic models, with genes like JUN and COL6A1 having no functional evidence and DVL3, MYH11, EIF4G1 at decisive evidence (\Supp{SF 16}), indicating that {cBVS is capable of identifying associations even in the absence of prior evidence}, as has been noted in the simulation studies before too (\Cref{sec: simulation}). The histogram indicates that a vast majority of the non-selected genes and proteins have little to no evidence from the mechanistic models, but 194 of them are at the strong and decisive levels of evidence and yet not selected by cBVS due to the absence of sufficient association with the outcome data (\Supp{SF 17}).
	
	For CORE, two genes - PEA15 and KDR are selected in the stemness cBVS model, both negatively associated with the stemness index as can be seen from the sign of the estimated regression coefficients (\Cref{fig:om_si}I). Both the genes selected are at the decisive level of functional evidence (\Supp{SF 18}). The histogram again indicates that a vast majority of the non-selected genes and proteins have little to no evidence from the mechanistic models - however, 179 of them are at the strong/decisive levels of evidence and are still not selected by cBVS (\Supp{SF 19}).  The survival cBVS results are discussed in \Supp{SM 4.2}.
 
	\subsection{Biological Findings and Implications}\label{subsec: biological} 

    While a majority of the proteogenomic biomarkers identified in our analyses have supporting evidence from past literature in terms of their mechanisms, we further delineate their functional roles in driving cancer progression and patient survival. The detailed summary of these biomarkers, along with relevant references from the literature is presented in \Supp{SM 4.3}. All associations discussed below are significant at a 10\% level of FDR control.
	
	\paragraph{RPS6KA1 Gene and p90RSK Kinases in Gynecological Cancer Progression} The gene RPS6KA1 (protein p90RSK) has decisive evidence for all but two mechanistic models in the pan-gyne cancers (\Cref{fig:mm_gyne_gi}A).
    The p90RSK protein is selected in the survival cBVS model for OV (\Supp{SF 22}). Both RPS6KA1 and p90RSK have previously shown significant association with risks of and outcomes in gynecological cancers (\Supp{SM 4.3}).
	
	\paragraph{YAP1 Gene and YAP Proteins in Gynecological Malignancies} The gene YAP1 (proteins YAP, YAPPS127) has decisive evidence in all mechanistic models for all gynecological cancers except UCS (smallest sample size in the group, $n = 57$) (\Cref{fig:mm_gyne_gi}A).
    Further, YAP1 is negatively associated with stemness for BRCA (\Cref{fig:om_si}A) and with survival for OV and CESC (\Supp{SF 21-22}). Over-activation of YAP is a known cause of malignant transformation in gynecological cancers, associated with poor prognosis (\Supp{SM 4.3}).
	
	\paragraph{DIABLO Gene as a Marker of Gynecological Tumors}
    We identify higher DIABLO expression to be associated with higher tumor stemness for BRCA and UCS (\Cref{fig:om_si}A and \Supp{SF 11}),
    and higher survival for another (OV, \Supp{SF 22}).
    While neither DIABLO nor its corresponding protein Smac are identified as top candidates in the mechanistic analyses, the findings of the cBVS model are in line with existing literature, where DIABLO has been proposed as a biomarker for breast and uterine cancers (\Supp{SM 4.3}).
	
	\paragraph{HER2 as a Therapeutic Target in Gastrointestinal Cancers} The gene ERBB2 (protein HER2) has decisive evidence for all mechanistic models in the pan-GI group, emerging as the top candidate (\Cref{fig:mm_gyne_gi}C).
    In our cBVS analyses, both ERBB2/HER2 are positively associated with stemness for ESCA and STAD (\Supp{SF 14-15}).
    Past knowledge on HER2 identifies its increased expression as an agent of poorer progression of gastric cancer, and it has been considered as a therapeutic target, in line with our findings (\Supp{SM 4.3}).
	
	
	\section{Discussion and Future Work}\label{sec: discussion}
	
	We propose fiBAG, a hierarchical Bayesian framework to perform integration of multi-omics data and outcomes. Using GPs, we quantify the functional roles of proteogenomic biomarkers along three axes of interest (driver gene, driver protein, cascading protein). Then, using a Bayesian variable selection procedure with a calibrated spike-and-slab prior, we incorporate this functional evidence in the outcome model to improve covariate selection and effect size estimation. The framework offers novelty and utility in multiple directions. First, it offers the user liberty in terms of multi-platform integration, in the sense that depending on the data available, the mechanistic and outcome layers can be appended/modified via additions/alterations of the complete set of parameters of interest. For example, we use DNA methylation and copy number alterations as upstream covariates, but other information such as microRNAs/epigenetic factors (e.g. histone modifications) could easily be incorporated. Second, the calibrated spike-and-slab prior addresses the more general statistical question of incorporating external information in a variable selection setting - by updating the mapping function used to calibrate evidences to a prior probability scale, the procedure can be adapted to settings where the numerical evidences are in a different scale and/or sourced from different models. cBVS is compared with multiple benchmarks via simulation studies in synthetic and real data-based settings. The benchmarks include uncalibrated Bayesian variable selection (EMVS), standard and grouped penalized regression (LASSO and IPF-LASSO), and Bayesian variable selection incorporating external information (graper and xtune). We use selection metrics like AUC and MCC to evaluate performances across a broad spectrum of $n/p$ ratios. cBVS outperforms all the uncalibrated methods across all $n/p$ settings, and performs comparably with the calibrated methods in the high $n/p$ settings. For low $n/p$, cBVS offers improvement in selection compared to other calibrated methods (\Cref{fig: sims}).
	
	Having established the improved utility of our method in an evidence-based setting, we analyze pan-cancer multiomics data from TCGA, across a total of four cancer groups (pan-gyne, pan-kidney, pan-squamous and pan-GI) and 14 cancers. Our real data analyses identify both known and novel associations at cancer-specific and pan-cancer levels, such as the identification of the roles of RPS6KA1 gene and p90RSK kinases in progression of gynecological cancers, and the potential utility of the EGFR gene and protein as therapeutic targets for ESCA where their expressions are negatively associated with survival outcomes. Our framework is cancer-specific (each mechanistic and outcome model is built for each cancer separately and then assessed in a pan-cancer fashion) for several reasons. First of all, non-genomic covariates ($\mathbf{B}$ - e.g., demographic information such as gender or age, and clinical information such as cancer stage) would potentially be entirely different across cancers, and may have widely different scales of measure. Further, even the outcomes may not always be normalized across cancers - while a scaled outcome like the mRNAsi does not pose this problem, survival, for example, may differ considerably across different cancers. Finally, the mechanistic models need to be run separately for each gene and protein, and depending on the cancer of interest, {the functional roles of these genes and proteins are potentially different}, effectively rendering the measures of evidence to be cancer-specific as well. However, our qualitative summarizations across relevant pan-cancer groups provide valuable insights into the shared biology among these cancers.

    The flexible covariate-specific mechanistic modelling approach offers the feasibility of utilizing platforms with larger expression pools and deeper sequencing panels such as the National Cancer Institute Clinical Proteomic Tumor Analysis Consortium (\citealp{CPTAC}). Further, more cancers or cancer groups can be included in the analysis pipeline as well, since the cBVS model fitting is cancer-specific and we offer three options in decreasing order of computational complexity (full MCMC, selection-only MCMC, and EMVS). All these features make the whole procedure highly parallelizable (covariate level for mechanistic models and cancer level for outcome models) - the computation times for the simulation and real data settings described in this paper reassure the utility of our pipeline in this direction. The median computation times, rounded to the nearest whole time unit, for each procedure in our integrative analyses are as follows: 2 minutes(mechanistic model), 5 hours (cBVS full MCMC implementation), 32 minutes (cBVS selection-only MCMC implementation), 4 minutes (cBVS EMVS implementation). All computations were performed on an online cluster with each node having 16G memory.
	
	An important novelty of our integrative analyses is the use of mRNA-based cancer stemness indices as outcomes. Indeed, tumor stemness is a pertinent determinant of cancer growth and prognostic outcomes, and only recently there have been efforts to quantify stemness based on cellular signatures (\citealt{malta2018machine}). To the best of our knowledge, our study is the first to look at potential associations of stemness with a pool of proteogenomic biomarkers in an oncological context. A relevant question, however, would be whether the associations identified are meaningful or artifacts of the construction of the index. We further investigated the stemness indices and the inferred associations to confirm the veracity of our findings.
	A summary of these investigations is presented in \Supp{SM 4.4}.
	
	This work indicates a number of potential avenues for future statistical research. First, the flexibility in choosing the calibration function suggests that the function can even potentially be data-driven (such as choosing the tuning parameters via some pre-hoc analysis of the data). The cBVS framework can also be adapted to settings slightly different from ours - such as modeling other types of outcomes by using a nonlinear link function for the mean of the outcome model, or using bivariate outcomes and so on.
	
	To offer seamless interactive visualization of our integrative analysis results, we have built an R Shiny app, the functionalities of which are presented in \Supp{SM 5}. All our model and figure codes will be made available publicly via the app. We  hope that fiBAG will be a  significant contribution to current armamentarium of integrative models for multiomic data and can be profitably employed to enhance prognostic/therapeutic utility in oncological treatment and research as well as other disease domains. 
    
    \bibliography{Bibs}

    \newpage

    \begin{center}
        {\Large \textbf{Supplementary Materials}}
    \end{center}

    \setcounter{section}{0}

\section{fiBAG Method}\label{sec: fiBAG}

\subsection{Bayes Factor Expression for Gaussian Process Mechanistic Models}

We describe the Bayes factor derivation for the Gaussian process specification corresponding to the driver gene mechanistic model in this section. The expressions for the other models can be derived similarly. To recall the notations introduced, main manuscript, the $j^{\textrm{th}}$ driver gene model is written as $G_{ij} = f_{1j}((\mathbf{C}_{ij}^T, \mathbf{M}_{ij}^T)^T) + e_{1ij}$, where we assume $e_{1ij} \overset{\textrm{iid}}{\sim} \textrm{N}(0, \tau_{1j}^2)$. Let us also denote $f_{1j}^{(i)} = f_{1j}((\mathbf{C}_{ij}^T, \mathbf{M}_{ij}^T)^T)$ for all $i$. Then, the GP prior on $f_{1j}$ is placed as follows:
    \vspace{-12pt}
	\small
    \begin{eqnarray}
	   \textbf{GP prior:} &&(f_{1j}^{(1)}, \ldots, f_{1j}^{(n)})^T \sim \mathbf{N}(\mathbf{0}, \mathbf{K}_{1j}), \nonumber \\
    \textbf{Covariance matrix:} &&\mathbf{K}_{1j(i,k)} = K_{1j}((\mathbf{C}_{ij}^T, \mathbf{M}_{ij}^T)^T, (\mathbf{C}_{kj}^T, \mathbf{M}_{kj}^T)^T), \nonumber \\
	    \textbf{Kernel function:} &&K_{1j}(\mathbf{u}, \mathbf{v}) = g\tau_{1j}^2\textrm{exp}\big( -\frac{||\mathbf{u} - \mathbf{v}||^2}{\lambda_{1j}^2} \big). \nonumber
	\end{eqnarray}
	\normalsize
	\vspace{-6pt}
    The hyperpriors specify $\tau_{1j}^2 \sim \textrm{Inverse-Gamma}(\frac{\nu_{01j}}{2}, \frac{\nu_{01j}\tau_{01j}^2}{2})$ and $\lambda_{1j} \sim \textrm{exp}(\lambda_{01j})$. For all our analyses, we set $g = n$. To test the hypothesis of interest in this setting, we need to compare this model with an intercept model, specified by $G_{ij} = \mu_j + e_{1ij}$. We assume a Zellner's g-prior on $\mu_j$ as $N\left(0, \tau_{1j}^{2} \frac{g}{1+g}\right)$. We use $g = n$ for all our models. Under this specification, the marginal likelihood is the following.

\vspace{-36pt}

\begin{eqnarray}
&& \int_0^{\infty} \int_{-\infty}^{\infty} \textrm{P}(\mathbf{G}_{\bullet j} | \mu_j, \tau_{1j}^2) \textrm{P}(\mu_j | g) \textrm{P}(\tau_{1j}^2 | \nu_{01j}, \tau_{01j}) d\mu_j d\tau_{1j}^2\  \nonumber\\
&=& \int_0^{\infty} \int_{-\infty}^{\infty} (2\pi)^{-n/2} (\tau_{1j}^2)^{-n/2} \exp(-\sum_{i=1}^{n}(G_{ij}-\mu_j)^2 / 2\tau_{1j}^2) . (2\pi)^{-1/2} (\tau_{1j}^2\frac{g}{1+g})^{-1/2} \nonumber\\
&& \exp(-\mu_j^2 / (2\tau_{1j}^2\frac{g}{1+g})) . \frac{1}{\Gamma(\nu_{01j}/2)} (\nu_{01j}\tau_{01j}^2/2)^{\nu_{01j}/2} (\tau_{01j}^2)^{-\nu_{01j}/2-1} \exp(-\nu_{01j}\tau_{01j}^2/2\tau_{1j}^2) d\mu_j d\tau_{1j}^2 \nonumber\\
&=& C . \int_0^{\infty} (\tau_{1j}^2)^{-(n+\nu_{01j}+3)/2} \exp(-\nu_{01j}\tau_{01j}^2/2\tau_{1j}^2) \int_{-\infty}^{\infty} \exp\big[-\big\{(n - \frac{1+g}{g})\mu_j^2 - 2\mu_j\sum_{i=1}^{n}G_{ij}\big\} / 2\tau_{1j}^2\big] d\mu_j \nonumber\\
&& \exp(-\sum_{i=1}^{n}G_{ij}^2 / 2\tau_{1j}^2) d\tau_{1j}^2 \nonumber\\
&=& C . \int_0^{\infty} (\tau_{1j}^2)^{-(n+\nu_{01j}+3)/2} \exp\big\{-(\nu_{01j}\tau_{01j}^2 + \sum_{i=1}^{n}G_{ij}^2)/2\tau_{1j}^2\big\} \sqrt{2\tau_{1j}^2\pi/(n-1-1/g)} \nonumber\\
&& \exp\big[(\sum_{i=1}^{n}G_{ij} / 2\tau_{1j}^2)^2 \big\{2\tau_{1j}^2 / (n - 1 - 1/g)\big\}\big] d\tau_{1j}^2 \nonumber\\
&=& C^{'} . \int_0^{\infty} (\tau_{1j}^2)^{-(n+\nu_{01j})/2 - 1} \exp(-\frac{\nu_{01j}\tau_{01j}^2 + \sum_{i=1}^{n}G_{ij}^2 - (\sum_{i=1}^{n}G_{ij})^2 / (n-1-1/g)}{2\tau_{1j}^2}) d\tau_{1j}^2 \nonumber\\
&=& C^{'} . \Gamma((n+\nu_{01j})/2) . \big\{\frac{\nu_{01j}\tau_{01j}^2 + \sum_{i=1}^{n}G_{ij}^2 - (\sum_{i=1}^{n}G_{ij})^2 / (n-1-1/g)}{2}\big\} ^ {\big\{-(n+\nu_{01j})/2\big\}}. \nonumber\\
\nonumber
\end{eqnarray}

\vspace{-36pt}

Then, for the intercept model, the expression for the log-marginal likelihood simplifies to the following: $a^{'}_n + b_n\ln\Big( a+\sum_{i=1}^{n}G_{ij}^2-\frac{(\sum_{i=1}^{n}G_{ij})^2}{c_n} \Big)$, where $a^{'}_n, b_n, c_n$, and $a$ are constants dependent on data dimensions and hyperparameters. Following the same procedures, we can derive the log-marginal likelihood for the Gaussian process model. The only difference in the calculation above occurs due to the fact that the joint distribution of $\mathbf{G}_{\bullet j}$ does not have a diagonal covariance matrix anymore. Therefore, the terms involving the data inside the double integral does not anymore simplify to functions of $\sum_{i=1}^{n}G_{ij}^2$ and $(\sum_{i=1}^{n}G_{ij})^2$. The integral with respect to $d\tau_{1j}^2$ can still be simplified as before, but the one involving $d\lambda_{01j}$ does not have a closed-form solution. The final expression for the log-marginal likelihood is then: $a^{''}_n + \ln\int_0^{\infty} \exp(-\lambda_{01j}\lambda_{1j})\frac{2^{b_n}|\mathbf{K}_{1j}/\tau_{1j}^2+I|^{-\frac{1}{2}}}{\big\{\mathbf{G}_{\bullet j}^{T}(\mathbf{K}_{1j}/\tau_{1j}^2+I)\mathbf{G}_{\bullet j}+a)\big\}^{b_n}} d\lambda_{1j}$. Subtracting the previous log-marginal likelihood from this expression provides us the expression of the log-marginal likelihood as presented, main manuscript.

\subsection{Choice of the Evidence Calibration Function}

We perform the following steps to obtain the shape parameters of the final Beta prior as $\mathcal{F}(\boldsymbol{\mathcal{E}}_j) = \mathcal{F}(s_j)$ and $1/\mathcal{F}(\boldsymbol{\mathcal{E}}_j) = 1/\mathcal{F}(s_j)$. Note that due to this formulation, the prior mean for $\omega_j$ takes the form $\mathcal{F}(s_j)\big\{\mathcal{F}(s_j) + 1/\mathcal{F}(s_j)\big\}^{-1}$, and the corresponding variance is $\big[\big\{\mathcal{F}(s_j) + 1/\mathcal{F}(s_j)\big\}^2\big\{\mathcal{F}(s_j) + 1/\mathcal{F}(s_j) + 1\big\}\big]^{-1}$. We first ensure that all small positive and non-positive lBFs from the mechanistic model are effectively truncated to zero (no evidence, to be mapped to a uniform prior), by setting $s^*_j = \max(s_j, 10^{-6})$. Then, our mapping function evaluates $\mathcal{G}(s_j) = \frac{1}{2}\big[\big\{1 + (s^*_j/3)^{-2.75}\big\}^{-1} + 1\big]$. Finally, we compute $\mathcal{F}(s_j) = (2\mathcal{G}(s_j))^4 = 16\mathcal{G}^4(s_j)$ to ensure sharp increase in prior mean probability of inclusion when lBF increases from 1 to 2. Effectively, $\mathcal{F}(\bullet)$ is a composition of three functions.

\begin{itemize}

    \item A four-parameter logistic ($ s^*_j \rightarrow \frac{1}{1 + (s^*_j/3)^{-2.75}}$) allowing us to control the shape and scale of the prior probability calibration curve along with its asymptotic behavior and lower bound.
    
    \item A linear map ($x \rightarrow \frac{1}{2}(x - 1)$) that takes $[0, 1]$ to $[0.5, 1]$ enabling us to make the prior noninformative when the lBF is small or negative.
    
    \item A polynomial map ($c \rightarrow (2c)^4$) allowing additional control over the steepness of the resulting prior probability calibrated means.

\end{itemize}

The tuning parameters with values 3 and 2.75 are chosen via computational checks across ranges of $[2, 4]$ and $[2, 3]$, respectively. For the specification chosen, the prior mean probability of inclusion for a covariate ranges from $0.5$ (when the lBF is $0$ or negative) to approximately $1$ (when the lBF is large, say $> 5$).

\subsection{Outcome Model Fitting Procedures}

As described in the main manuscript, the complete set of parameters to be estimated by cBVS is $(\boldsymbol{\beta}^T, \boldsymbol{\gamma}^T, \boldsymbol{\omega}^T, \sigma)^T$. $\boldsymbol{\beta}^T$ provides estimates of the effect sizes of the proteogenomic covariates on the outcome, and the $(\boldsymbol{\gamma}^T, \boldsymbol{\omega}^T)^T$ guides their selection. The parameter estimation is focused on the posterior of  $(\boldsymbol{\beta}^T, \boldsymbol{\gamma}^T, \boldsymbol{\omega}^T, \sigma)^T$. For large proteogenomic panels, this posterior will be computationally resource-intensive to directly sample from. The question, therefore, is of a trade-off between computational simplicity and estimation accuracy. Due to this reason, we offer three implementations of cBVS in increasing order of computational efficiency -- a standard Markov chain Monte Carlo (MCMC) using Gibbs sampler to sample from the complete posterior, a selection-only MCMC to sample from the marginal posterior of $\boldsymbol{\gamma}$, and an expectation-maximization based variable selection (EMVS) procedure to approximate the posterior modes of the parameters. Briefly, the selection-only MCMC focuses on estimating $\boldsymbol{\gamma}$ first and then estimates $\boldsymbol{\beta}$ using a Bayesian model averaging-type procedure (\citealt{hinne2020conceptual}), and the EMVS sacrifices the full posterior along with error estimates to achieve fast point estimation (\citealt{rovckova2014emvs}). The exact details of each are as follows.

\begin{enumerate}

    \item \textbf{Full MCMC:} The simplest approach is to perform a complete Markov chain Monte Carlo (MCMC) procedure to sample from the joint posterior -- we utilize a Gibbs sampler for this. For both continuous and survival outcomes, we use the \texttt{rjags} (\citealt{plummer2016package}) package in R. The rjags model descriptions are available in the code archive included in this submission.
    
    \item \textbf{Selection-only MCMC:} One way to mitigate the time- and resource-intensiveness of the full MCMC is to focus on $\boldsymbol{\gamma}$ only - by integrating $(\boldsymbol{\beta}^T, \boldsymbol{\omega}^T, \sigma)^T$ out of the joint posterior. This results in $\Pi(\boldsymbol{\gamma} | \textrm{Data, Hyperparameters})$, to be then approximated via MCMC. The final estimates for $\boldsymbol{\omega}$ are computed by taking the average of the traversed MCMC path for $\boldsymbol{\gamma}$ post burn-in. The components of $\boldsymbol{\beta}$ are estimated using a Bayesian model averaging-type procedure (\citealt{hinne2020conceptual}), taking a convex combination of draws from their conditional posterior at every step, weighted proportionally to the negative log posterior evaluated at that step. Since this MCMC only performs a search across a lattice space of size $2^p$, it results in significant improvements in computation times. The overview of the selection-only MCMC procedure is presented in Supplementary Algorithm 1.1. The codes are available in the code archive included in this submission.
    
    \item \textbf{E-M based variable selection (EMVS):} The fastest alternative to a full MCMC is to sacrifice approximating the full posterior (along with error margins) and to only focus on point estimates of the parameters of interest. For this purpose, we adapt the EMVS procedure by \cite{rovckova2014emvs} for our continuous and survival settings. The EMVS procedure estimates the posterior mode instead of approximating the full posterior, resulting in faster iterations. The model codes for the implementation of EMVS in our setting are available in the code archive included in this submission.

\end{enumerate}

\begin{tcolorbox}[colframe = blue!36, colback = white, coltitle = black, title = Supplementary Algorithm 1.1. Selection-only MCMC algorithm] \label{alg: fiBAG_SMCMC}

\begin{enumerate}

\setcounter{enumi}{-1}

    \item Initiate the MCMC with $\boldsymbol{\gamma} = \boldsymbol{\gamma}_{\textrm{old}}$, where $\gamma_{\textrm{old,}j} \sim Ber(\frac{a(s_j)}{a(s_j) + \frac{1}{a(s_j)}}), \forall j$.
    
    \item With probability $1/3$ each, select one of the following steps and perform the corresponding task.
    
    \begin{enumerate}
    
        \item \textbf{Add:} Randomly pick one of the $\boldsymbol{\gamma}_{\textrm{old}}$ indices with value $0$, change it to $1$, and call the resulting vector $\boldsymbol{\gamma}_{\textrm{new}}$. If there is no index $j$ where $\gamma_{\textrm{old,}j} = 0$, perform 1.2.
        
        \item \textbf{Delete:} Randomly pick one of the $\boldsymbol{\gamma}_{\textrm{old}}$ indices with value $1$, change it to $0$, and call the resulting vector $\boldsymbol{\gamma}_{\textrm{new}}$. If there is no index $j$ where $\gamma_{\textrm{old,}j} = 1$, perform 1.1.
        
        \item \textbf{Swap:} Randomly pick one of the $\boldsymbol{\gamma}_{\textrm{old}}$ indices with value $0$, and independently, randomly pick one of the $\boldsymbol{\gamma}_{\textrm{old}}$ indices with value $1$. Swap the values of these two indices, and call the resulting vector $\boldsymbol{\gamma}_{\textrm{new}}$. If there is no index $j$ where $\gamma_{\textrm{old,}j} = 1$, perform 1.1. If there is no index $j$ where $\gamma_{\textrm{old,}j} = 0$, perform 1.2.
    
    \end{enumerate}
    
    \item Compute $\log(p^*) := \min\{0, \log(\Pi(\boldsymbol{\gamma}_{\textrm{new}} | .)) - \log(\Pi(\boldsymbol{\gamma}_{\textrm{old}} | .))\}$. The proposed $\boldsymbol{\gamma}_{\textrm{new}}$ is then accepted with probability $p^*$.
    
    \item Iterate between 1-2 until convergence.

\end{enumerate}

\end{tcolorbox}

\subsection{FDR Control for Variable Selection using Posterior Inclusion Probabilities}

The MCMC/EMVS procedures provide us with posterior inclusion probabilities (PIPs) $\hat{\omega_j}$ and regression coefficient estimates $\hat{\beta_j}$ for each covariate in the model. A cut-off on the PIPs is computed using a false discovery rate adjustment procedure at a specified level of significance, treating the $1 - \hat{\omega_j}$ as p-value type quantities. Suppose the estimated posterior probabilities of inclusion are denoted by $\{\hat{\omega}_j, j \in \{1, \ldots, q_g + q _p\}\}$. Then, we define p-value type quantities $p_j = 1 - \hat{\omega}_j, \forall j \in \{1, \ldots, q_g + q _p\}$, and sort them in the increasing order of magnitude as $\{p^*_j, j \in \{1, \ldots, q_g + q _p\}\}$, with the understanding that for each $j$, $p^*_j = p_{k_j}$ for some $k_j \in \{1, \ldots, q_g + q _p\}$. Let the cumulative sums of these ordered quantities be denoted by $r_j = \sum_{l = 1}^{j}p^*_j$ for each $j$. Let $j^* := \min\{j: r_j \geq \alpha\}$, where $\alpha$ is a pre-specified level for the false discovery rate control. Then we infer that the covariates with indices $k_1, \ldots, k_{j^*}$ are selected in the outcome model.

\section{Simulation Studies}\label{sec: simulations}

\subsection{Data Generation for Simulation 1}

To compare performances across a grid of varying sample size/number of covariates ($n/p$) ratios, we fix the number of proteogenomic biomarkers generated in the simulated datasets at $p = 200$ and vary the sample size across $n = 50, 100, 200, 400, 800$, covering a range of $n/p = 1/4$ to $4$. For simplicity, we assume that there is only one upstream covariate for each biomarker. 100 replicates are generated for each n. We follow the steps described below to generate one such replicate.

\begin{itemize}

    \item Upstream covariates $\boldsymbol{U}_{n \times p}$ (comparable to copy number/methylation data in the biological setting) are generated such that each $U_{ij} \overset{\textrm{iid}}{\sim} \textrm{N}(0, 1)$.
    
    \item Let $\boldsymbol{X}_{n \times p}$ denote the design matrix of the proteogenomic expression data for the outcome model. Then the generating model for the $j^{\textrm{th}}$ expression is $X_{ij} \overset{\textrm{ind}}{\sim} \textrm{N}(\xi_j U_{ij}, 1)$, $\forall i \in \{1, \ldots, n\}$. The mechanistic effect size $\xi_j$ controls the level of evidence reflected by an lBF for the mechanistic model of $\boldsymbol{X}_{\bullet j}$ on $\boldsymbol{U}_{\bullet j}$. The correspondence between values of $\xi_j$ and the four levels of $\textrm{lBF}$: no evidence, substantial, strong, and decisive, is established numerically.
    
    Among the 200 $\boldsymbol{X}_{\bullet j}$s, the first 60 are distributed into four groups of 15 each, in the order of no evidence, substantial, strong, and decisive, followed by two groups of five at the levels strong and decisive, respectively. Those first 60 are distributed further into varying outcome effect sizes in the later steps to cover the complete range of combinations of prior evidence and effect size. The latter two groups will be assigned no true effects to include an in-built checkpoint for false positive evidences.
    
    \item Among each group of 15 for the four levels of prior functional evidence, we put three groups of five $\boldsymbol{X}_{\bullet j}$s with effect sizes $\beta_j$s generated respectively at the low ($U(0, 0.2)$), medium ($U(0.4, 0.6)$), and high ($U(0.9, 1.1)$) levels. All the other $\beta_j$s are set $= 0$.
    
    This results in 12 groups of covariates of size five each, because there are 12 possible combinations from the three levels of effect sizes and the four levels of evidences. Additionally, we have two groups of size five each with strong/decisive level of evidence but no true effect. All other 130 covariates have no evidence and no true effect.
    
    \item Finally, we generate the continuous outcome data as $\mathbf{Y}_{i} \overset{\textrm{ind}}{\sim} \mathcal{N}_p(\boldsymbol{\beta}^T \mathbf{X}_{i \bullet}, 1)$.

\end{itemize}

\subsection{Comparing Linear Models and Gaussian Processes in Capturing Non-linear Evidence}

We perform a simple simulation study to compare the performances of Gaussian process models and linear models in capturing significance based on data generated from varying orders of non-linear associations. Hypothetically, with increasing proportion of non-linearity in the generating model, the Bayes factors from the Gaussian process models should be better able to quantify significance than those from the linear models. To perform this, our generating models are defined as the following.

\begin{itemize}

    \item For a given sample size $n = 100$ and number of covariates $p = 5$, we generate $\mathbf{X}_{n \times p}$ where each $X_{ij} \overset{iid}{\sim} U(0, 1)$.
    
    \item The fully nonlinear model is defined as the following.
    
    $Y_i \sim N(10\cos(X_{i1}) - 15X_{i2}^2 + 10\exp(-X_{i3})X_{i2} - 8\sin(X_{i3})\cos(X_{i4}) + 20X_{i1}X_{i5}, 1)$,
    i.e., $Y_i \sim N(\sum_{j = 1}^p \beta_j f_{ij}, 1)$. For $l \in \{0, 1, \ldots, 5\}$, a $20l\%$ nonlinear model is then generated from the following distribution.
    
    $Y_i \sim N(\sum_{j = 1}^l \beta_j f_{ij} + \sum_{j = l + 1}^p \beta_j X_{ij}, 1)$.
    
    \item For each non-linearity scenario, 100 independent replicates are generated. For each replicate, the Bayes factors corresponding to the Gaussian process model (as described in Section 2.2, main manuscript) and the linear model are computed.

\end{itemize}

As seen in Supplementary \Cref{fig: Sim_GPLM}, for 0\% and 20\% levels of non-linearity, the median lBF from the linear models is larger than that from the Gaussian processes, whereas this pattern is reversed for all the higher levels of non-linearity. Noticeably, the difference between the two sets of lBFs increase steadily with increasing non-linearity. This supports our claim that the \textit{Gaussian process models are better equipped in identifying evidence} {based on data originating from nonlinear generating models}.

\subsection{Comparing Coefficient Estimation from MCMC and EMVS Algorithms}

To perform a comparison between the estimated coefficients from the full MCMC and EMVS-based cBVS implementations, we used the data generated in Simulation 1. For each sample size and each replicate in Simulation 1, we estimate the coefficients using both procedures. For a given pool of $p$ covariates ($p = 200$) in Simulation 1, let us call these two sets of coefficient estimates $\hat{\boldsymbol{\beta}}_M$ and $\hat{\boldsymbol{\beta}}_E$, each having dimension $p \times 1$. For each replicate within each sample size scenario then, we compute the mean squared difference between these as $||\hat{\boldsymbol{\beta}}_M - \hat{\boldsymbol{\beta}}_E|| ^ 2 / p$. These quantities are summarized via box and violin plots in Supplementary \Cref{fig: Sim_ME}.

As can be seen in Supplementary \Cref{fig: Sim_ME}, the highest mean squared difference between the two sets of estimates for the smallest sample size ($n = 50$) is lower than $1$. The mean difference across $100$ replicates decreases consistently with increase in the sample size. In particular, all differences are in the order (or less) of $10 ^ {-1}$ for the sample sizes $\geq p = 200$. These summaries indicate that the compromise in approximating the full posterior in lieu of estimating the posterior mode via EMVS does not affect the estimation performance of cBVS substantially.

\section{Integrative Proteogenomic Analyses}\label{sec: pancan}

\subsection{List of TCGA Cancer Groups and Cancers}

\begin{itemize}

    \item \textbf{Pan-gyne:} breast invasive carcinoma (BRCA), cervical squamous cell carcinoma and endocervical adenocarcinoma (CESC), ovarian serous cystadenocarcinoma (OV), uterine corpus endometrial carcinoma (UCEC), uterine carcinosarcoma (UCS).
    
    \item \textbf{Pan-kidney:} kidney chromophobe (KICH), kidney renal clear cell carcinoma (KIRC), kidney renal papillary cell carcinoma (KIRP).
    
    \item \textbf{Pan-squamous:} esophageal carcinoma (ESCA, squamous), head and neck squamous cell carcinoma (HNSC), lung squamous cell carcinoma (LUSC).
    
    \item \textbf{Pan-gastrointestinal (pan-GI):} colon and rectum adenocarcinoma (CORE), esophageal carcinoma (ESCA, adeno), stomach adenocarcinoma (STAD).

\end{itemize}

\subsection{Survival Outcome Model Results}

We summarize the survival cBVS outputs for each TCGA cancer using plots showing $-log_{10}(1 - \hat{\omega})$ on the y-axis and $\hat{\beta}$ on the x-axis for each cancer, in Supplementary Figures \ref{fig: OM_SV_1}-\ref{fig: OM_SV_14}. Further, we present bar diagrams exhibiting the lBFs for the selected genes and proteins and histograms with density plots of lBFs for the predictors not selected. These figures are available in the shiny app for interactive exploration. Here, similar to the stemness outcome model results, main manuscript, we briefly discuss the results from the BRCA and CORE analyses.

For BRCA, no gene is selected using the 10\% FDR cut-off (Supplementary \Cref{fig: OM_SV_1}). The only protein selected is Collagen VI, which has decisive functional evidence. Again, a vast majority of the non-selected predictors had little to no functional evidence. For CORE, no gene/protein is selected at the 10\% FDR threshold (Supplementary \Cref{fig: OM_SV_12}).

\subsection{Past Evidences Related to Biological Findings}

\paragraph{RPS6KA1 gene and p90RSK kinases in gynecological cancer progression} The gene RPS6KA1 (protein p90RSK) has decisive evidence for all but two mechanistic models in the pan-gyne cancers, as seen in Figure 5A, main manuscript. The gene has been known to be differentially expressed in endometrial cancer tissue as opposed to benign endometrial tissue (\citealt{mamoor2021over}). Specifically, it is known to have a favorable prognostic effect on clinical outcomes in endometrial cancers (\citealt{bradfield2020investigating}). The p90RSK protein is selected in the survival cBVS model for OV (Supplementary \Cref{fig: OM_SV_OV}). p90RSK has been known to impact metastatic seeding of ovarian cancer cells, effecting the invasiveness of the cancer via activating a self-reinforcing cell autonomous circuit (\citealt{torchiaro2016peritoneal}). The RPS6KA1 gene is also significantly associated with the increased risk of breast cancer (\citealt{shareefi2020pathway}).

\paragraph{YAP1 gene and YAP proteins in gynecological malignancies} The gene YAP1 (proteins YAP, YAPPS127) has decisive evidence in all mechanistic models for all gynecological cancers except UCS (smallest sample size in the group, $n = 57$), as seen in Figure 5A, main manuscript. YAP is a crucial agent impacting gynecological cancers. As a transcriptional co-activator within the Hippo pathway, over-activation of YAP leads to uncontrolled cell growth and malignant transformation in gynecological malignancies, including cervical, ovarian, and endometrial cancers (\citealt{wang2020hippo}). Further, YAP expression is associated with a poor prognosis for gynecological cancers - activation of YAP induces cancer cell proliferation and migration (breast: \cite{guo2019yap}, cervical: \cite{he2015hippo}, endometrial: \cite{tsujiura2014yes}). This aligns with our identification of YAP1 as negatively associated with stemness for BRCA (Figure 6A, main manuscript) and with survival for OV and CESC (Supplementary Figures \ref{fig: OM_SV_CESC}-\ref{fig: OM_SV_OV}).

\paragraph{DIABLO gene as a marker of gynecological tumors} Another interesting candidate emerging from the mechanistic models is the gene DIABLO (protein Smac), which has been proposed as a biomarker for gynecological tumors, so far with little knowledge about its cellular mechanism. A recent study shows some evidence in favor of a positive association of Smac/DIABLO expression levels with estrogen receptor positivity in breast cancer (\citealt{espinosa2021coexpression}). Our cBVS analyses identify higher DIABLO expression to be associated with higher tumor stemness for two gynecological cancers (BRCA and UCS, Figure 6, main manuscript), which is in line with prior knowledge (\citealt{arellano2006high, arbiser2018diablo}). On the other hand, DIABLO is associated with higher survival for OV (Supplementary \Cref{fig: OM_SV_OV}), supported by earlier evidence - higher expression of DIABLO is a good prognostic sign for ovarian and endometrial cancers (\citealt{dobrzycka2010prognostic, dobrzycka2015prognostic}). However, neither DIABLO nor its corresponding protein Smac are identified as top candidates in the mechanistic analyses.

\paragraph{HER2 as a therapeutic target in gastrointestinal cancers} The gene ERBB2 (protein HER2) has decisive evidence for all mechanistic models in the pan-GI group, emerging as the top candidate (Figure 5C, main manuscript). HER2 overexpression has been known to be a frequent molecular abnormality in gastric cancers via gene amplification (\citealt{gravalos2008her2}). HER2 has also been considered as a molecular therapeutic target for patients with advanced gastric cancer (\citealt{abrahao2016her2}). In our cBVS analyses, both ERBB2/HER2 are positively associated with stemness for ESCA and STAD (Supplementary Figures \ref{fig: OM_SI_ESCA}-\ref{fig: OM_SI_STAD}), which aligns with past knowledge - increased expression of HER2 leads to quicker growth and poorer progression of gastric and esophageal cancers (\citealt{malaguti2015erbb2, lee2019detection}).

\paragraph{EGFR exhibiting contradictory associations with gastrointestinal cancers} The EGFR gene/protein emerge jointly on top in the pan-GI mechanistic models, with consistent decisive evidence (Figure 5C, main manuscript). Among gastric cancer patients, 2–35\% are reported to have EGFR protein overexpression and/or gene amplification (\citealt{adashek2020therapeutic}). However, the utility of EGFR as a therapeutic target has been questionable at best so far, with no general consensus on its prognostic value in gastric cancer (\citealt{arienti2019epidermal}). While some studies have indicated that high EGFR gene amplification is associated with poor outcome (\citealt{kandel2014association}), others have suggested the opposite (\citealt{aydin2014effect}). In line with this, cBVS found contradictory associations - EGFR gene expression is negatively associated with both stemness and survival for ESCA but positively associated with both for STAD (Supplementary Figures \ref{fig: OM_SI_ESCA}-\ref{fig: OM_SI_STAD}, \ref{fig: OM_SV_ESCA}-\ref{fig: OM_SV_14}).

\subsection{Investigating Biomarker Associations with Stemness Indices}

One concern related to the stemness outcome analysis arises due to the fact that the mRNA-based stemness indices (mRNAsi) are constructed by \cite{malta2018machine} using a one-class logistic regression (OCLR) model that uses a large pool of genes as the covariates. Potentially, these genes could overlap with the panel of genes used in our integrative proteogenomic analyses and resultingly, some of the associations detected by our stemness cBVS models may be spurious and caused only by the bias towards these genes in the stemness index training models. To reassure that this is not the case, we investigate the overlap of the gene panel in the OCLR training model with that of our cBVS models, and assess the weights of the common genes in the mRNAsi formula.

All the $160+$ genes that are included in our cBVS models are included in the training phase of the OCLR model for stemness, which uses a total of $12,945$ genes. We first ordered the $12,945$ genes in decreasing order of the absolute value of their weights in mRNAsi construction. None of the cBVS genes feature in the top 50 of this list, only two of them feature in the top 100, with three more each in the top 500 and the top 1000. For the overlapping genes, the largest absolute value of the weights is $7.1 \times 10 ^ {-3}$ and the median weight is $3.7 \times 10 ^ {-5}$. This indicates that there is not a considerable effect of these genes in the mRNAsi. Further, we assessed the correlations of mRNAsi (logit-transformed, as used in cBVS) for BRCA (the largest sample size among the cancers of our interest) with the corresponding gene expressions from the TCGA data. In summary, the largest absolute value among these correlations is $<0.65$ and the median correlation is $0.03$. Again, as before, this suggests that the mRNAsi is not spuriously correlated with the expressions of the genes of our interest, and that the associations detected by the stemness cBVS models are reliable.

\section{Interactive R Shiny Dashboard}\label{sec: shiny}

We provide an interactive R shiny dashboard summarizing all the results from our pan-cancer integrative proteogenomic analyses. The main page of the dashboard provides a brief overview of the fiBAG framework, as seen in Supplementary \Cref{fig:shiny1}. Broadly, the shiny dashboard contains pages providing interactive visualizations on the sample sizes, mechanistic model results, and stemness and survival outcome model results for the different cancers. The sample size summary page enables a user to choose a specific cancer, alongside an outcome data type, in order to provide summaries in the form of Venn diagrams presenting shared sample sizes (Supplementary \Cref{fig:shiny2}). The mechanistic model summary page allows the user to explore the results for each of the four cancer groups via heatmaps, upset plots, or wordclouds based on the lBFs (Supplementary \Cref{fig:shiny3}). The stemness and survival outcome model results are presented in two separate tabs, which work in similar fashion. For example, the stemness outcome models are summarized for each cancer using volcano-type plots, upset plots, and histograms of the evidence quantities (Supplementary \Cref{fig:shiny4}). In its online de-anonymized version, the shiny dashboard will have an additional page providing the user with the options to (1) download the software codes used for the analyses and (2) visit the GitHub page of the project. Due to the anonymity requirements of this submission, we remove the said tab from the shiny dashboard and separately submit the codes and the processed datasets in a compressed file.

\section{Supplementary Figures}\label{sec: figs}


\setcounter{figure}{0}
\renewcommand{\figurename}{Supplementary Figure}
\renewcommand{\thefigure}{\arabic{figure}}

\begin{figure}[!htbp]
\centering
\includegraphics[scale = 0.30]{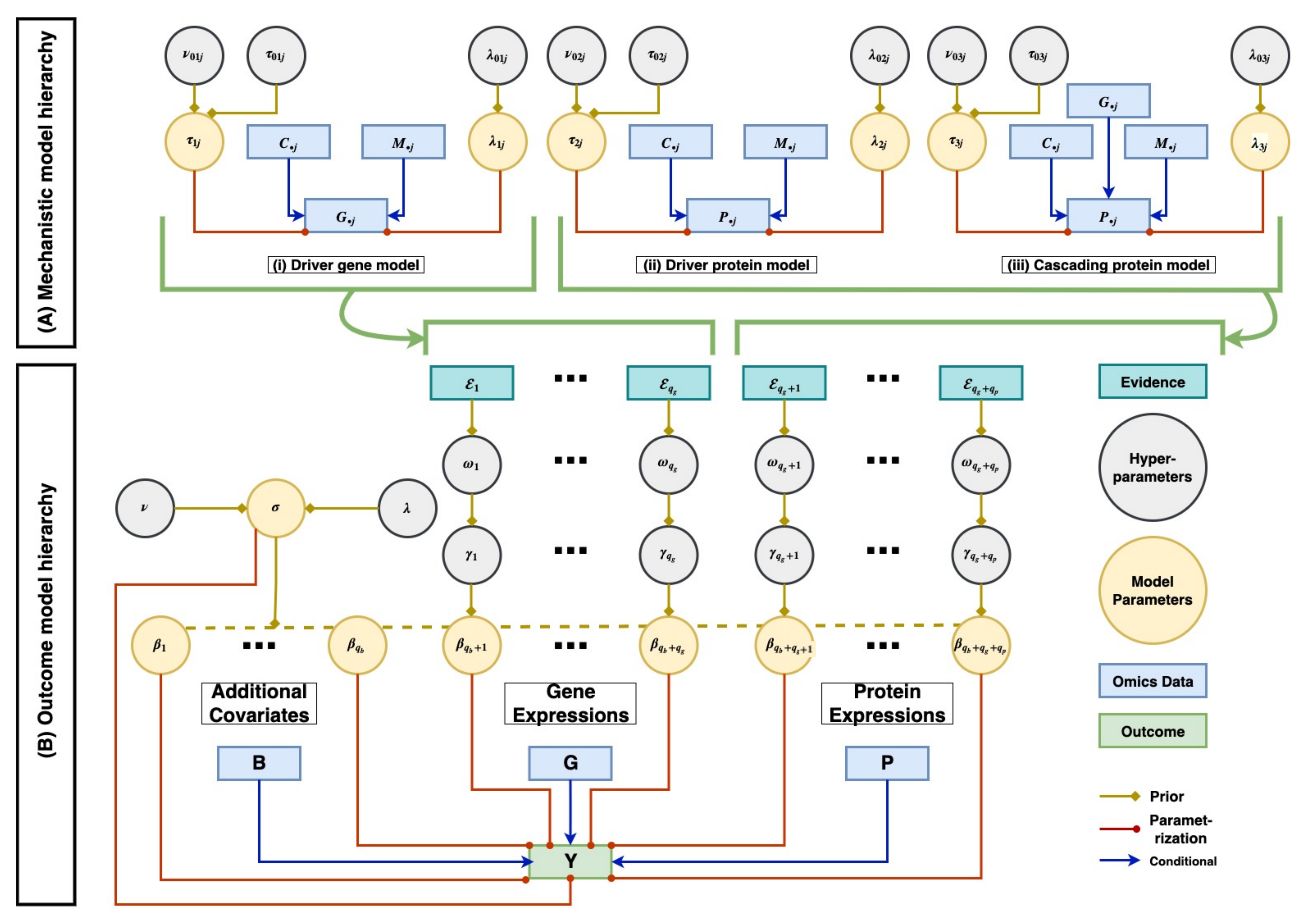}
\caption{\textbf{Plate diagram summarizing the dependence structure across the variables, parameters, and hyperparameters in fiBAG.} Panels (A) and (B) respectively summarize the mechanistic and outcome models. $\mathbf{C}, \mathbf{M}, \mathbf{B}, \mathbf{G}$, and $\mathbf{P}$ each have $n$ rows and are matched across samples, along with the outcome vector $\mathbf{Y}$. $\mathbf{B}, \mathbf{G}$, and $\mathbf{P}$ respectively have $q_b$, $q_g$ and $q_p$ columns. $\mathbf{C}$ and $\mathbf{M}$ respectively represent copy number and methylation data. All the parameters and data structures are described in full detail in Section 2 in the main manuscript.}
\label{fig: dependence}
\end{figure}

\begin{figure}[!htbp]
    
    \centering
    \includegraphics[scale = 0.69]{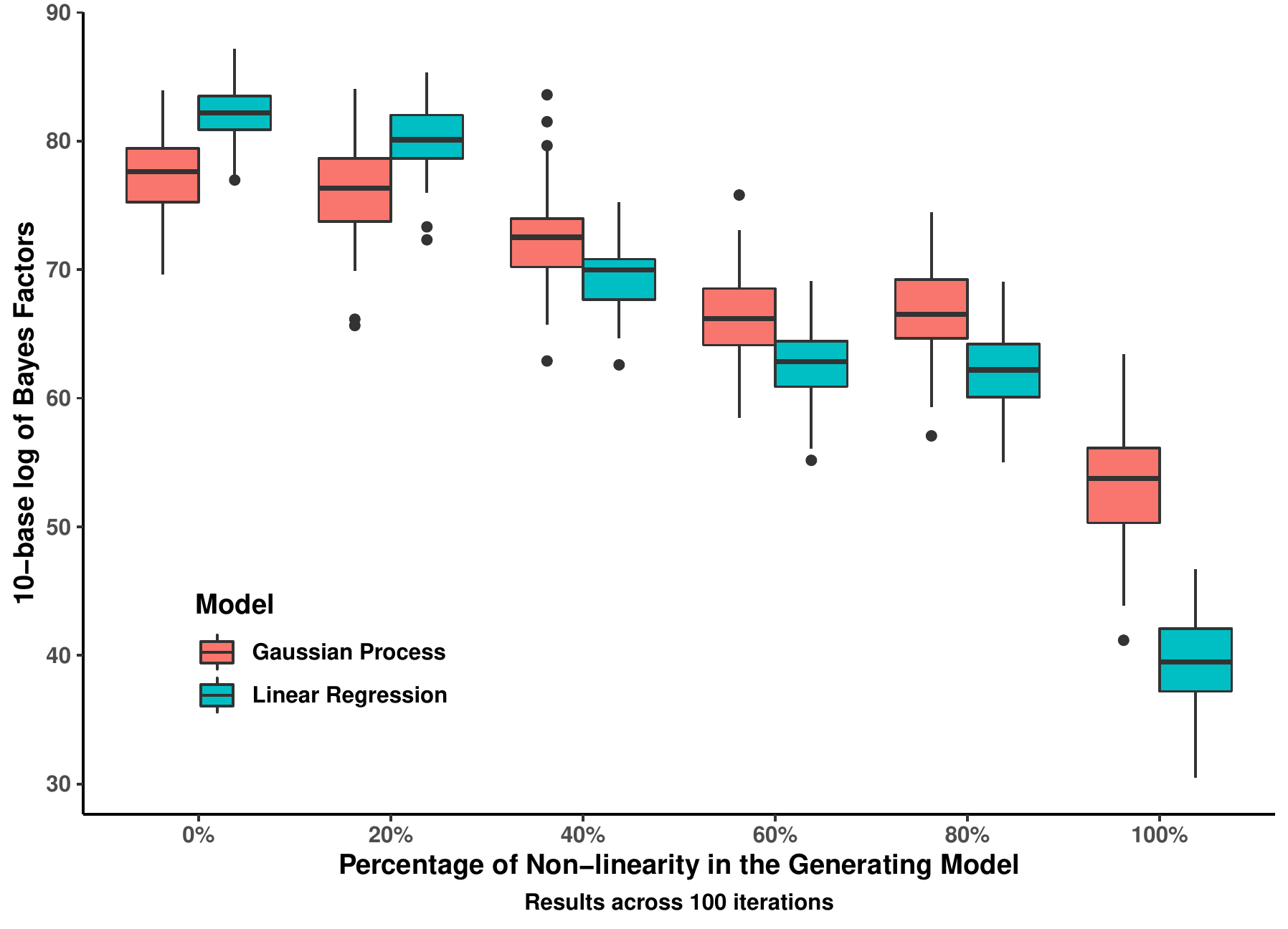}
    
    \caption[Results from Simulation comparing Gaussian processes and linear models.]{\textbf{Results from Simulation comparing Gaussian processes and linear models.} The x-axis is in increasing order of non-linearity (in terms of the mean function of the outcome based on the covariates) in the generating model. The y-axis presents box plots across 100 replicates of the lBFs from Gaussian processes and linear models.}
    \label{fig: Sim_GPLM}
    
\end{figure}

\begin{figure}[!htbp]
    
    \centering
    \includegraphics[scale = 0.54]{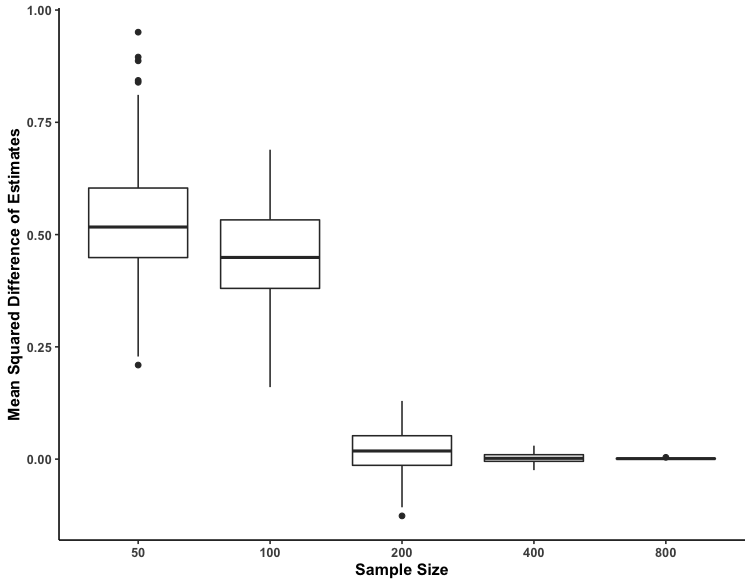}
    
    \caption[Results from Simulation comparing MCMC and EMVS implementations of cBVS.]{\textbf{Results from Simulation comparing MCMC and EMVS implementations of cBVS.} The x-axis presents the sample sizes for generating the replicates, with 200 covariates in each case. The y-axis presents box plots across 100 replicates of the mean squared differences of the coefficients estimated via the two implementations.}
    \label{fig: Sim_ME}
    
\end{figure}

\begin{sidewaysfigure}[!htbp]
    
    \centering
    \includegraphics[scale = 0.08]{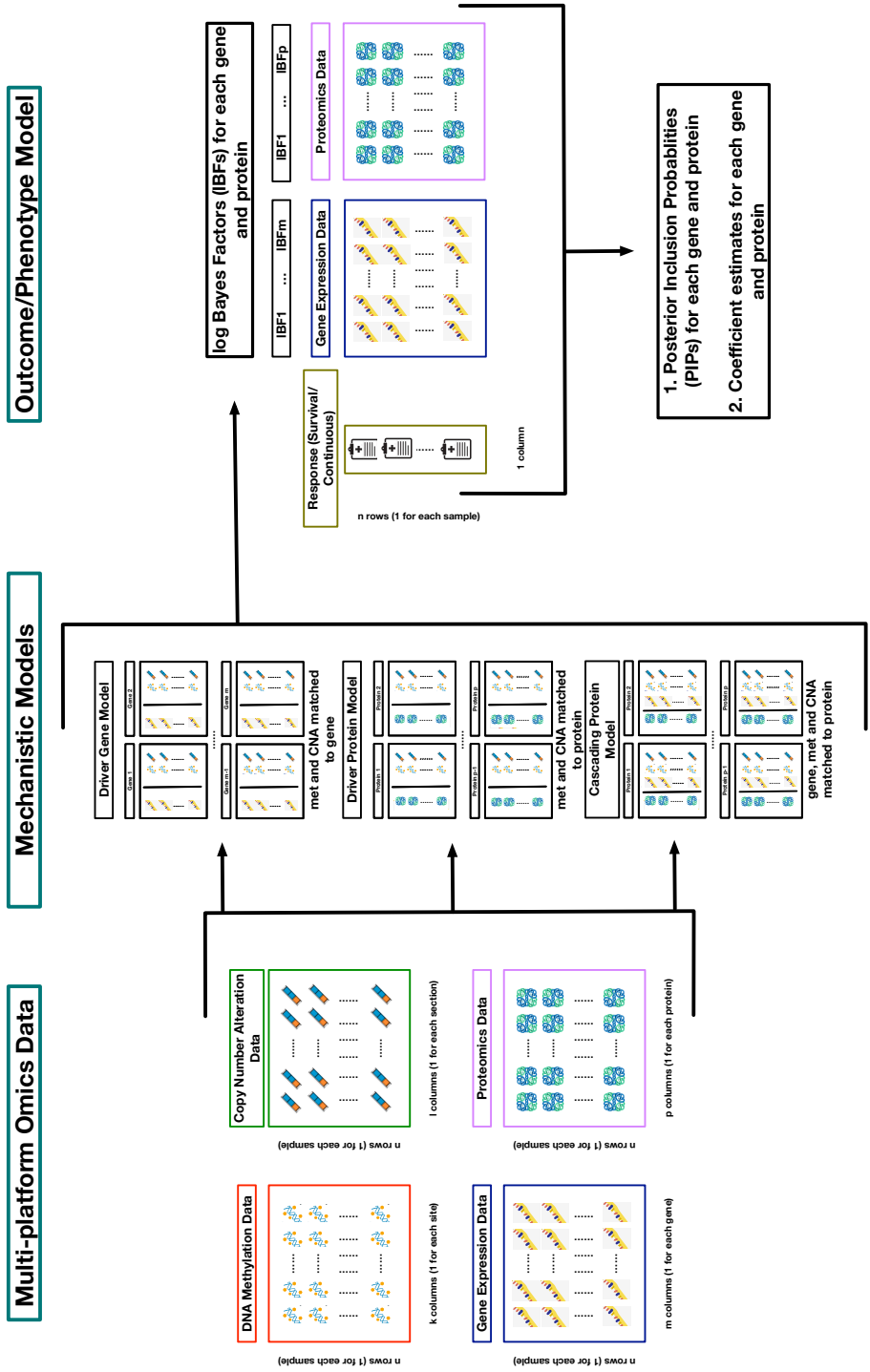}
    
    \caption[Integrative analysis using the fiBAG pipeline.]{\textbf{Integrative analysis using the fiBAG pipeline.} In each mechanistic model presented in the middle panel, upstream covariates are matched to each proteogenomic biomarker. The lBFs computed from each model are used in the cBVS outcome models to perform calibrated variable selection, as outlined in the right panel. The outcome models provide posterior inclusion probabilities for each biomarker along with estimated coefficient for association with the outcome of choice. The entire pipeline is independently applied on each cancer. We build one survival and one stemness outcome model for each cancer.}
    \label{fig: Pipeline2}
    
\end{sidewaysfigure}

\begin{figure}[!htbp]
    
    \centering
    \includegraphics[scale = 0.42]{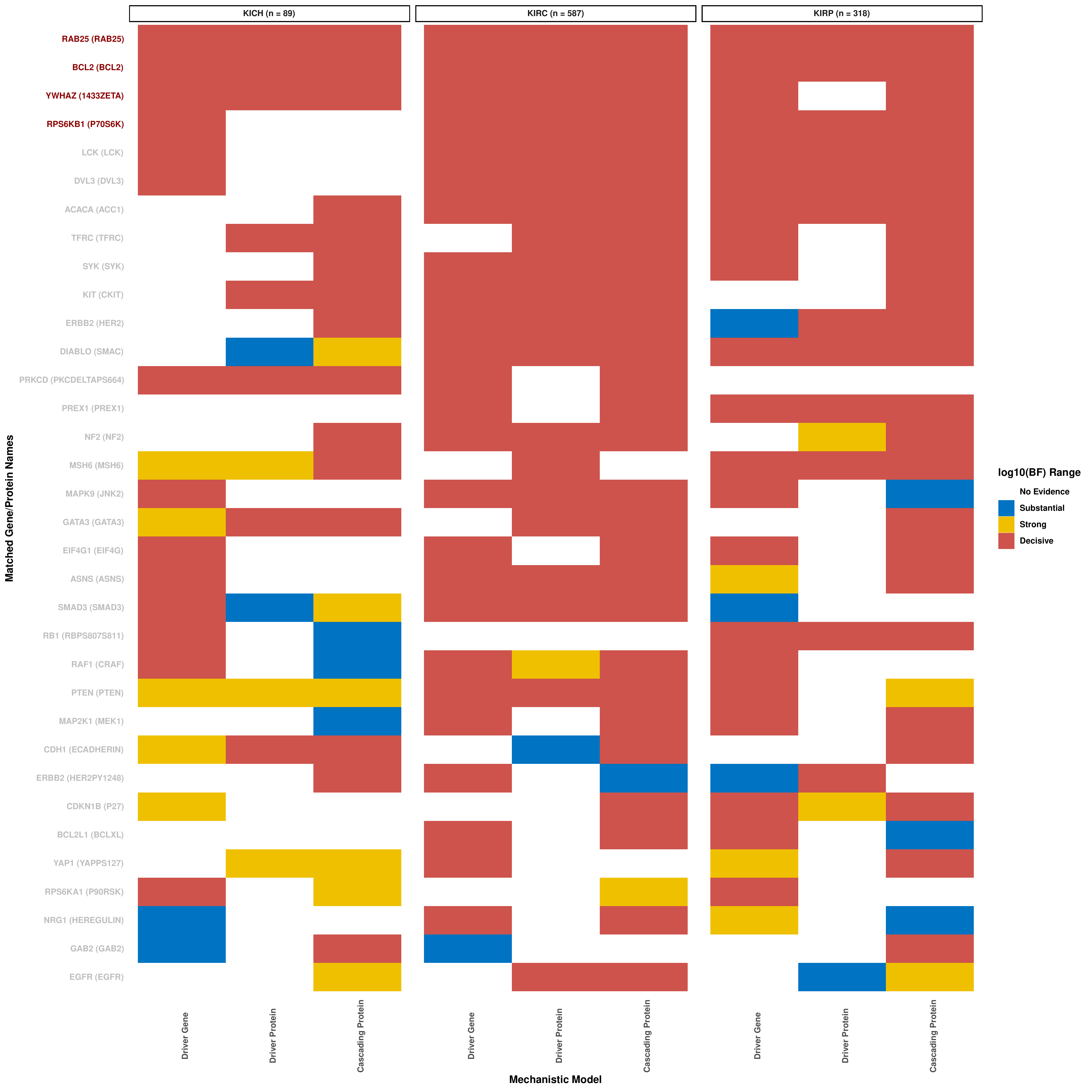}
    
    \caption[Mechanistic model heatmaps for pan-kidney cancers.]{\textbf{Mechanistic model heatmaps for the pan-kidney cancers.} Each cancer column consists of three sub-columns, one each for the three mechanistic models (driver gene, driver protein and cascading protein). The lBF ranges are defined as: $< 0.5$ (no evidence), $0.5 - 1$ (substantial), $1 - 2$ (strong), $> 2$ (decisive). Only the gene-protein pairs which are at the decisive level of significance across all four cancers in at least two out of the three mechanistic model types are shown here.}
    \label{fig: fiBAG_Kidney_MM2}
    
\end{figure}

\begin{figure}[!htbp]
    
    \centering
    \includegraphics[scale = 0.42]{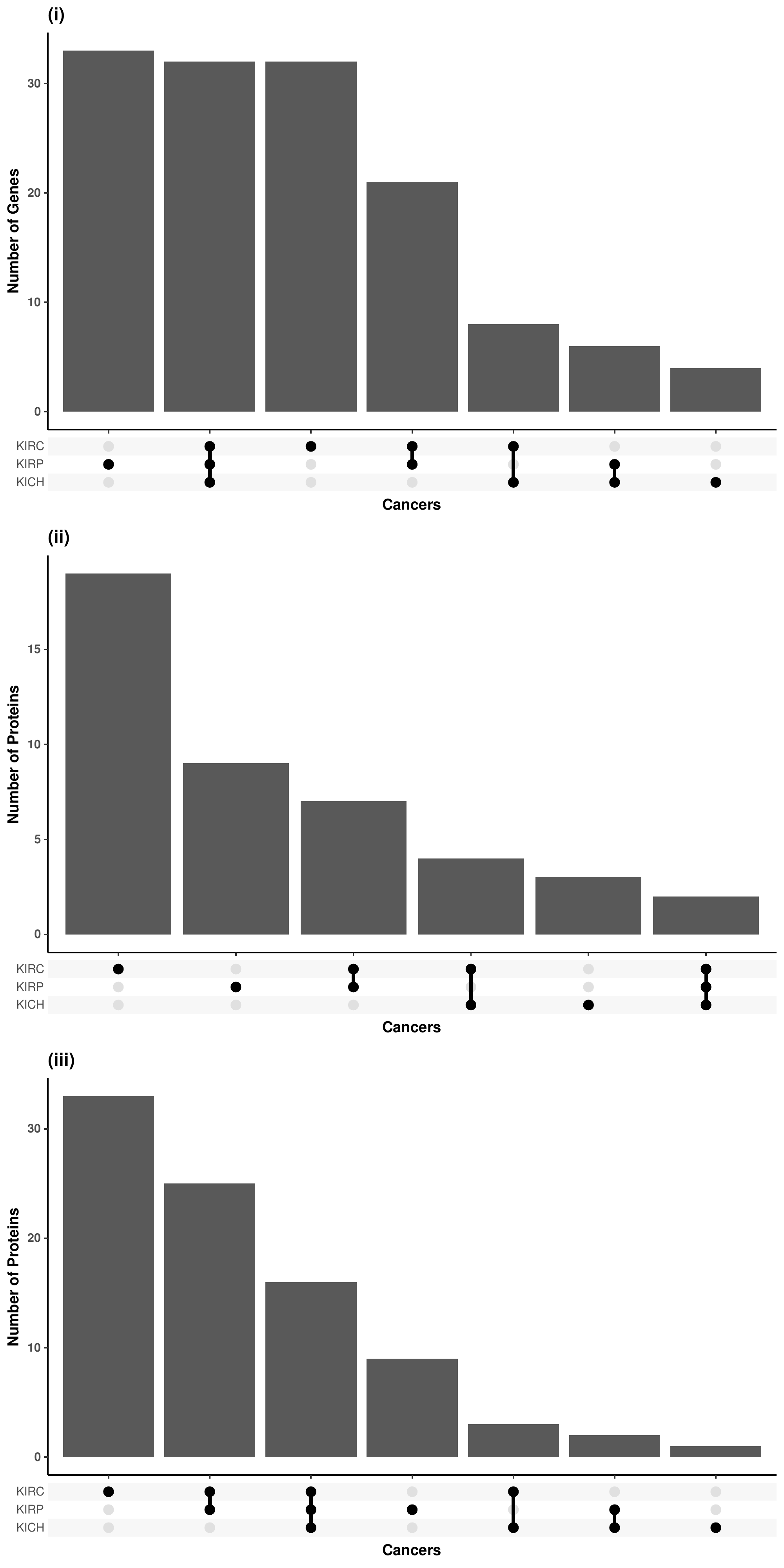}
    
    \caption[Mechanistic model upset plots for pan-kidney cancers.]{\textbf{Mechanistic model upset plots for the pan-kidney cancers.} Upset plots exhibit the number of genes (panel A) or proteins (panels B-C) that are at the decisive level of significance (lBF $> 2$) for the (A) driver gene, (B) driver protein, and (C) cascading protein mechanistic models respectively, stratified by intersections across cancers.}
    \label{fig: fiBAG_Kidney_MM3}
    
\end{figure}

\begin{figure}[!htbp]
    
    \centering
    \includegraphics[scale = 0.42]{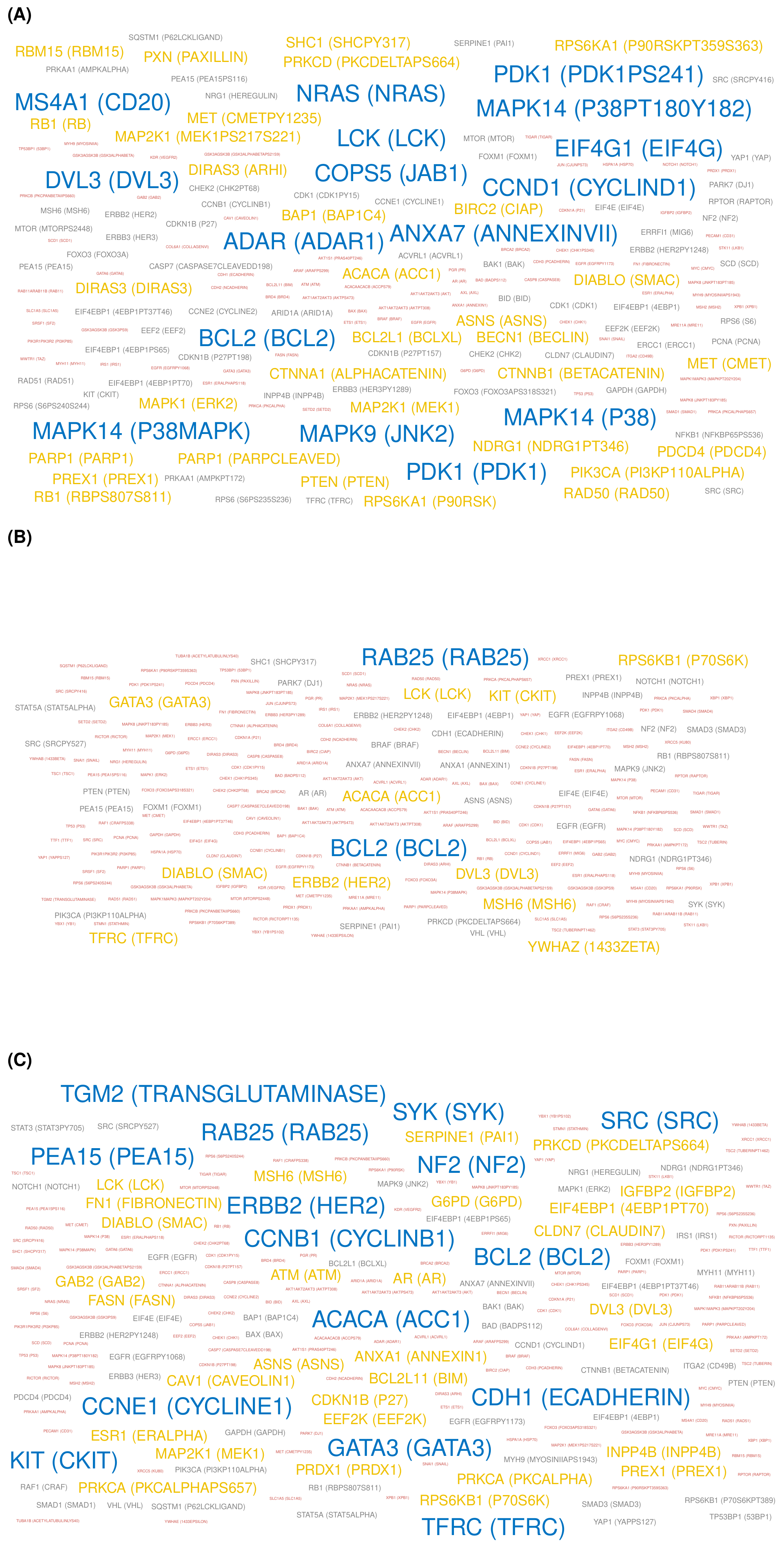}
    
    \caption[Mechanistic model word clouds for pan-kidney cancers.]{\textbf{Mechanistic model word clouds for the pan-kidney cancers.} Word clouds summarize pan-cancer mechanistic model results for genes (panel A) or proteins (panels B-C) for the (A) driver gene, (B) driver protein, and (C) cascading protein mechanistic models. The size of the gene/protein names are proportional to $(\textrm{no. of cancers where the gene/protein is at the decisive level of significance})^3$. Here, decisive is defined as lBF $> 2$.}
    \label{fig: fiBAG_Kidney_MM4}
    
\end{figure}

\begin{figure}[!htbp]
    
    \centering
    \includegraphics[scale = 0.42]{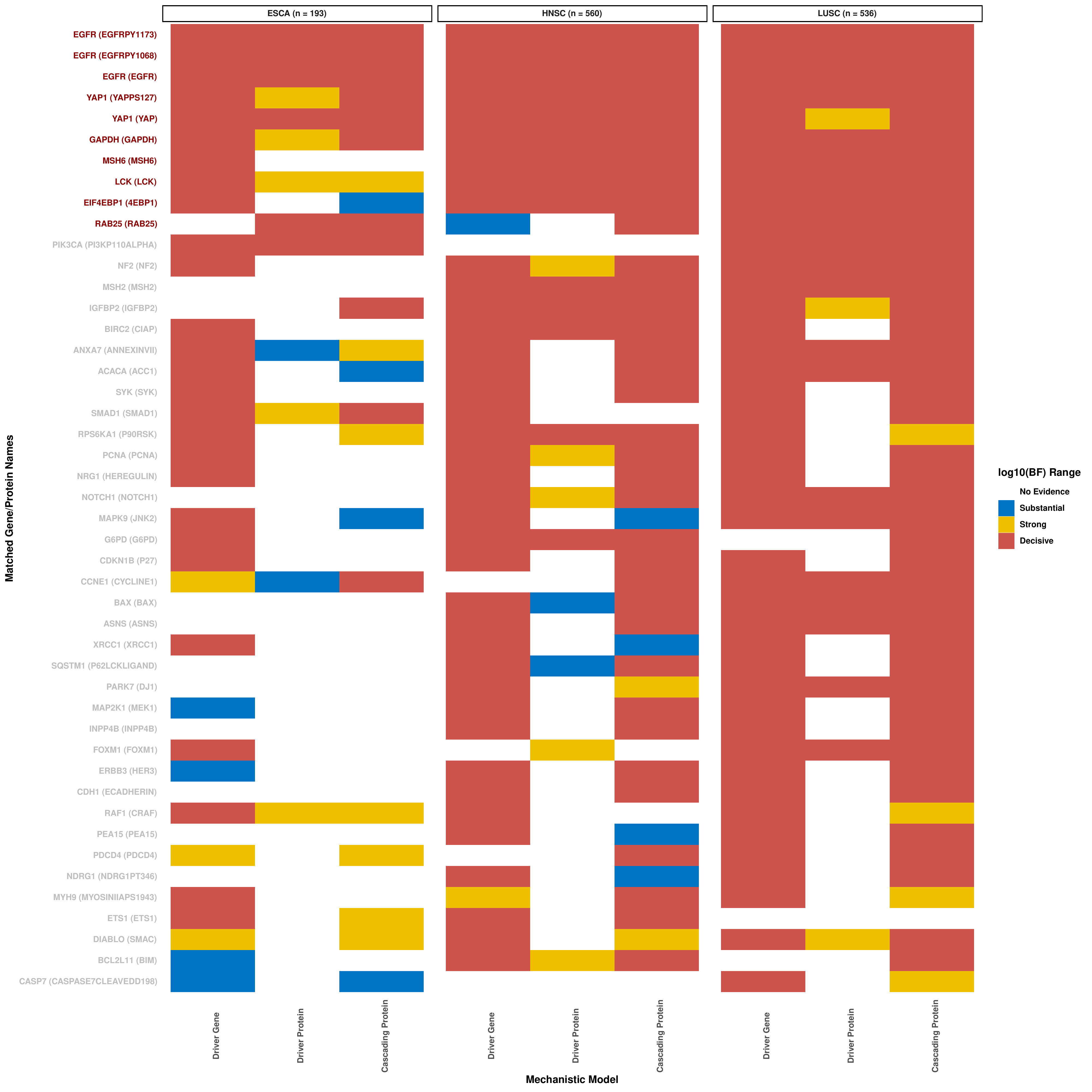}
    
    \caption[Mechanistic model heatmaps for pan-squamous cancers.]{\textbf{Mechanistic model heatmaps for the pan-squamous cancers.} Each cancer column consists of three sub-columns, one each for the three mechanistic models (driver gene, driver protein and cascading protein). The lBF ranges are defined as: $< 0.5$ (no evidence), $0.5 - 1$ (substantial), $1 - 2$ (strong), $> 2$ (decisive). Only the gene-protein pairs which are at the decisive level of significance across all four cancers in at least two out of the three mechanistic model types are shown here.}
    \label{fig: fiBAG_Squ_MM2}
    
\end{figure}

\begin{figure}[!htbp]
    
    \centering
    \includegraphics[scale = 0.42]{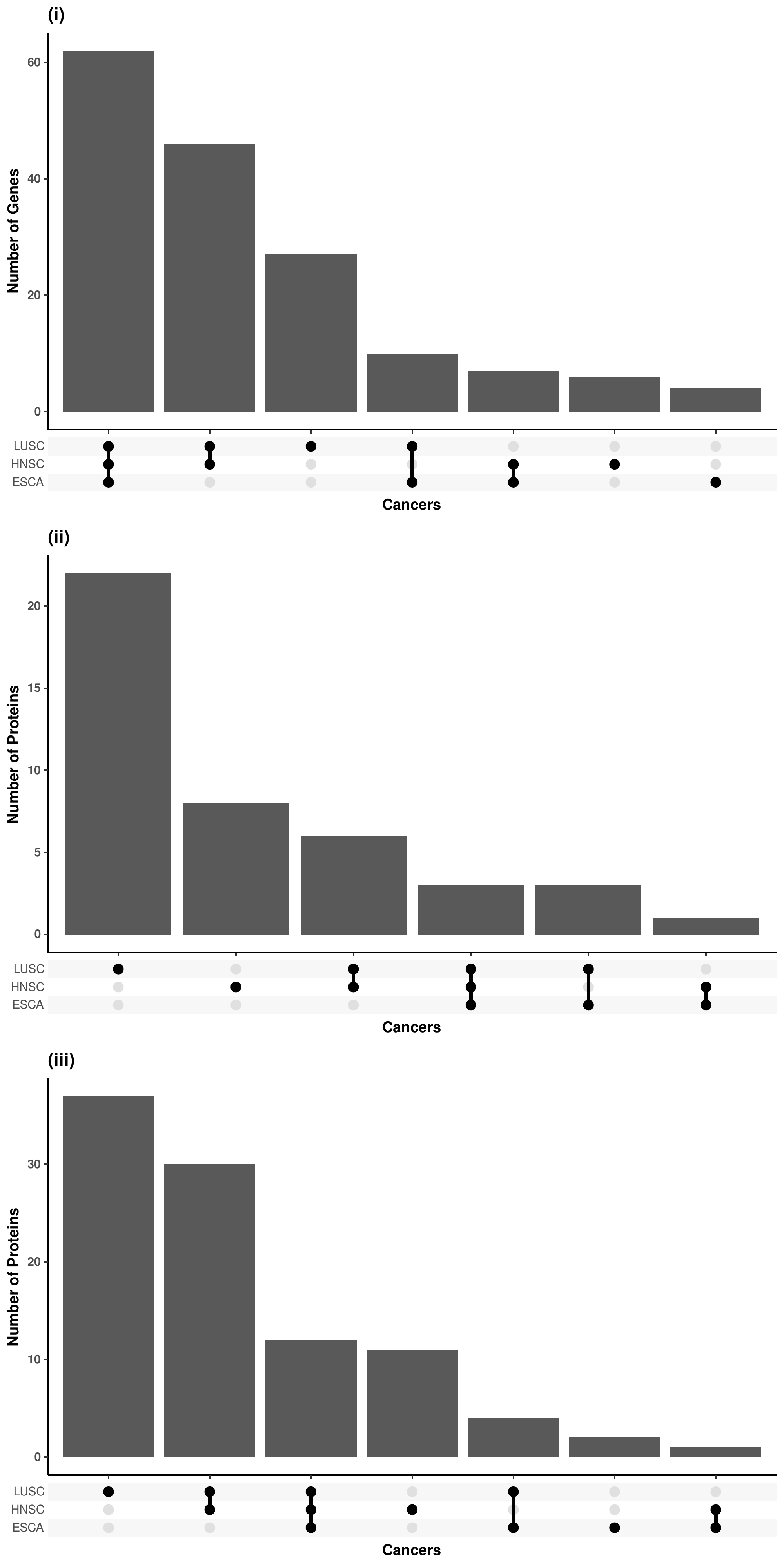}
    
    \caption[Mechanistic model upset plots for pan-squamous cancers.]{\textbf{Mechanistic model upset plots for the pan-squamous cancers.} Upset plots exhibit the number of genes (panel A) or proteins (panels B-C) that are at the decisive level of significance (lBF $> 2$) for the (A) driver gene, (B) driver protein, and (C) cascading protein mechanistic models respectively, stratified by intersections across cancers.}
    \label{fig: fiBAG_Squ_MM3}
    
\end{figure}

\begin{figure}[!htbp]
    
    \centering
    \includegraphics[scale = 0.42]{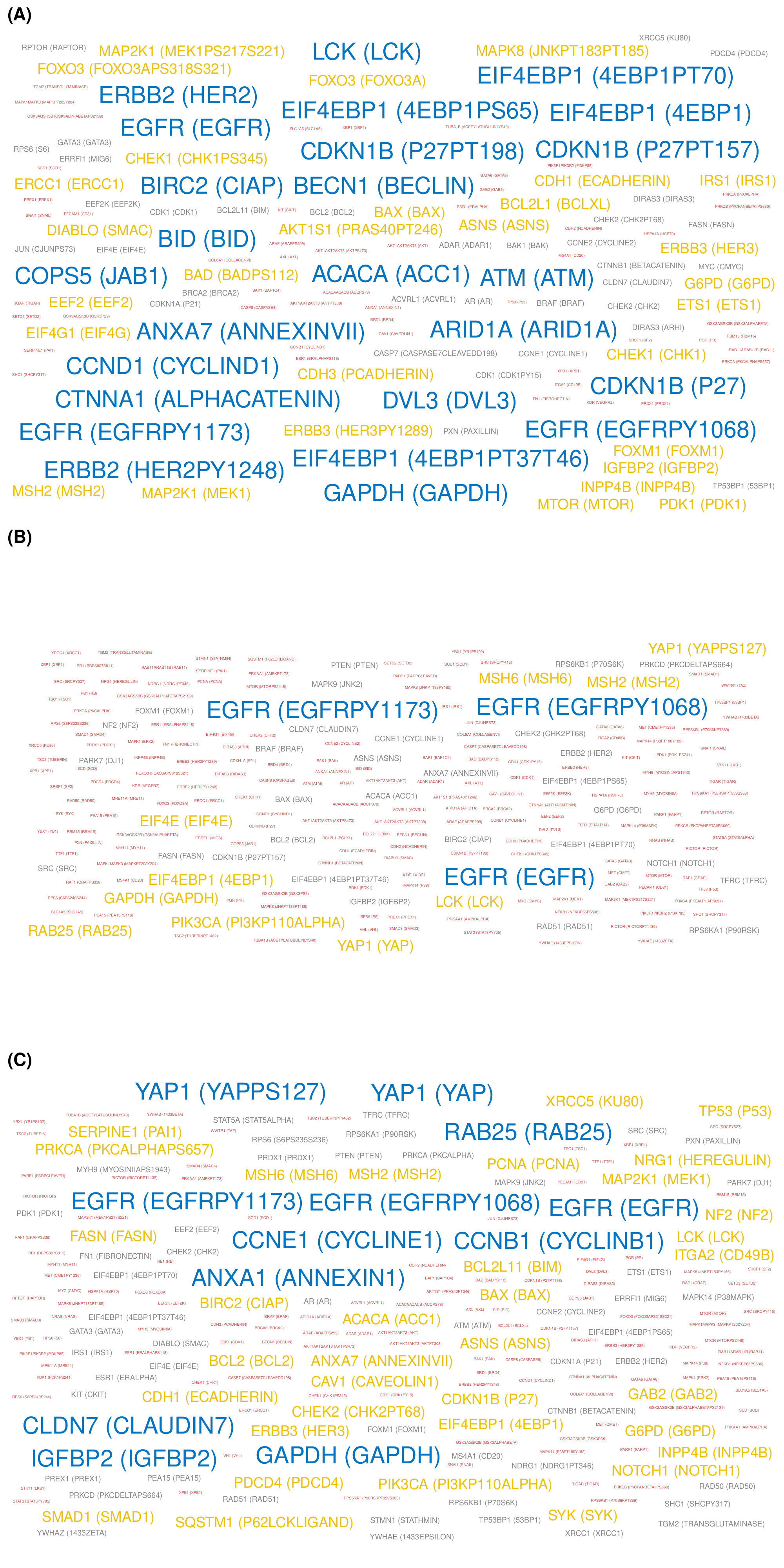}
    
    \caption[Mechanistic model word clouds for pan-squamous cancers.]{\textbf{Mechanistic model word clouds for the pan-kidney cancers.} Word clouds summarize pan-cancer mechanistic model results for genes (panel A) or proteins (panels B-C) for the (A) driver gene, (B) driver protein, and (C) cascading protein mechanistic models. The size of the gene/protein names are proportional to $(\textrm{no. of cancers where the gene/protein is at the decisive level of significance})^3$. Here, decisive is defined as lBF $> 2$.}
    \label{fig: fiBAG_Squ_MM4}
    
\end{figure}

\begin{figure}[hbt!]
\centering
\includegraphics[scale = 0.63]{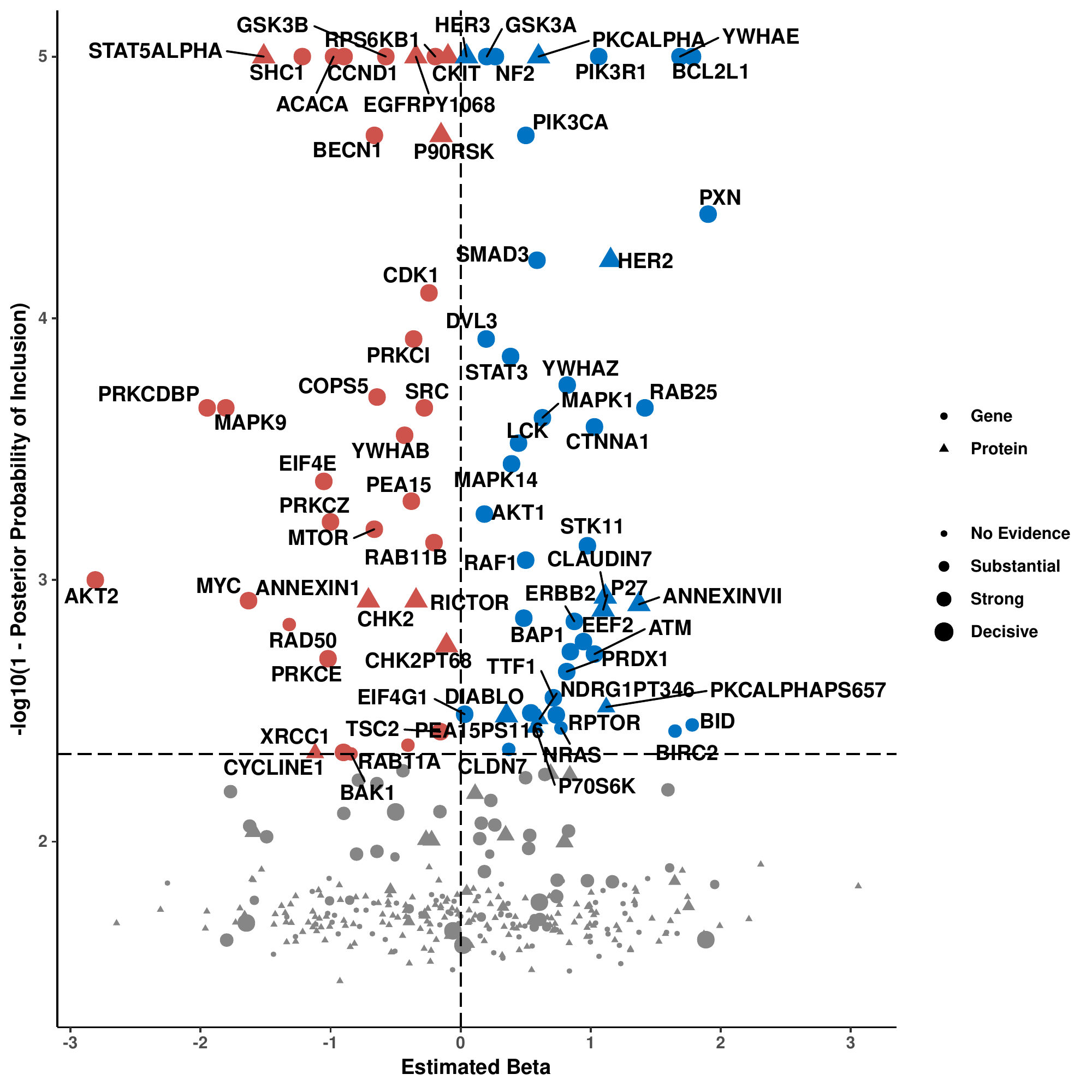}
\caption[Plot summarizing outcome model results based on stemness indices (SI) for TCGA cancer UCS.]{\textbf{Plot summarizing outcome model results based on stemness indices (SI) for TCGA cancer UCS.} Proteins are represented by triangles and genes by circles. The shapes are colored red if the estimated $\hat{\beta}_j$ from fiBAG is negative, and green if positive. The x-axis shows the $\hat{\beta}_j$s, and the y-axis shows the $-\log_{10}(1 - \hat{\omega}_j)$s. An FDR check to adjust for multiple comparisons is performed treating $1 - \hat{\omega}_j$ as a p-value type quantity at the 10\% FDR level. Only the selected biomarkers are marked in non-gray colors and labeled. The sizes of the points are in the increasing order of evidence from the mechanistic models: lBF ranges are defined as: $< 0.5$ (no evidence), $0.5 - 1$ (substantial), $1 - 2$ (strong), $> 2$ (decisive).}
\end{figure}

\begin{figure}[hbt!]
\centering
\includegraphics[scale = 0.63]{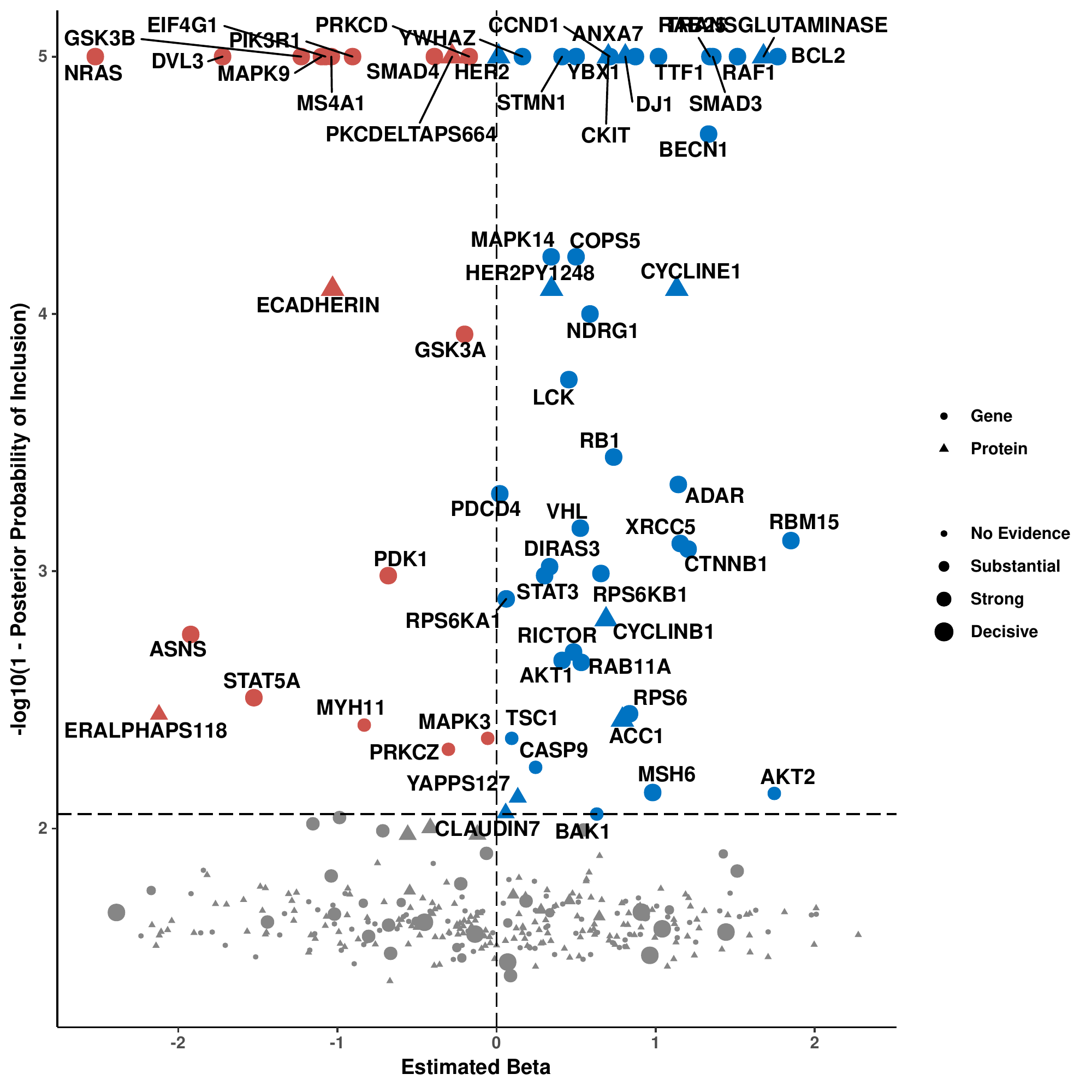}
\caption[Plot summarizing outcome model results based on stemness indices (SI) for TCGA cancer KICH.]{\textbf{Plot summarizing outcome model results based on stemness indices (SI) for TCGA cancer KICH.} Proteins are represented by triangles and genes by circles. The shapes are colored red if the estimated $\hat{\beta}_j$ from fiBAG is negative, and green if positive. The x-axis shows the $\hat{\beta}_j$s, and the y-axis shows the $-\log_{10}(1 - \hat{\omega}_j)$s. An FDR check to adjust for multiple comparisons is performed treating $1 - \hat{\omega}_j$ as a p-value type quantity at the 10\% FDR level. Only the selected biomarkers are marked in non-gray colors and labeled. The sizes of the points are in the increasing order of evidence from the mechanistic models: lBF ranges are defined as: $< 0.5$ (no evidence), $0.5 - 1$ (substantial), $1 - 2$ (strong), $> 2$ (decisive).}
\end{figure}

\begin{figure}[hbt!]
\centering
\includegraphics[scale = 0.63]{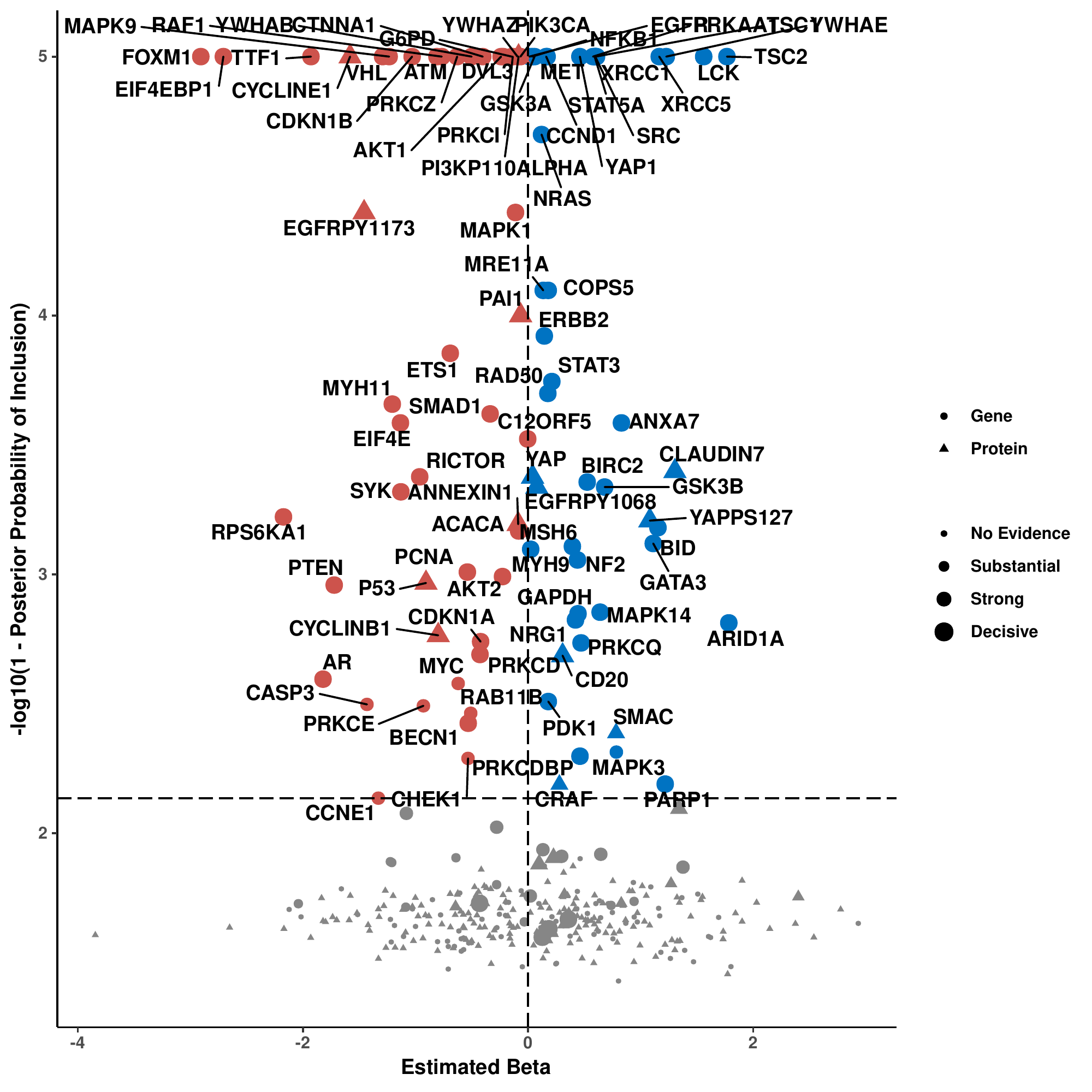}
\caption[Plot summarizing outcome model results based on stemness indices (SI) for TCGA cancer ESCA-squamous.]{\textbf{Plot summarizing outcome model results based on stemness indices (SI) for TCGA cancer ESCA-squamous.} Proteins are represented by triangles and genes by circles. The shapes are colored red if the estimated $\hat{\beta}_j$ from fiBAG is negative, and green if positive. The x-axis shows the $\hat{\beta}_j$s, and the y-axis shows the $-\log_{10}(1 - \hat{\omega}_j)$s. An FDR check to adjust for multiple comparisons is performed treating $1 - \hat{\omega}_j$ as a p-value type quantity at the 10\% FDR level. Only the selected biomarkers are marked in non-gray colors and labeled. The sizes of the points are in the increasing order of evidence from the mechanistic models: lBF ranges are defined as: $< 0.5$ (no evidence), $0.5 - 1$ (substantial), $1 - 2$ (strong), $> 2$ (decisive).}
\end{figure}

\begin{figure}[hbt!]
\centering
\includegraphics[scale = 0.63]{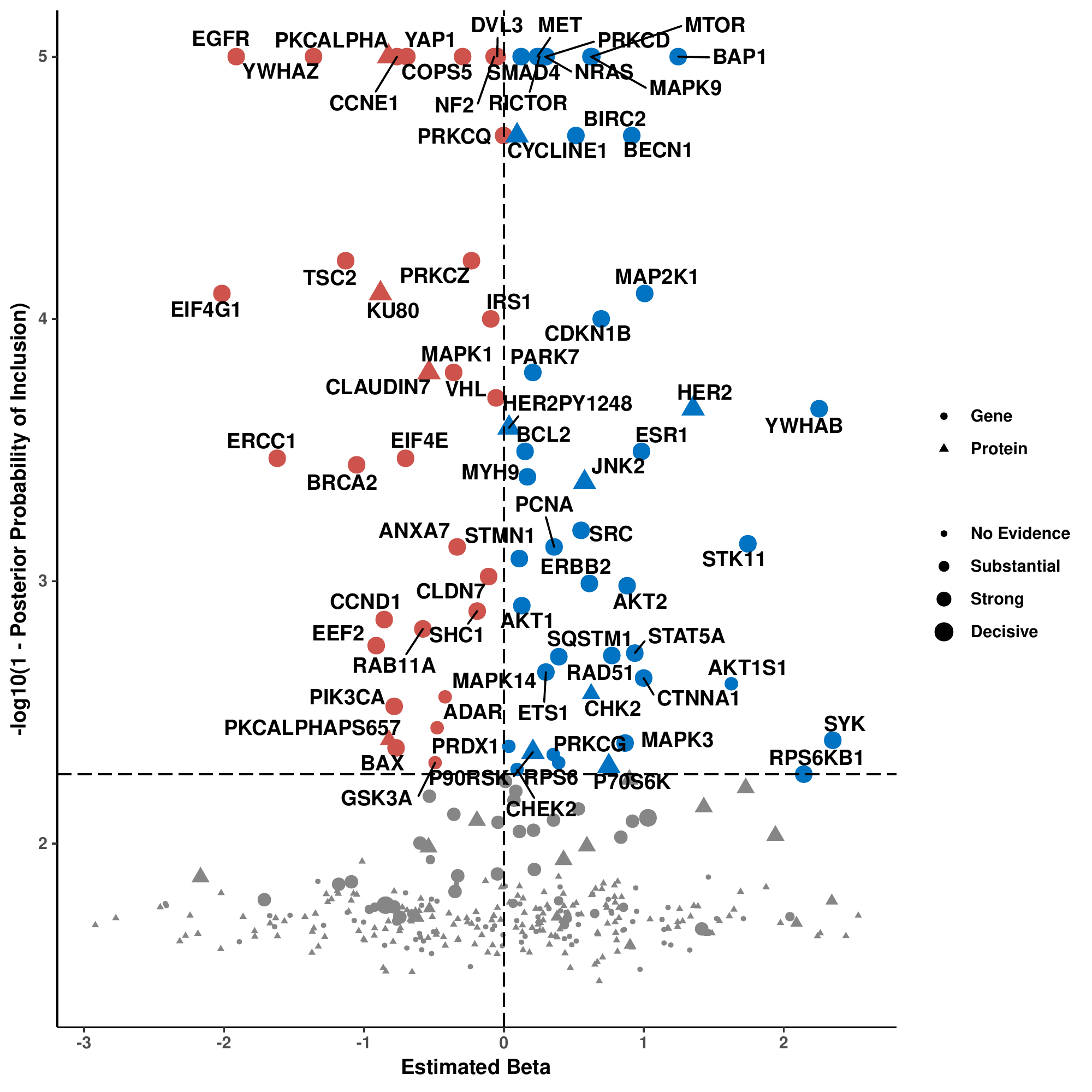}
\caption[Plot summarizing outcome model results based on stemness indices (SI) for TCGA cancer ESCA-adeno.]{\textbf{Plot summarizing outcome model results based on stemness indices (SI) for TCGA cancer ESCA-adeno.} Proteins are represented by triangles and genes by circles. The shapes are colored red if the estimated $\hat{\beta}_j$ from fiBAG is negative, and green if positive. The x-axis shows the $\hat{\beta}_j$s, and the y-axis shows the $-\log_{10}(1 - \hat{\omega}_j)$s. An FDR check to adjust for multiple comparisons is performed treating $1 - \hat{\omega}_j$ as a p-value type quantity at the 10\% FDR level. Only the selected biomarkers are marked in non-gray colors and labeled. The sizes of the points are in the increasing order of evidence from the mechanistic models: lBF ranges are defined as: $< 0.5$ (no evidence), $0.5 - 1$ (substantial), $1 - 2$ (strong), $> 2$ (decisive).} \label{fig: OM_SI_ESCA}
\end{figure}

\begin{figure}[hbt!]
\centering
\includegraphics[scale = 0.63]{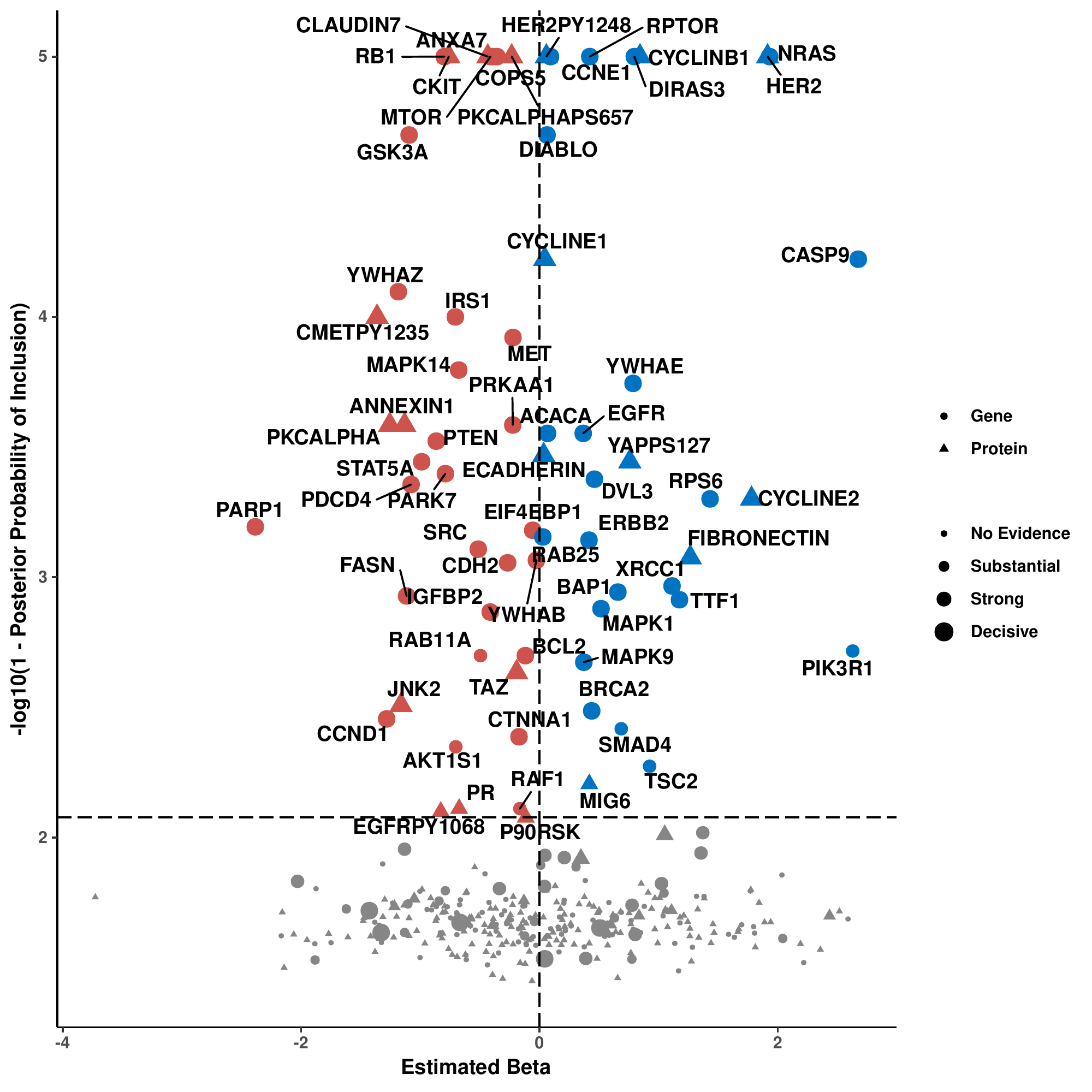}
\caption[Plot summarizing outcome model results based on stemness indices (SI) for TCGA cancer STAD.]{\textbf{Plot summarizing outcome model results based on stemness indices (SI) for TCGA cancer STAD.} Proteins are represented by triangles and genes by circles. The shapes are colored red if the estimated $\hat{\beta}_j$ from fiBAG is negative, and green if positive. The x-axis shows the $\hat{\beta}_j$s, and the y-axis shows the $-\log_{10}(1 - \hat{\omega}_j)$s. An FDR check to adjust for multiple comparisons is performed treating $1 - \hat{\omega}_j$ as a p-value type quantity at the 10\% FDR level. Only the selected biomarkers are marked in non-gray colors and labeled. The sizes of the points are in the increasing order of evidence from the mechanistic models: lBF ranges are defined as: $< 0.5$ (no evidence), $0.5 - 1$ (substantial), $1 - 2$ (strong), $> 2$ (decisive).}\label{fig: OM_SI_STAD}
\end{figure}

\begin{figure}[hbt!]
\centering
\includegraphics[scale = 0.63]{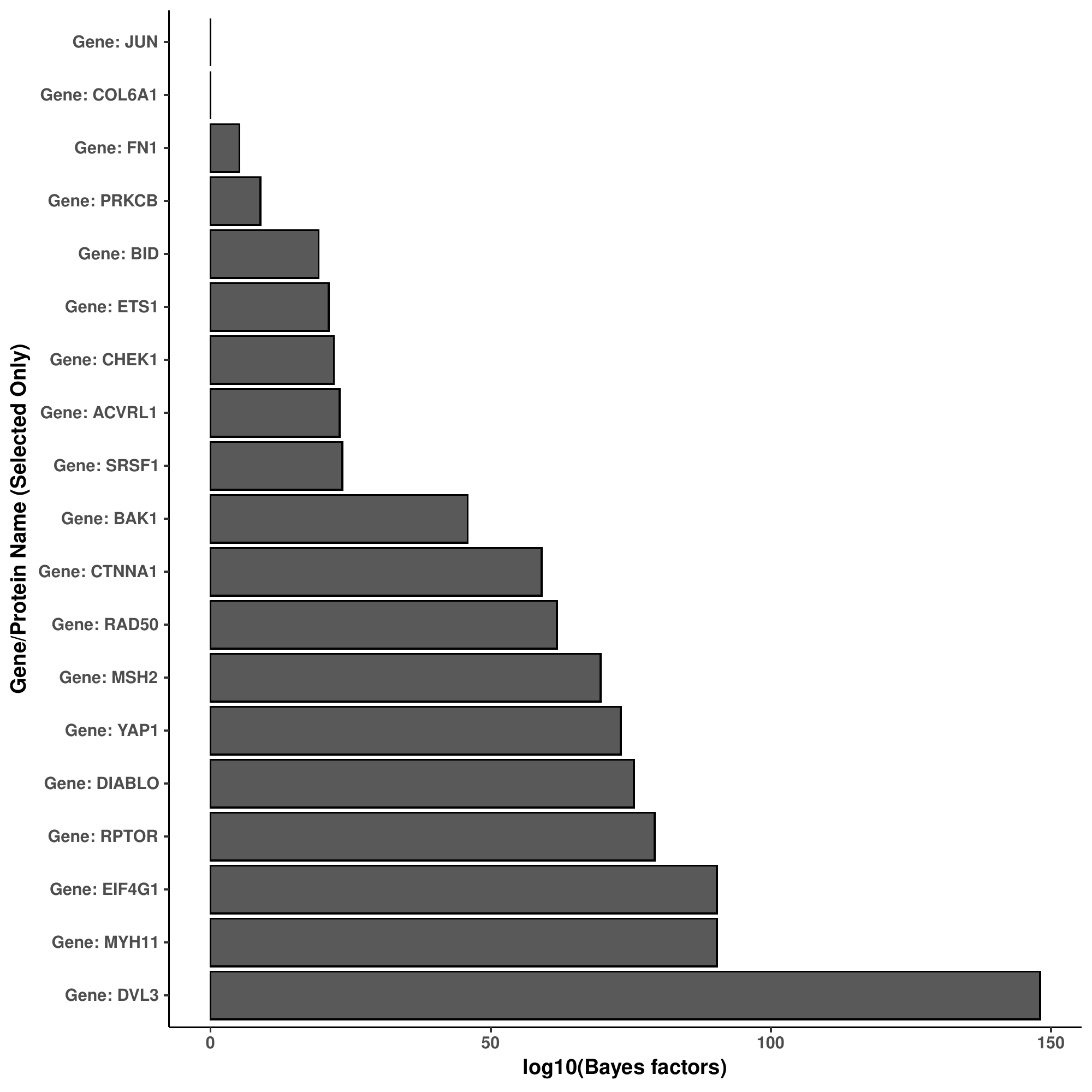}
\caption[Bar diagram summarizing lBFs for the selected proteogenomic biomarkers in the stemness outcome model for TCGA cancer BRCA.]{\textbf{Bar diagram summarizing lBFs for the selected proteogenomic biomarkers in the stemness outcome model for TCGA cancer BRCA.} An FDR check to adjust for multiple comparisons is performed treating $1 - \hat{\omega}_j$ as a p-value type quantity at the 10\% FDR level. Only the selected biomarkers are presented.}
\end{figure}

\begin{figure}[hbt!]
\centering
\includegraphics[scale = 0.63]{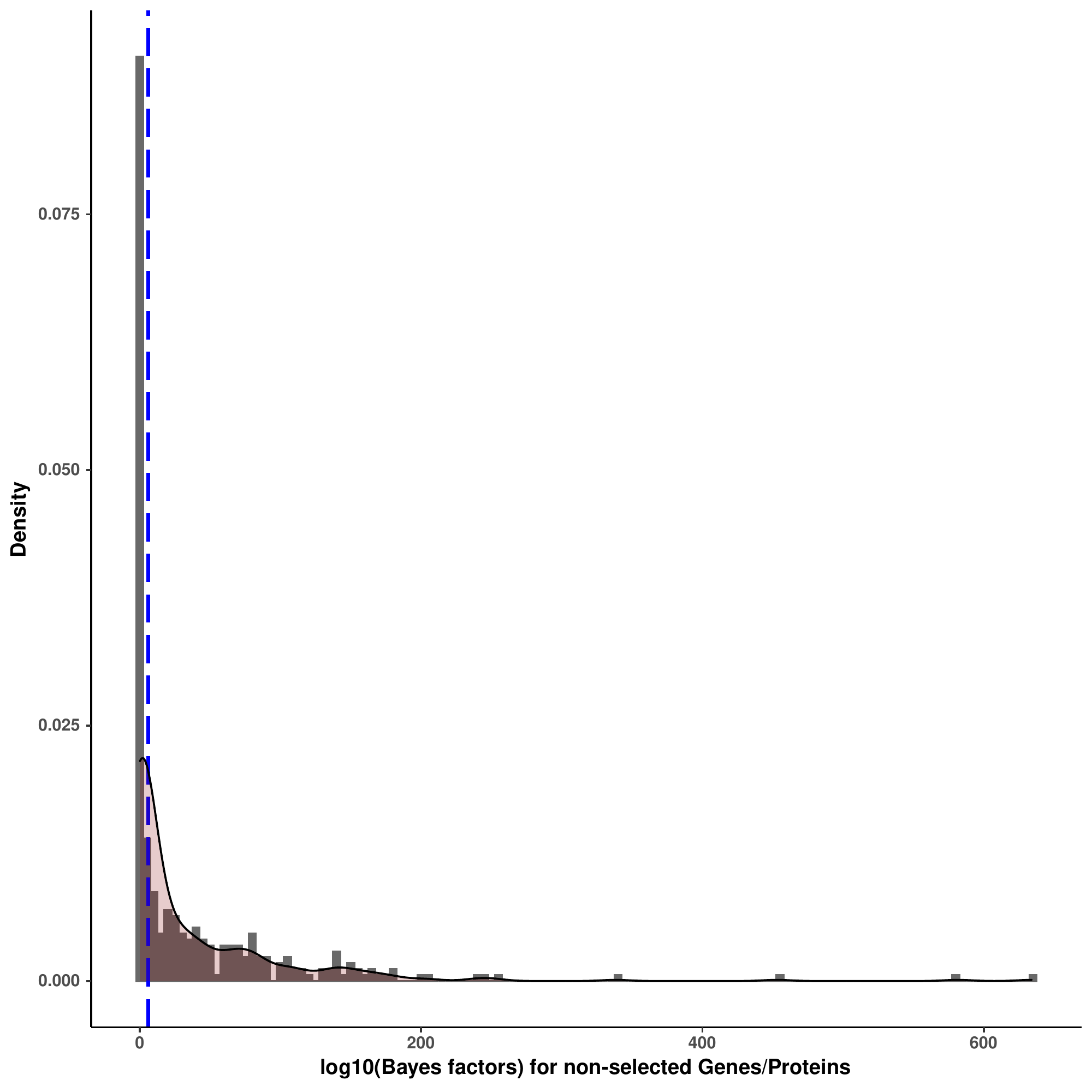}
\caption[Histogram of lBFs for the non-selected proteogenomic biomarkers in the stemness outcome model for TCGA cancer BRCA.]{\textbf{Histogram of lBFs for the non-selected proteogenomic biomarkers in the stemness outcome model for TCGA cancer BRCA.} An FDR check to adjust for multiple comparisons is performed treating $1 - \hat{\omega}_j$ as a p-value type quantity at the 10\% FDR level. Only the non-selected biomarkers are included.}
\end{figure}

\begin{figure}[hbt!]
\centering
\includegraphics[scale = 0.63]{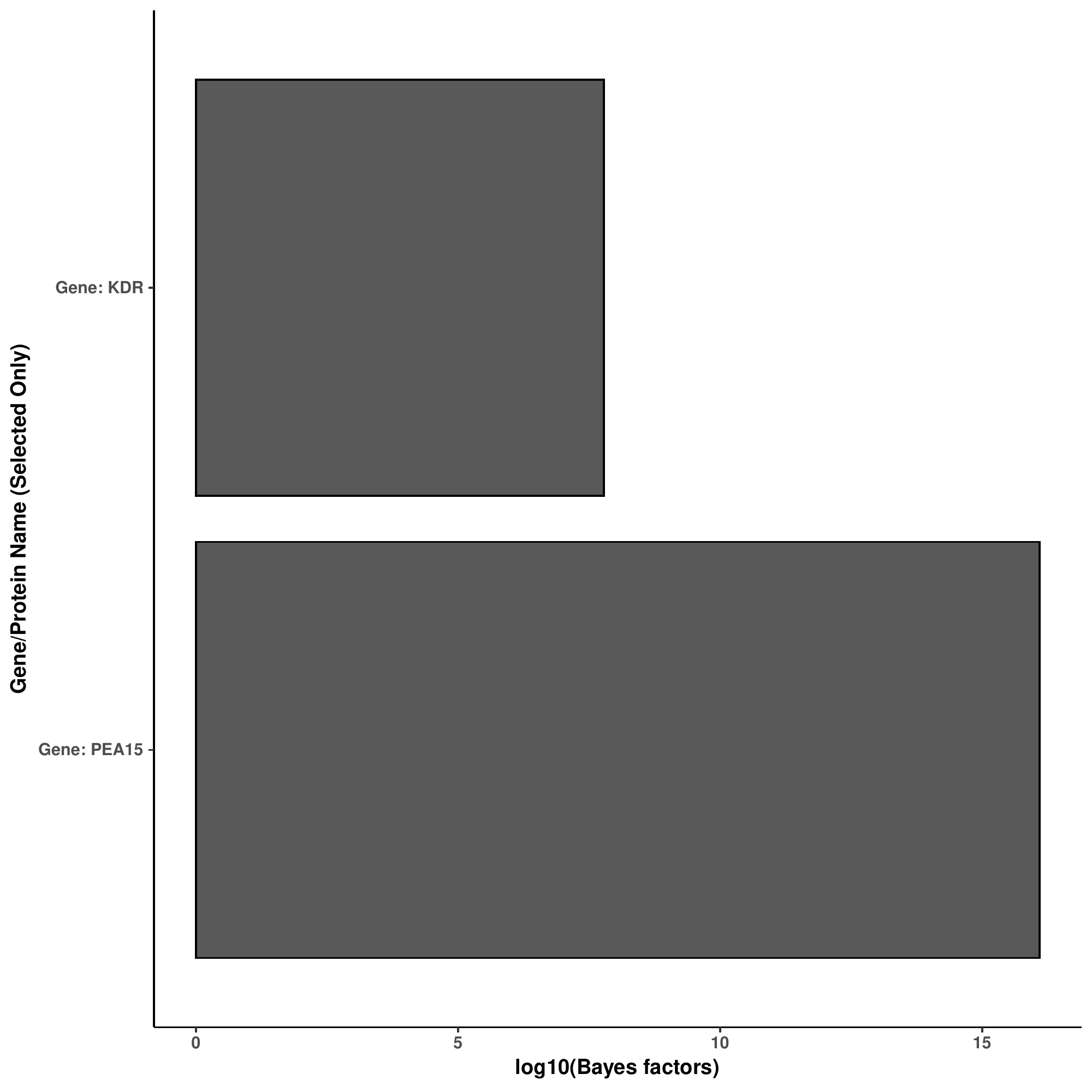}
\caption[Bar diagram summarizing lBFs for the selected proteogenomic biomarkers in the stemness outcome model for TCGA cancer CORE.]{\textbf{Bar diagram summarizing lBFs for the selected proteogenomic biomarkers in the stemness outcome model for TCGA cancer CORE.} An FDR check to adjust for multiple comparisons is performed treating $1 - \hat{\omega}_j$ as a p-value type quantity at the 10\% FDR level. Only the selected biomarkers are presented.}
\end{figure}

\begin{figure}[hbt!]
\centering
\includegraphics[scale = 0.63]{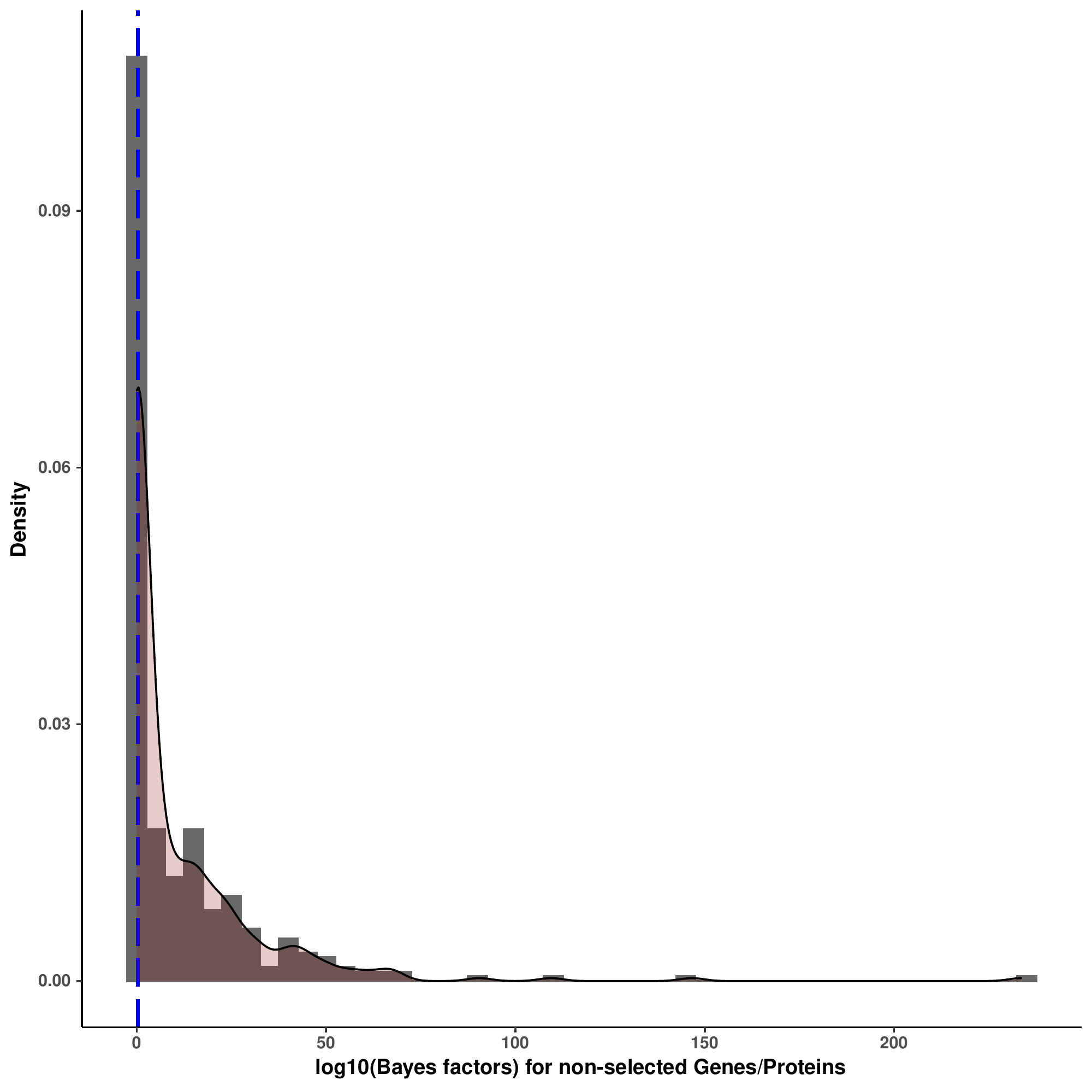}
\caption[Histogram of lBFs for the non-selected proteogenomic biomarkers in the stemness outcome model for TCGA cancer CORE.]{\textbf{Histogram of lBFs for the non-selected proteogenomic biomarkers in the stemness outcome model for TCGA cancer CORE.} An FDR check to adjust for multiple comparisons is performed treating $1 - \hat{\omega}_j$ as a p-value type quantity at the 10\% FDR level. Only the non-selected biomarkers are included.}
\end{figure}

\begin{figure}[hbt!]
\centering
\includegraphics[scale = 0.63]{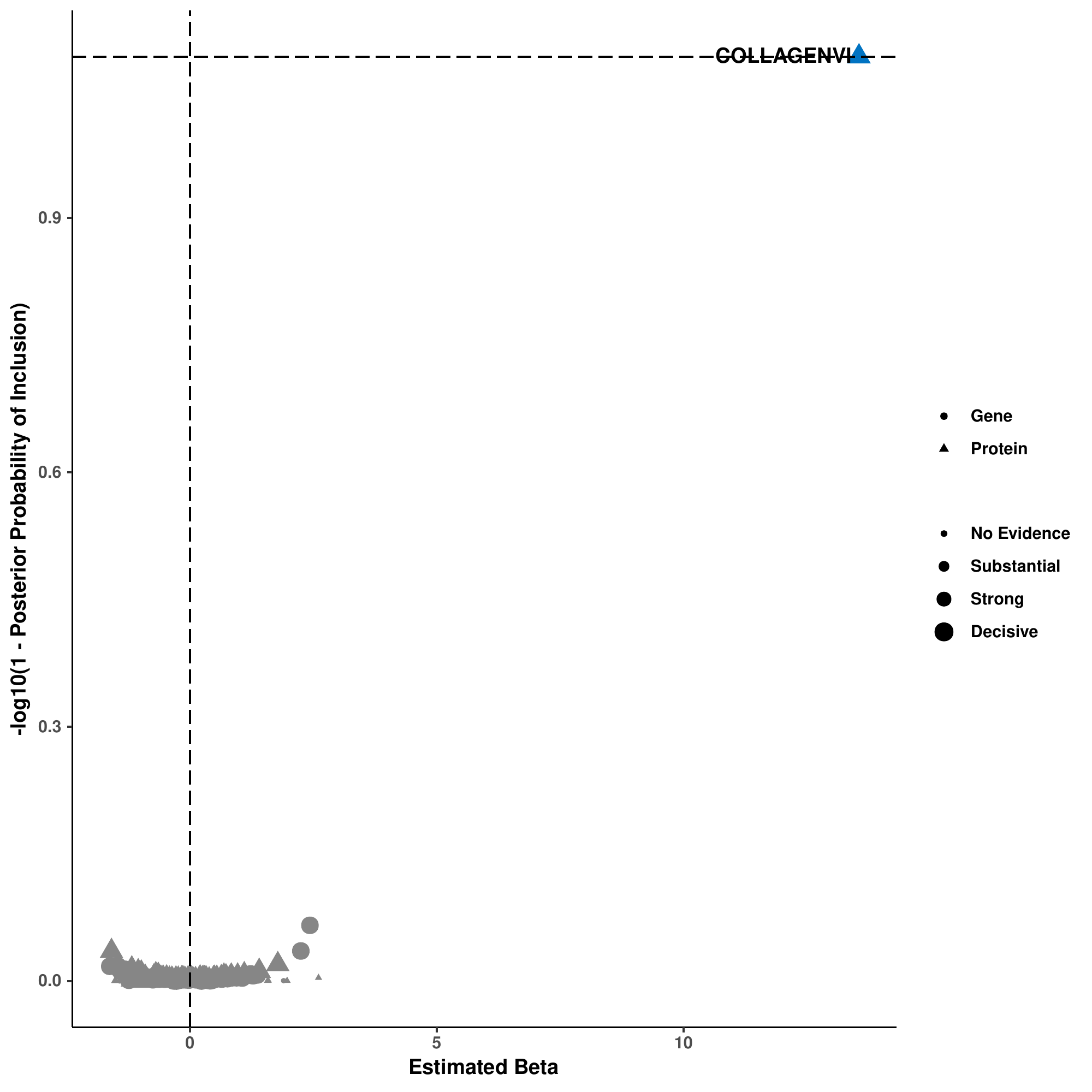}
\caption[Plot summarizing outcome model results based on overall survival for TCGA cancer BRCA.]{\textbf{Plot summarizing outcome model results based on overall survival for TCGA cancer BRCA.} Proteins are represented by triangles and genes by circles. The shapes are colored red if the estimated $\hat{\beta}_j$ from fiBAG is negative, and green if positive. The x-axis shows the $\hat{\beta}_j$s, and the y-axis shows the $-\log_{10}(1 - \hat{\omega}_j)$s. An FDR check to adjust for multiple comparisons is performed treating $1 - \hat{\omega}_j$ as a p-value type quantity at the 10\% FDR level. Only the selected biomarkers are marked in non-gray colors and labeled. The sizes of the points are in the increasing order of evidence from the mechanistic models: lBF ranges are defined as: $< 0.5$ (no evidence), $0.5 - 1$ (substantial), $1 - 2$ (strong), $> 2$ (decisive).} \label{fig: OM_SV_1}
\end{figure}

\begin{figure}[hbt!]
\centering
\includegraphics[scale = 0.63]{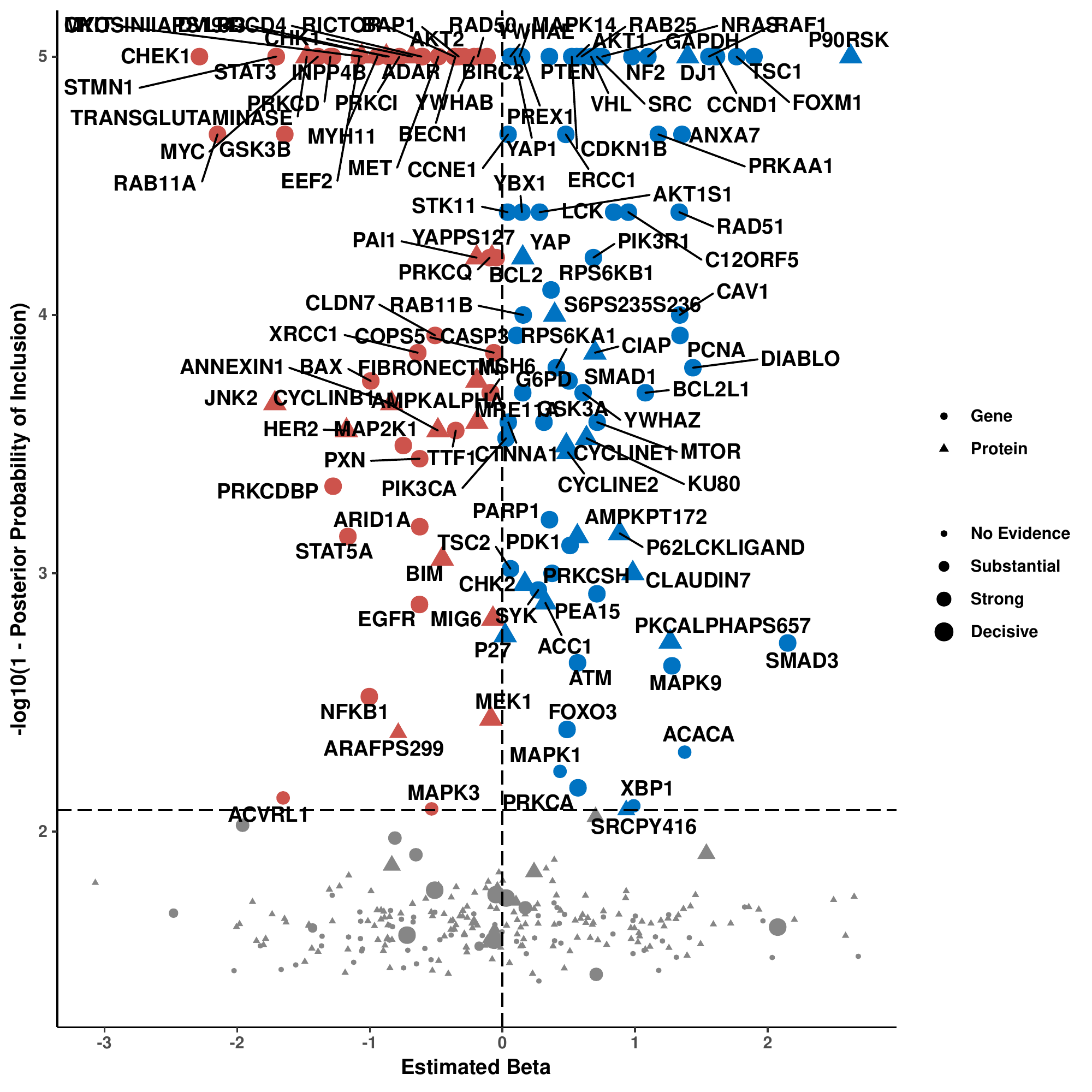}
\caption[Plot summarizing outcome model results based on overall survival for TCGA cancer CESC.]{\textbf{Plot summarizing outcome model results based on overall survival for TCGA cancer CESC.} Proteins are represented by triangles and genes by circles. The shapes are colored red if the estimated $\hat{\beta}_j$ from fiBAG is negative, and green if positive. The x-axis shows the $\hat{\beta}_j$s, and the y-axis shows the $-\log_{10}(1 - \hat{\omega}_j)$s. An FDR check to adjust for multiple comparisons is performed treating $1 - \hat{\omega}_j$ as a p-value type quantity at the 10\% FDR level. Only the selected biomarkers are marked in non-gray colors and labeled. The sizes of the points are in the increasing order of evidence from the mechanistic models: lBF ranges are defined as: $< 0.5$ (no evidence), $0.5 - 1$ (substantial), $1 - 2$ (strong), $> 2$ (decisive).} \label{fig: OM_SV_CESC}
\end{figure}

\begin{figure}[hbt!]
\centering
\includegraphics[scale = 0.63]{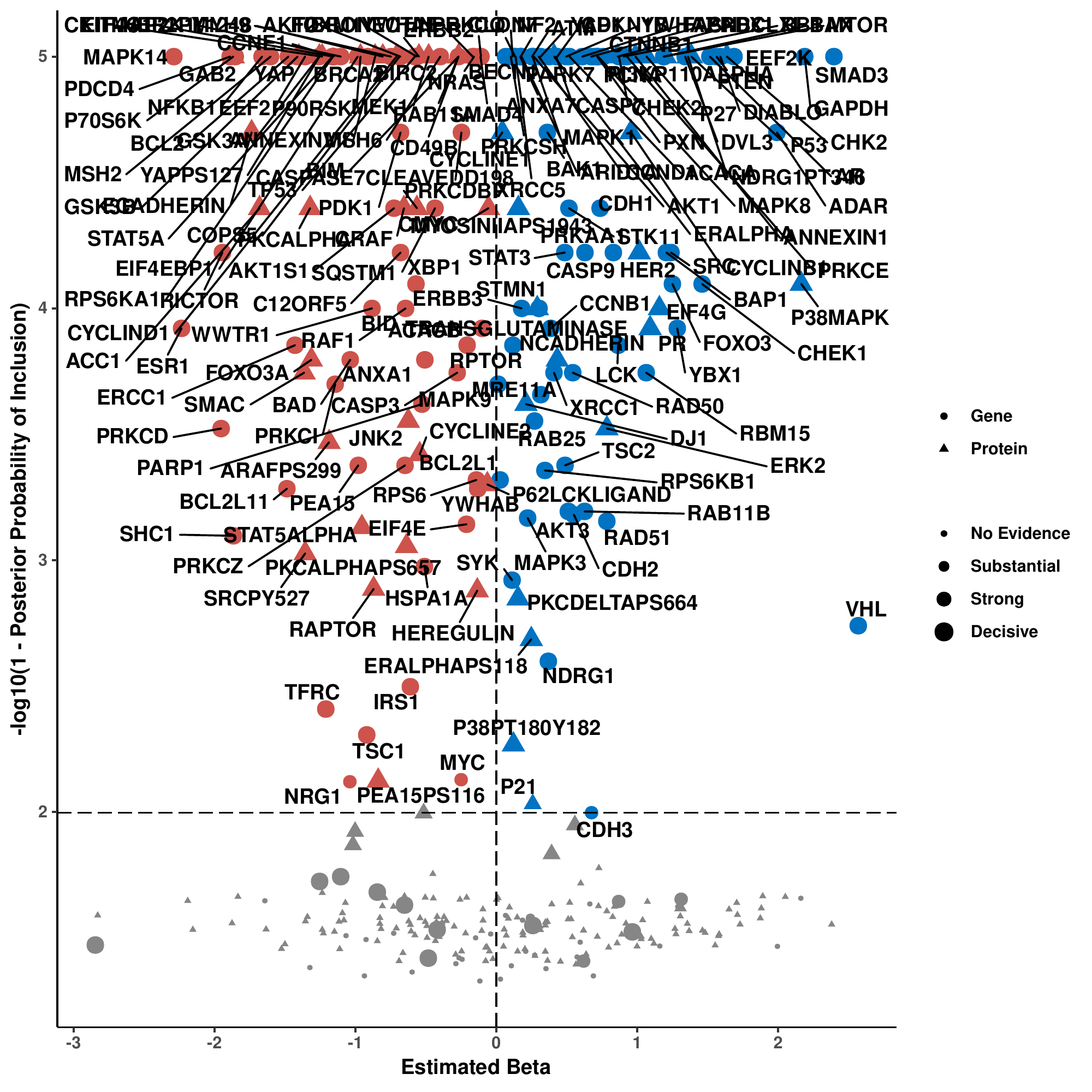}
\caption[Plot summarizing outcome model results based on overall survival for TCGA cancer OV.]{\textbf{Plot summarizing outcome model results based on overall survival for TCGA cancer OV.} Proteins are represented by triangles and genes by circles. The shapes are colored red if the estimated $\hat{\beta}_j$ from fiBAG is negative, and green if positive. The x-axis shows the $\hat{\beta}_j$s, and the y-axis shows the $-\log_{10}(1 - \hat{\omega}_j)$s. An FDR check to adjust for multiple comparisons is performed treating $1 - \hat{\omega}_j$ as a p-value type quantity at the 10\% FDR level. Only the selected biomarkers are marked in non-gray colors and labeled. The sizes of the points are in the increasing order of evidence from the mechanistic models: lBF ranges are defined as: $< 0.5$ (no evidence), $0.5 - 1$ (substantial), $1 - 2$ (strong), $> 2$ (decisive).} \label{fig: OM_SV_OV}
\end{figure}

\begin{figure}[hbt!]
\centering
\includegraphics[scale = 0.63]{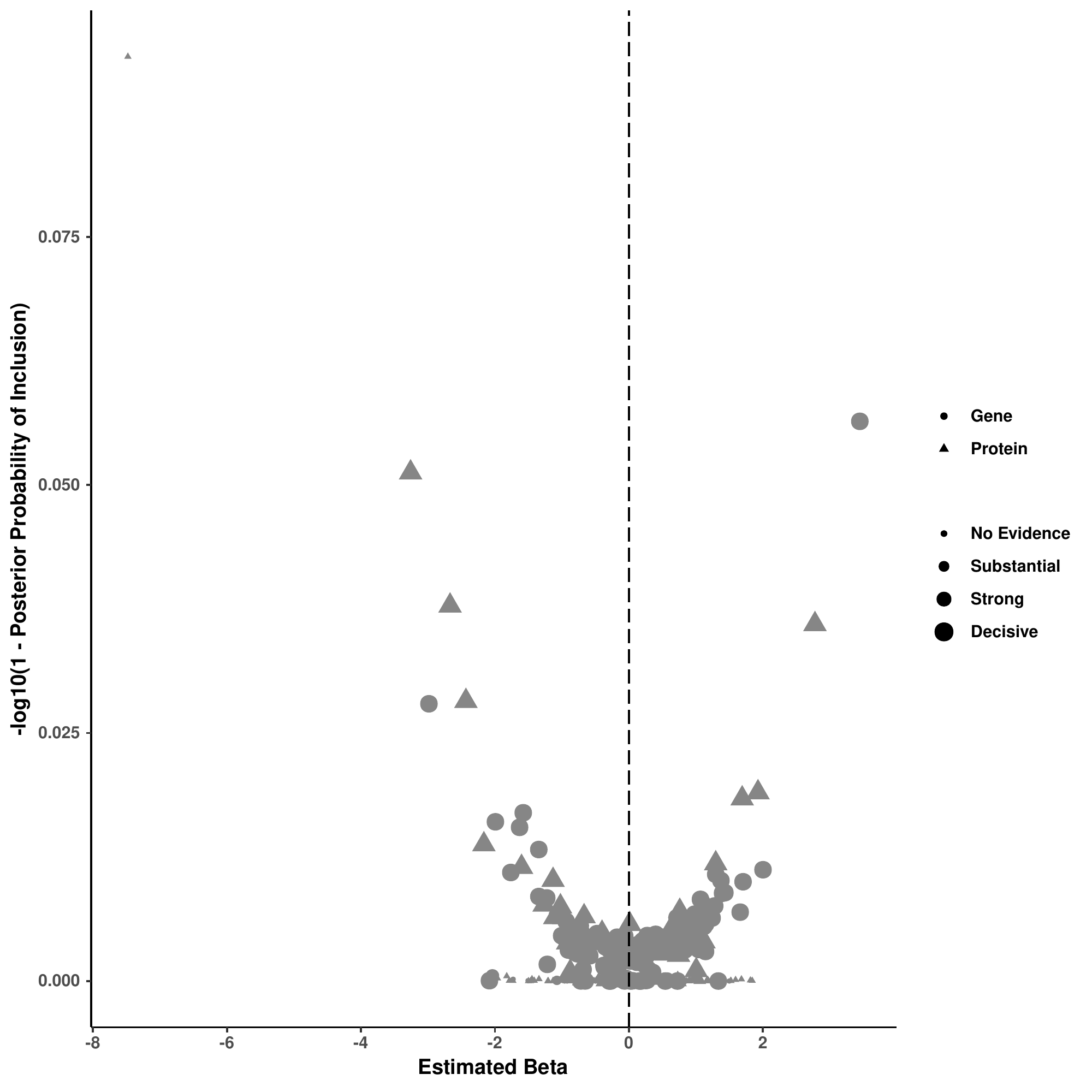}
\caption[Plot summarizing outcome model results based on overall survival for TCGA cancer UCEC.]{\textbf{Plot summarizing outcome model results based on overall survival for TCGA cancer UCEC.} Proteins are represented by triangles and genes by circles. The shapes are colored red if the estimated $\hat{\beta}_j$ from fiBAG is negative, and green if positive. The x-axis shows the $\hat{\beta}_j$s, and the y-axis shows the $-\log_{10}(1 - \hat{\omega}_j)$s. An FDR check to adjust for multiple comparisons is performed treating $1 - \hat{\omega}_j$ as a p-value type quantity at the 10\% FDR level. Only the selected biomarkers are marked in non-gray colors and labeled. The sizes of the points are in the increasing order of evidence from the mechanistic models: lBF ranges are defined as: $< 0.5$ (no evidence), $0.5 - 1$ (substantial), $1 - 2$ (strong), $> 2$ (decisive).}
\end{figure}

\begin{figure}[hbt!]
\centering
\includegraphics[scale = 0.63]{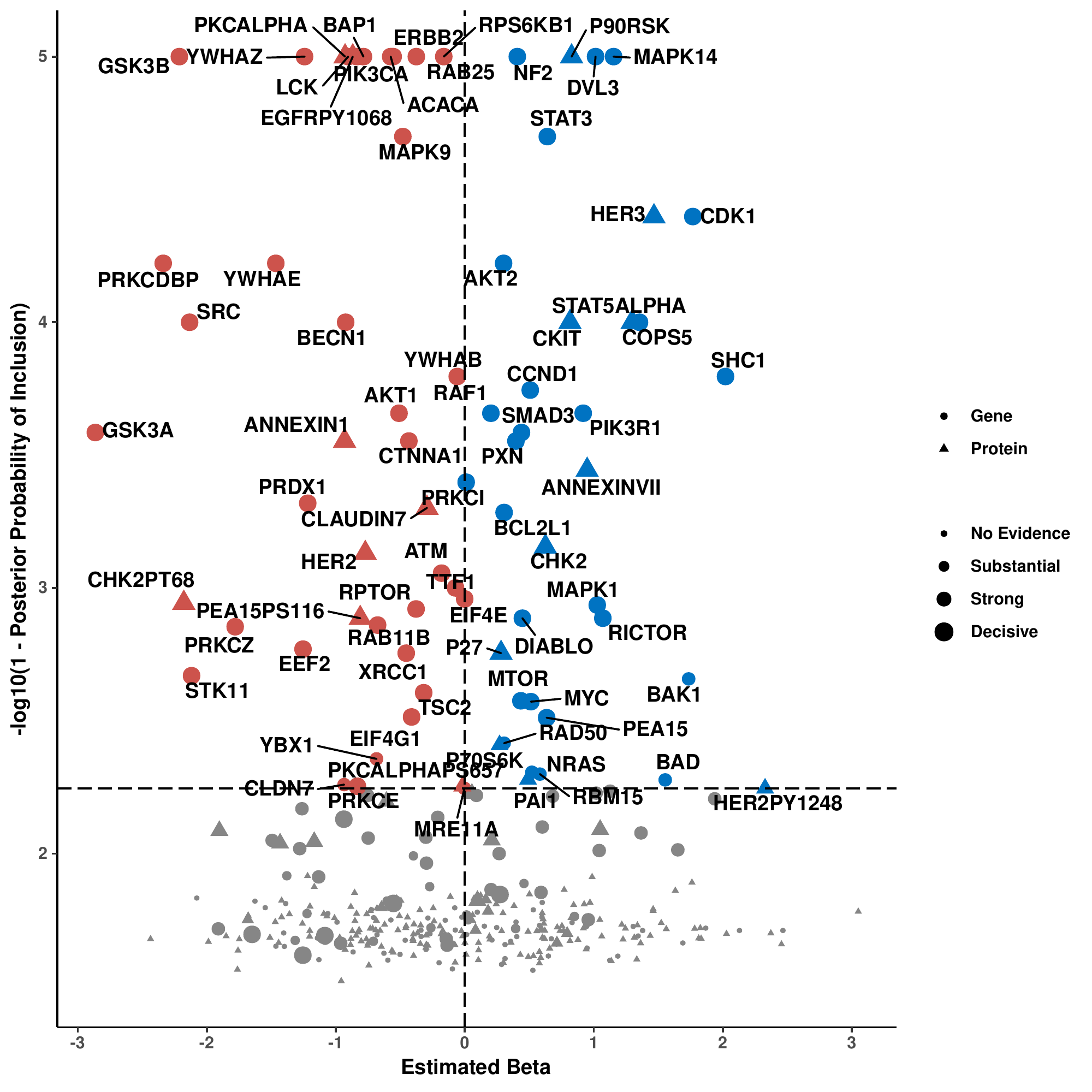}
\caption[Plot summarizing outcome model results based on overall survival for TCGA cancer UCS.]{\textbf{Plot summarizing outcome model results based on overall survival for TCGA cancer UCS.} Proteins are represented by triangles and genes by circles. The shapes are colored red if the estimated $\hat{\beta}_j$ from fiBAG is negative, and green if positive. The x-axis shows the $\hat{\beta}_j$s, and the y-axis shows the $-\log_{10}(1 - \hat{\omega}_j)$s. An FDR check to adjust for multiple comparisons is performed treating $1 - \hat{\omega}_j$ as a p-value type quantity at the 10\% FDR level. Only the selected biomarkers are marked in non-gray colors and labeled. The sizes of the points are in the increasing order of evidence from the mechanistic models: lBF ranges are defined as: $< 0.5$ (no evidence), $0.5 - 1$ (substantial), $1 - 2$ (strong), $> 2$ (decisive).}
\end{figure}

\begin{figure}[hbt!]
\centering
\includegraphics[scale = 0.63]{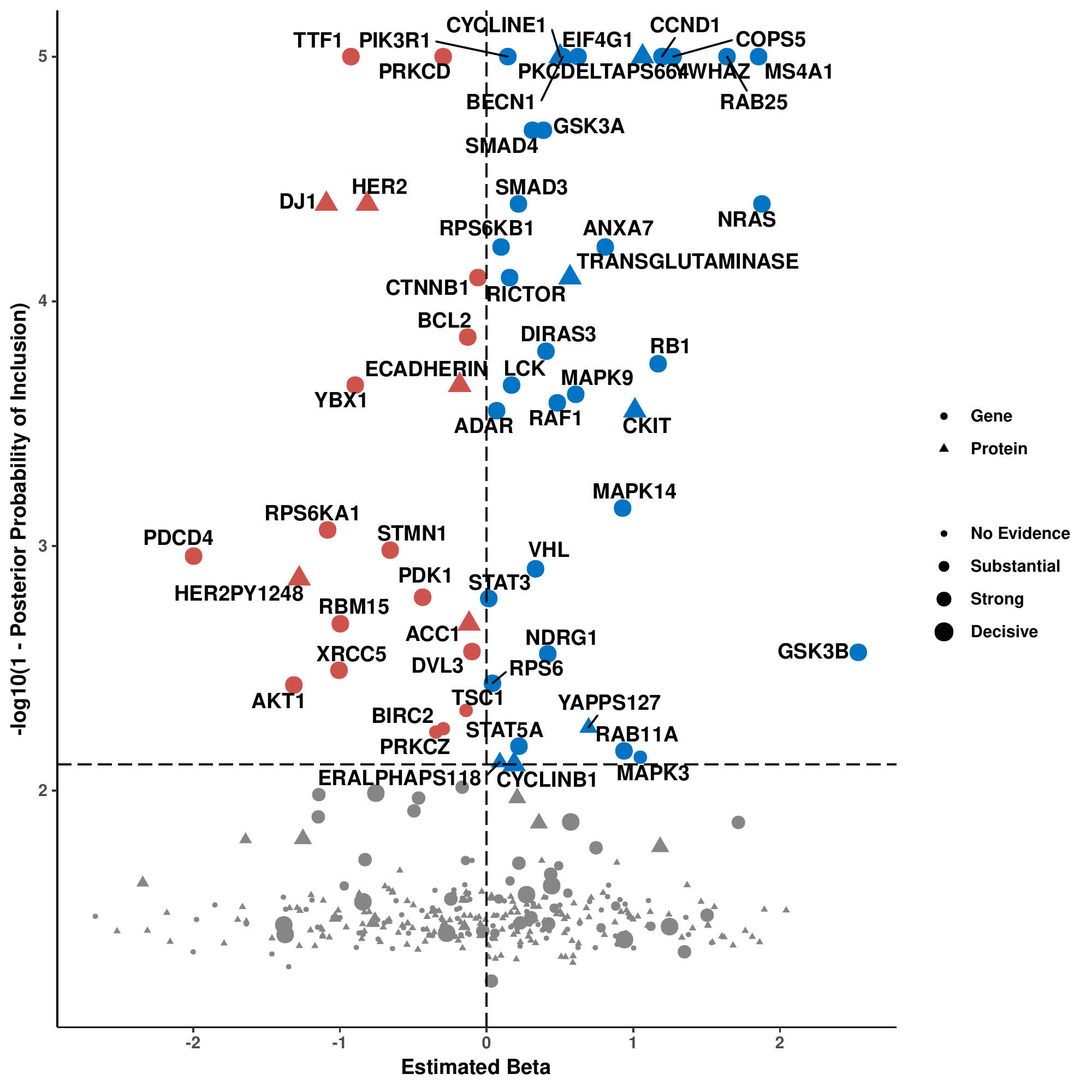}
\caption[Plot summarizing outcome model results based on overall survival for TCGA cancer KICH.]{\textbf{Plot summarizing outcome model results based on overall survival for TCGA cancer KICH.} Proteins are represented by triangles and genes by circles. The shapes are colored red if the estimated $\hat{\beta}_j$ from fiBAG is negative, and green if positive. The x-axis shows the $\hat{\beta}_j$s, and the y-axis shows the $-\log_{10}(1 - \hat{\omega}_j)$s. An FDR check to adjust for multiple comparisons is performed treating $1 - \hat{\omega}_j$ as a p-value type quantity at the 10\% FDR level. Only the selected biomarkers are marked in non-gray colors and labeled. The sizes of the points are in the increasing order of evidence from the mechanistic models: lBF ranges are defined as: $< 0.5$ (no evidence), $0.5 - 1$ (substantial), $1 - 2$ (strong), $> 2$ (decisive).}
\end{figure}

\begin{figure}[hbt!]
\centering
\includegraphics[scale = 0.63]{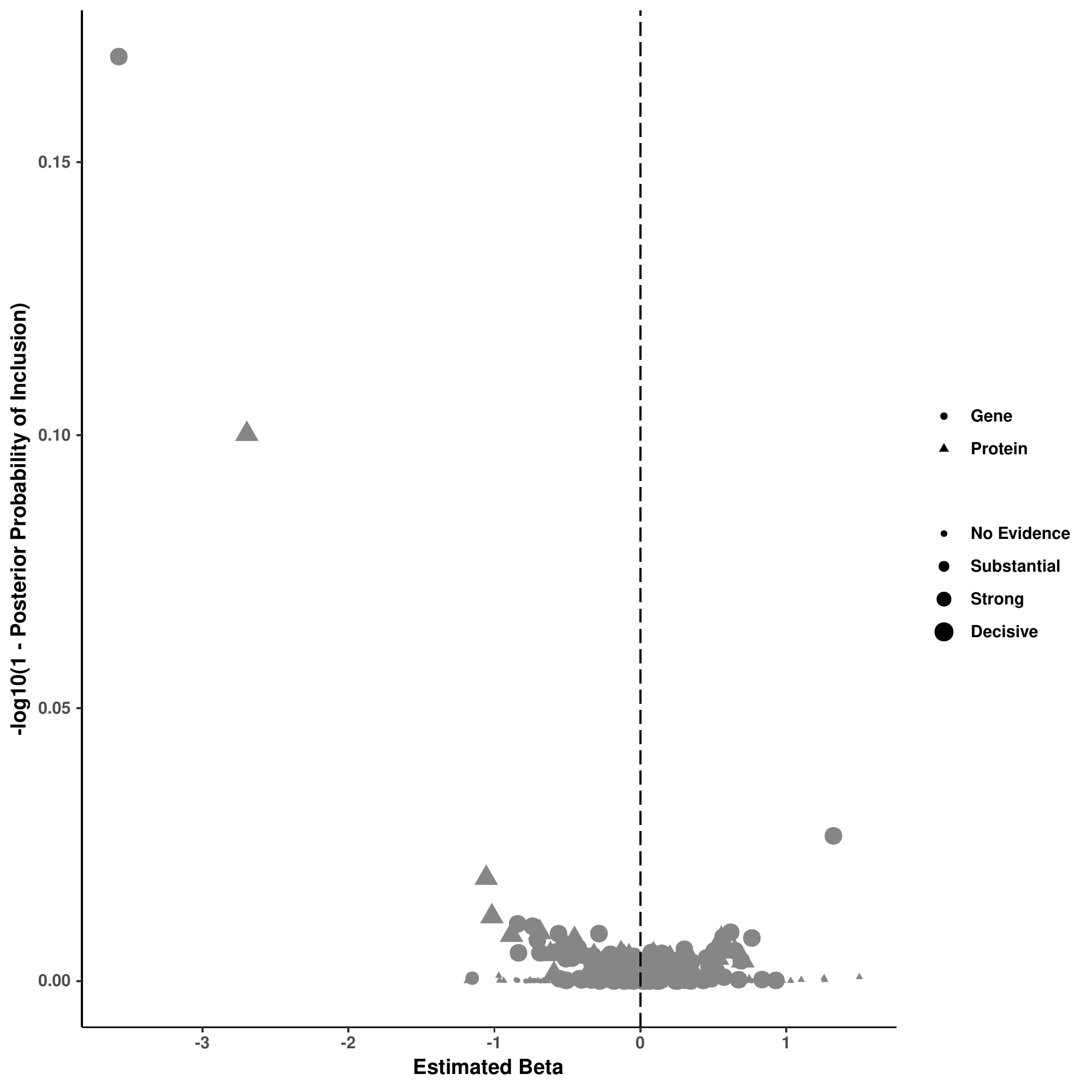}
\caption[Plot summarizing outcome model results based on overall survival for TCGA cancer KIRC.]{\textbf{Plot summarizing outcome model results based on overall survival for TCGA cancer KIRC.} Proteins are represented by triangles and genes by circles. The shapes are colored red if the estimated $\hat{\beta}_j$ from fiBAG is negative, and green if positive. The x-axis shows the $\hat{\beta}_j$s, and the y-axis shows the $-\log_{10}(1 - \hat{\omega}_j)$s. An FDR check to adjust for multiple comparisons is performed treating $1 - \hat{\omega}_j$ as a p-value type quantity at the 10\% FDR level. Only the selected biomarkers are marked in non-gray colors and labeled. The sizes of the points are in the increasing order of evidence from the mechanistic models: lBF ranges are defined as: $< 0.5$ (no evidence), $0.5 - 1$ (substantial), $1 - 2$ (strong), $> 2$ (decisive).}
\end{figure}

\begin{figure}[hbt!]
\centering
\includegraphics[scale = 0.63]{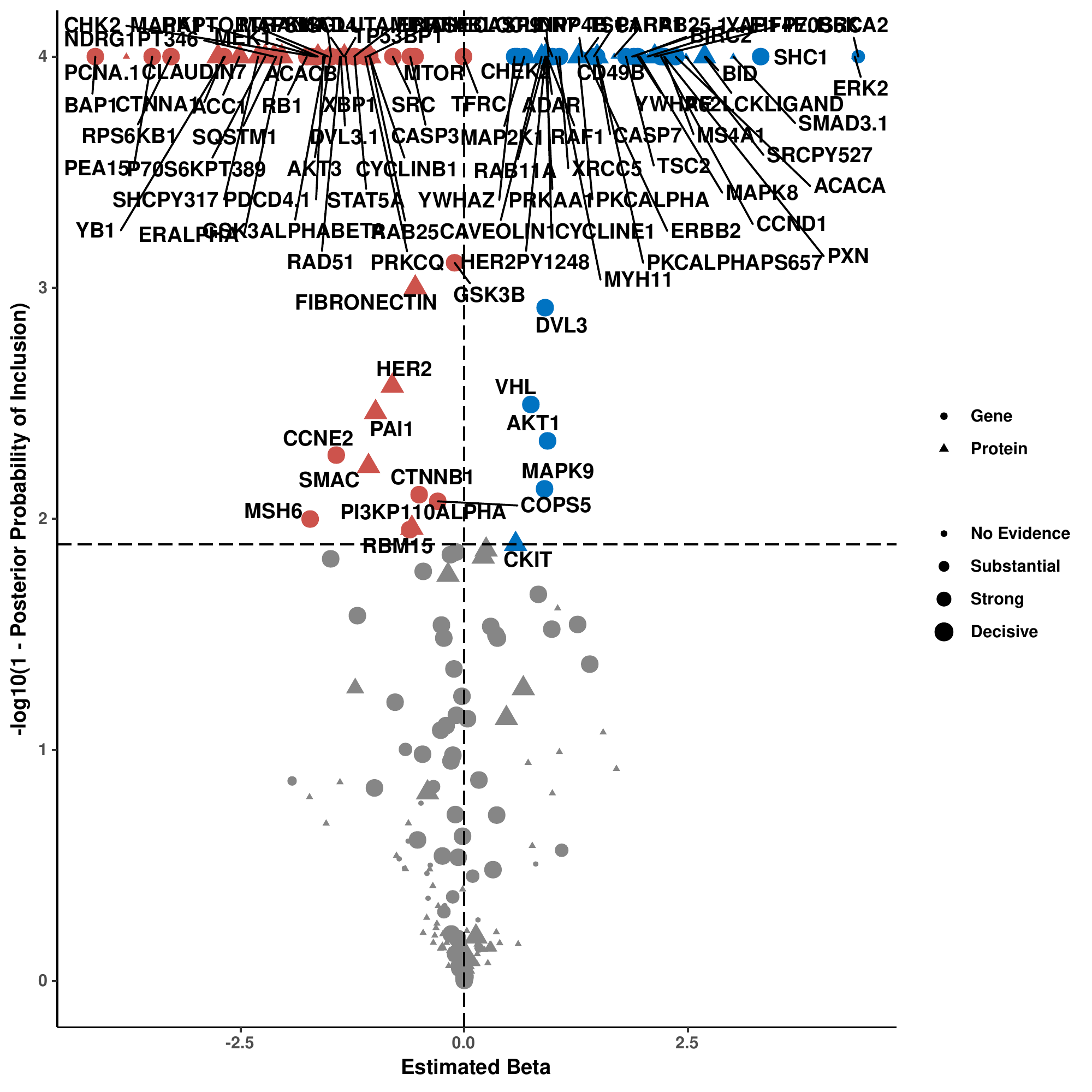}
\caption[Plot summarizing outcome model results based on overall survival for TCGA cancer KIRP.]{\textbf{Plot summarizing outcome model results based on overall survival for TCGA cancer KIRP.} Proteins are represented by triangles and genes by circles. The shapes are colored red if the estimated $\hat{\beta}_j$ from fiBAG is negative, and green if positive. The x-axis shows the $\hat{\beta}_j$s, and the y-axis shows the $-\log_{10}(1 - \hat{\omega}_j)$s. An FDR check to adjust for multiple comparisons is performed treating $1 - \hat{\omega}_j$ as a p-value type quantity at the 10\% FDR level. Only the selected biomarkers are marked in non-gray colors and labeled. The sizes of the points are in the increasing order of evidence from the mechanistic models: lBF ranges are defined as: $< 0.5$ (no evidence), $0.5 - 1$ (substantial), $1 - 2$ (strong), $> 2$ (decisive).}
\end{figure}

\begin{figure}[hbt!]
\centering
\includegraphics[scale = 0.63]{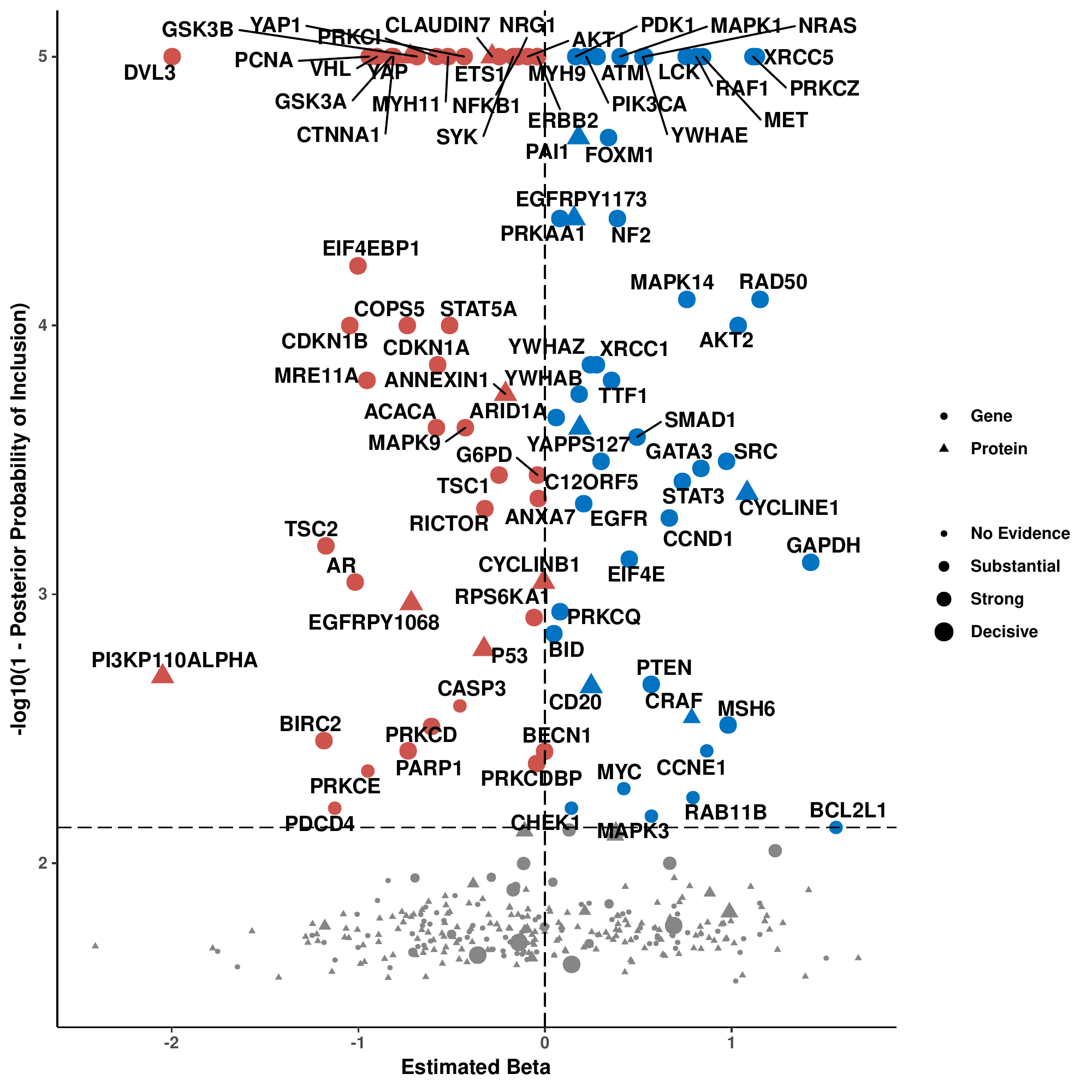}
\caption[Plot summarizing outcome model results based on overall survival for TCGA cancer ESCA-squamous.]{\textbf{Plot summarizing outcome model results based on overall survival for TCGA cancer ESCS-squamous.} Proteins are represented by triangles and genes by circles. The shapes are colored red if the estimated $\hat{\beta}_j$ from fiBAG is negative, and green if positive. The x-axis shows the $\hat{\beta}_j$s, and the y-axis shows the $-\log_{10}(1 - \hat{\omega}_j)$s. An FDR check to adjust for multiple comparisons is performed treating $1 - \hat{\omega}_j$ as a p-value type quantity at the 10\% FDR level. Only the selected biomarkers are marked in non-gray colors and labeled. The sizes of the points are in the increasing order of evidence from the mechanistic models: lBF ranges are defined as: $< 0.5$ (no evidence), $0.5 - 1$ (substantial), $1 - 2$ (strong), $> 2$ (decisive).}
\end{figure}

\begin{figure}[hbt!]
\centering
\includegraphics[scale = 0.63]{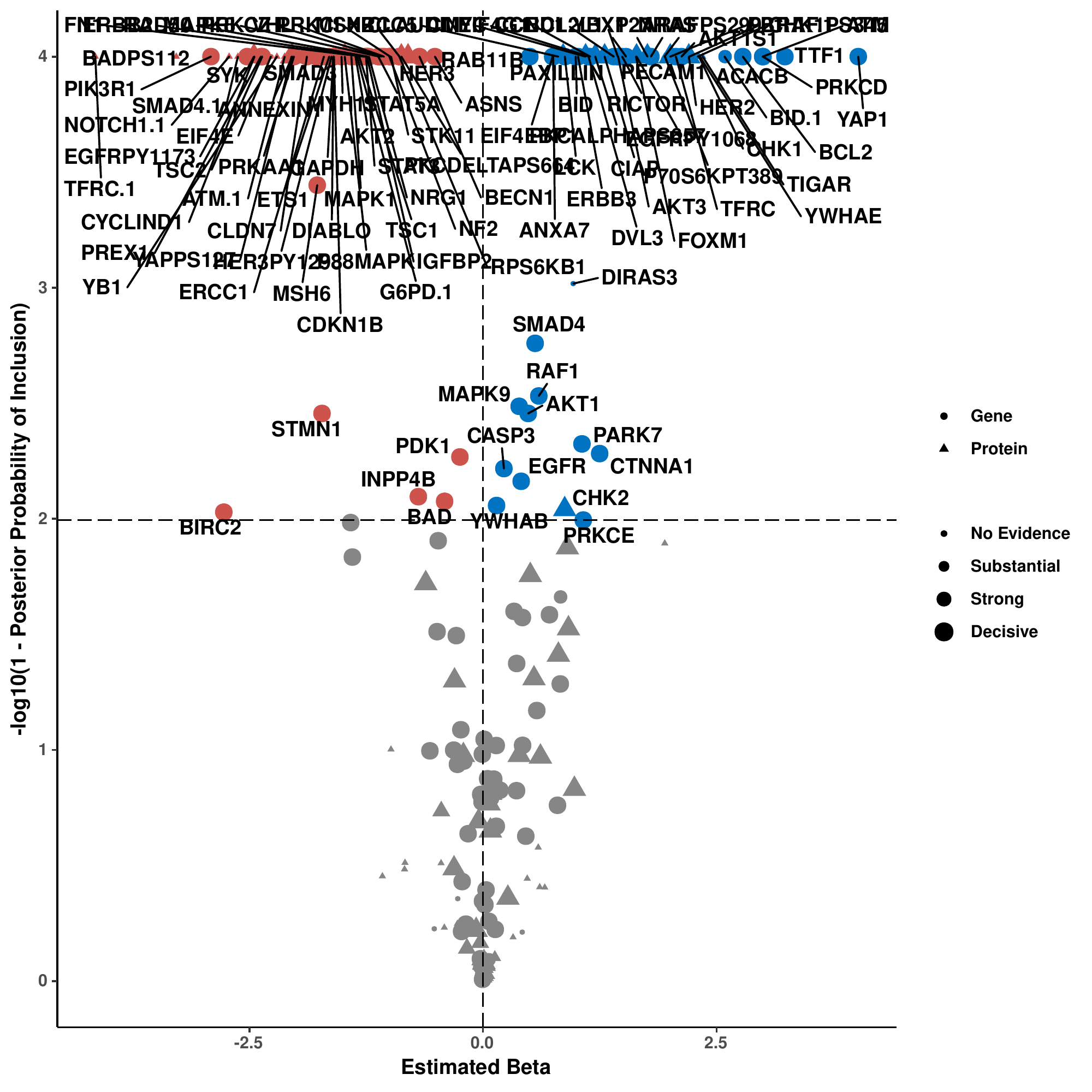}
\caption[Plot summarizing outcome model results based on overall survival for TCGA cancer HNSC.]{\textbf{Plot summarizing outcome model results based on overall survival for TCGA cancer HNSC.} Proteins are represented by triangles and genes by circles. The shapes are colored red if the estimated $\hat{\beta}_j$ from fiBAG is negative, and green if positive. The x-axis shows the $\hat{\beta}_j$s, and the y-axis shows the $-\log_{10}(1 - \hat{\omega}_j)$s. An FDR check to adjust for multiple comparisons is performed treating $1 - \hat{\omega}_j$ as a p-value type quantity at the 10\% FDR level. Only the selected biomarkers are marked in non-gray colors and labeled. The sizes of the points are in the increasing order of evidence from the mechanistic models: lBF ranges are defined as: $< 0.5$ (no evidence), $0.5 - 1$ (substantial), $1 - 2$ (strong), $> 2$ (decisive).}
\end{figure}

\begin{figure}[hbt!]
\centering
\includegraphics[scale = 0.63]{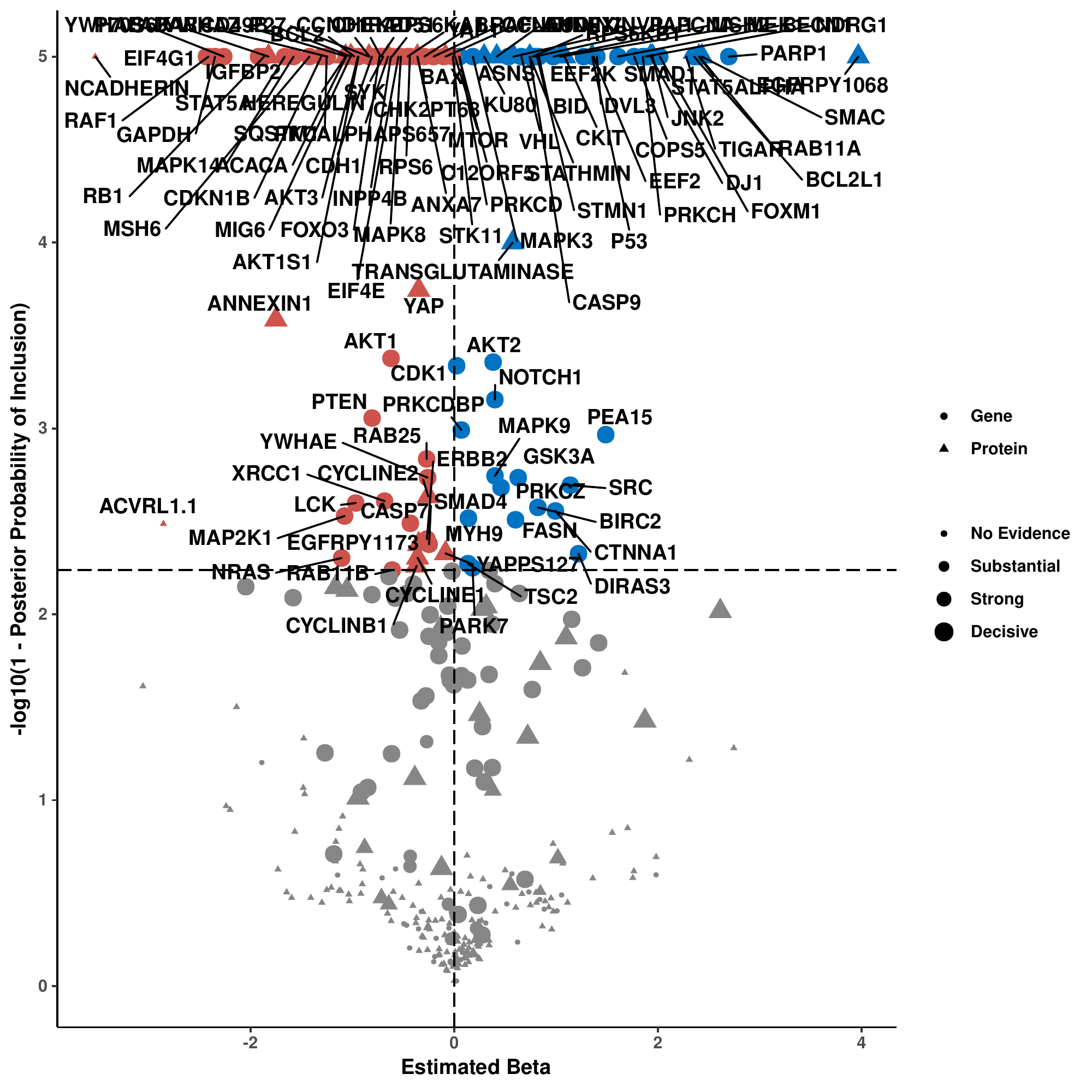}
\caption[Plot summarizing outcome model results based on overall survival for TCGA cancer LUSC.]{\textbf{Plot summarizing outcome model results based on overall survival for TCGA cancer LUSC.} Proteins are represented by triangles and genes by circles. The shapes are colored red if the estimated $\hat{\beta}_j$ from fiBAG is negative, and green if positive. The x-axis shows the $\hat{\beta}_j$s, and the y-axis shows the $-\log_{10}(1 - \hat{\omega}_j)$s. An FDR check to adjust for multiple comparisons is performed treating $1 - \hat{\omega}_j$ as a p-value type quantity at the 10\% FDR level. Only the selected biomarkers are marked in non-gray colors and labeled. The sizes of the points are in the increasing order of evidence from the mechanistic models: lBF ranges are defined as: $< 0.5$ (no evidence), $0.5 - 1$ (substantial), $1 - 2$ (strong), $> 2$ (decisive).}
\end{figure}

\begin{figure}[hbt!]
\centering
\includegraphics[scale = 0.63]{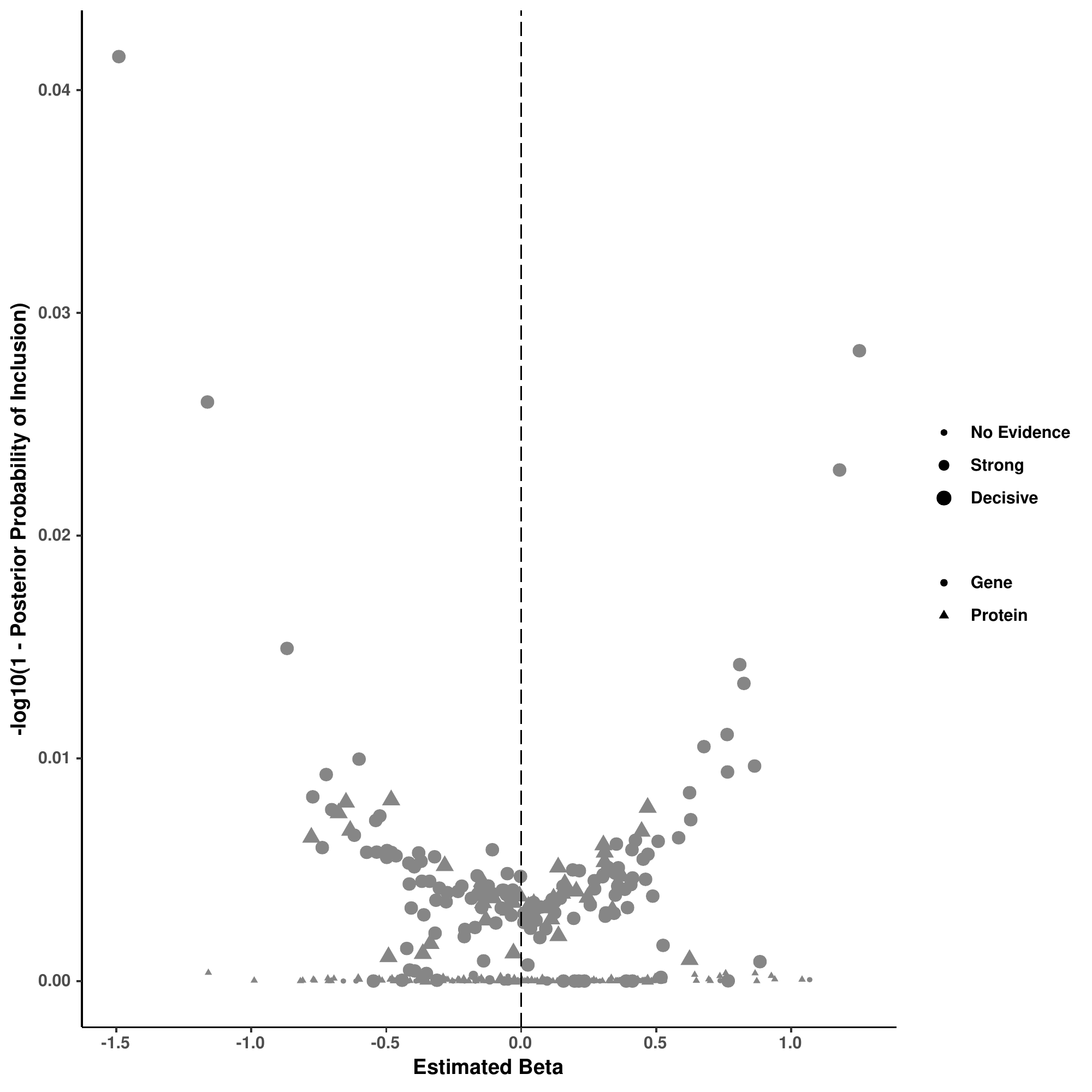}
\caption[Plot summarizing outcome model results based on overall survival for TCGA cancer CORE.]{\textbf{Plot summarizing outcome model results based on overall survival for TCGA cancer CORE.} Proteins are represented by triangles and genes by circles. The shapes are colored red if the estimated $\hat{\beta}_j$ from fiBAG is negative, and green if positive. The x-axis shows the $\hat{\beta}_j$s, and the y-axis shows the $-\log_{10}(1 - \hat{\omega}_j)$s. An FDR check to adjust for multiple comparisons is performed treating $1 - \hat{\omega}_j$ as a p-value type quantity at the 10\% FDR level. Only the selected biomarkers are marked in non-gray colors and labeled. The sizes of the points are in the increasing order of evidence from the mechanistic models: lBF ranges are defined as: $< 0.5$ (no evidence), $0.5 - 1$ (substantial), $1 - 2$ (strong), $> 2$ (decisive).} \label{fig: OM_SV_12}
\end{figure}

\begin{figure}[hbt!]
\centering
\includegraphics[scale = 0.63]{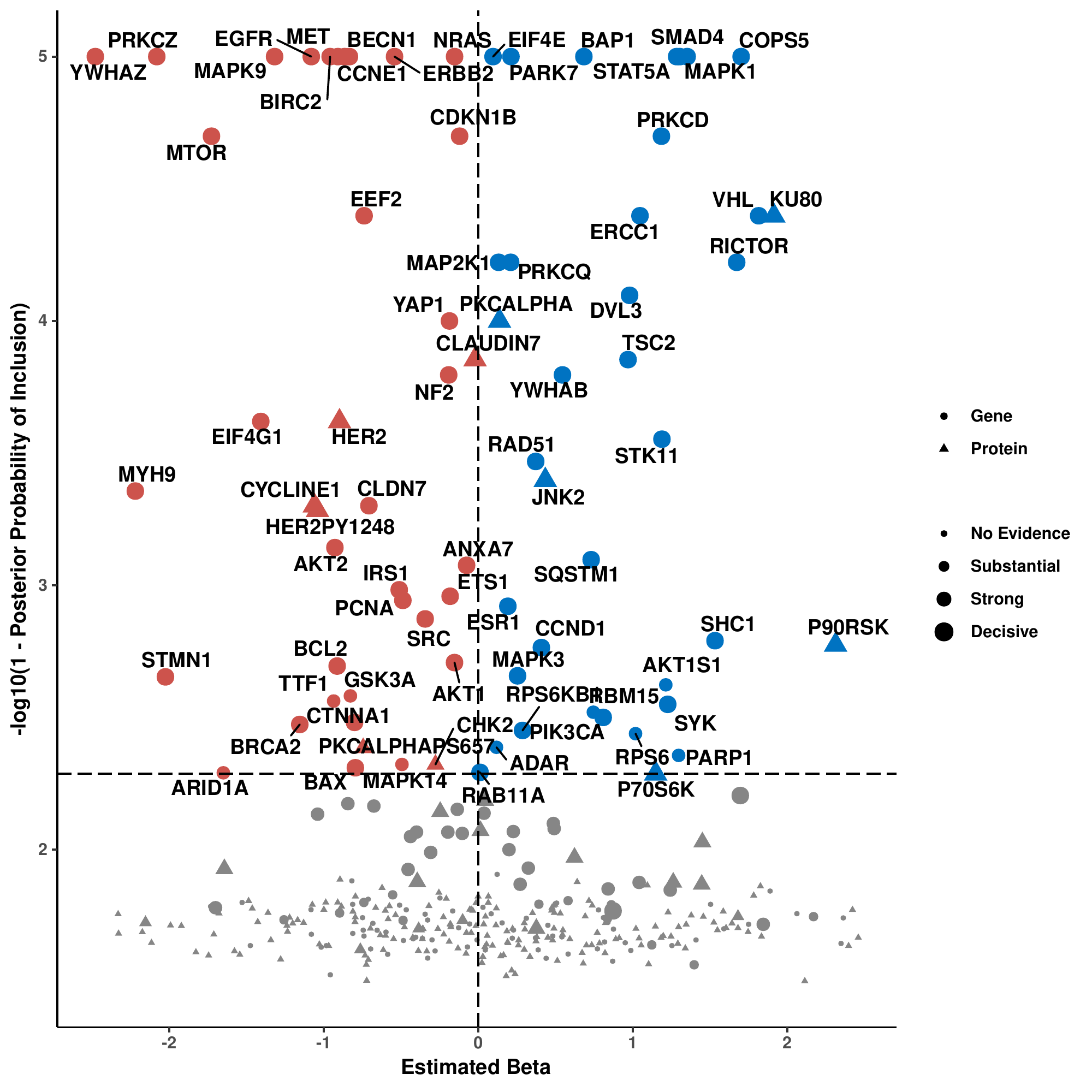}
\caption[Plot summarizing outcome model results based on overall survival for TCGA cancer ESCA-adeno.]{\textbf{Plot summarizing outcome model results based on overall survival for TCGA cancer ESCA-adeno.} Proteins are represented by triangles and genes by circles. The shapes are colored red if the estimated $\hat{\beta}_j$ from fiBAG is negative, and green if positive. The x-axis shows the $\hat{\beta}_j$s, and the y-axis shows the $-\log_{10}(1 - \hat{\omega}_j)$s. An FDR check to adjust for multiple comparisons is performed treating $1 - \hat{\omega}_j$ as a p-value type quantity at the 10\% FDR level. Only the selected biomarkers are marked in non-gray colors and labeled. The sizes of the points are in the increasing order of evidence from the mechanistic models: lBF ranges are defined as: $< 0.5$ (no evidence), $0.5 - 1$ (substantial), $1 - 2$ (strong), $> 2$ (decisive).} \label{fig: OM_SV_ESCA}
\end{figure}

\begin{figure}[hbt!]
\centering
\includegraphics[scale = 0.63]{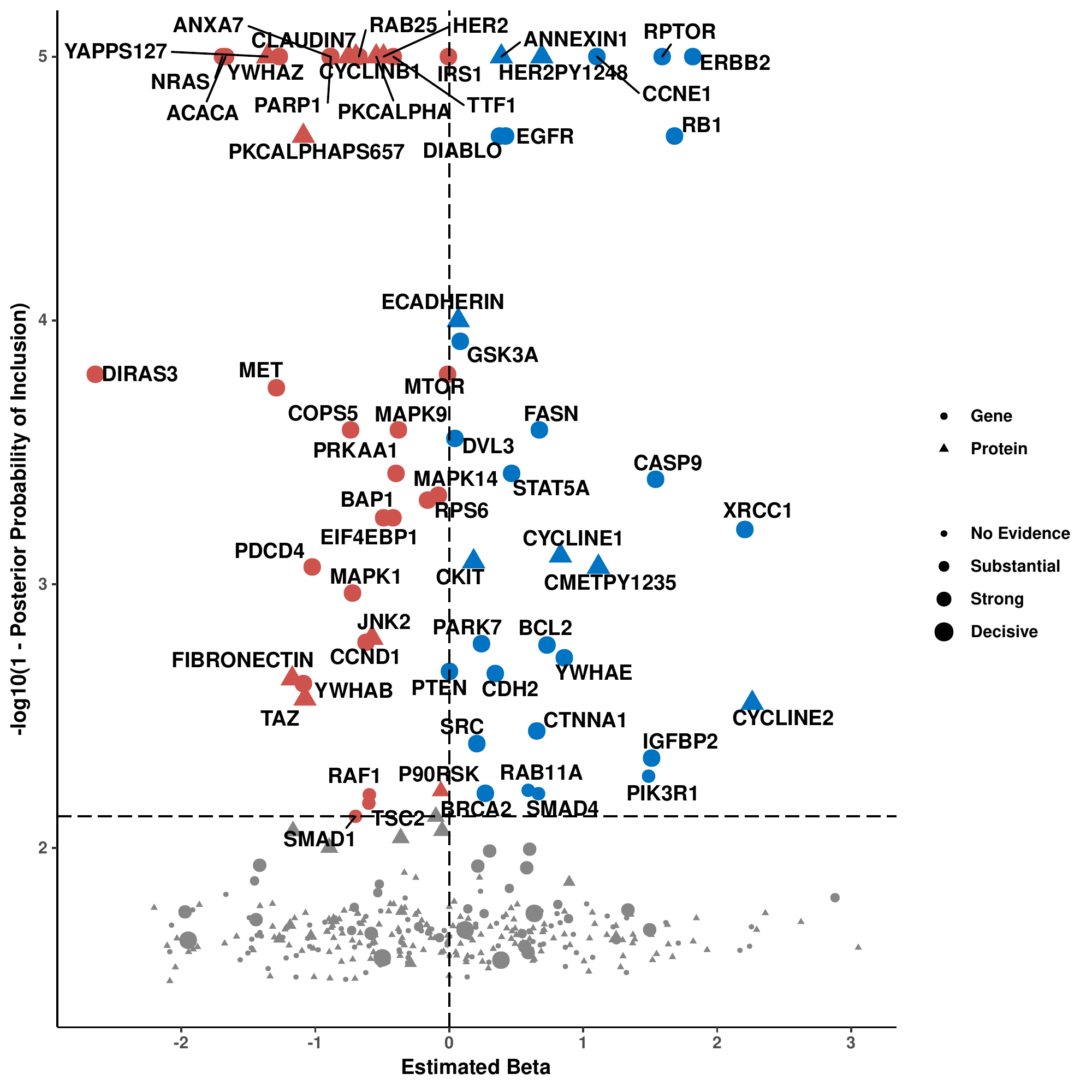}
\caption[Plot summarizing outcome model results based on overall survival for TCGA cancer STAD.]{\textbf{Plot summarizing outcome model results based on overall survival for TCGA cancer STAD.} Proteins are represented by triangles and genes by circles. The shapes are colored red if the estimated $\hat{\beta}_j$ from fiBAG is negative, and green if positive. The x-axis shows the $\hat{\beta}_j$s, and the y-axis shows the $-\log_{10}(1 - \hat{\omega}_j)$s. An FDR check to adjust for multiple comparisons is performed treating $1 - \hat{\omega}_j$ as a p-value type quantity at the 10\% FDR level. Only the selected biomarkers are marked in non-gray colors and labeled. The sizes of the points are in the increasing order of evidence from the mechanistic models: lBF ranges are defined as: $< 0.5$ (no evidence), $0.5 - 1$ (substantial), $1 - 2$ (strong), $> 2$ (decisive).} \label{fig: OM_SV_14}
\end{figure}

\begin{figure}[hbt!]
\centering
\includegraphics[scale = 0.495]{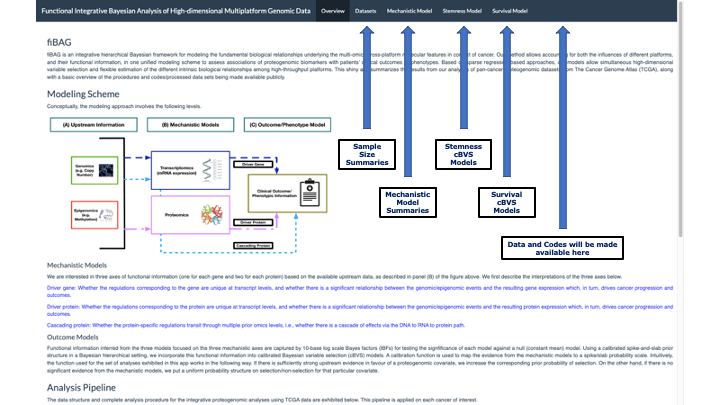}
\caption[Homepage of the interactive R shiny dashboard.]{\textbf{Homepage of the interactive R shiny dashboard.}}
\label{fig:shiny1}
\end{figure}

\begin{figure}[hbt!]
\centering
\includegraphics[scale = 0.495]{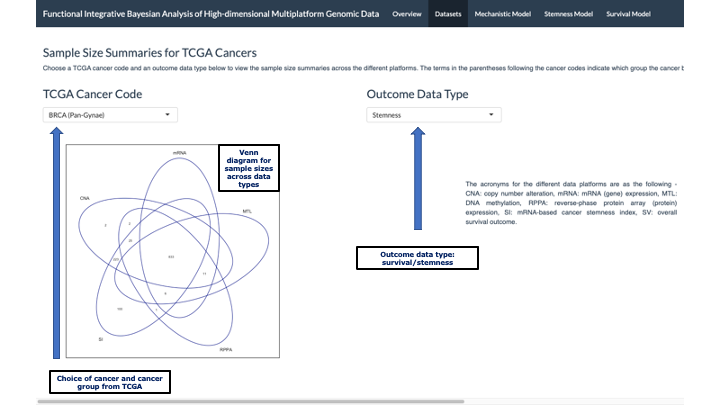}
\caption[Interactive sample size summaries across the data platforms and cancers.]{\textbf{Interactive sample size summaries across the data platforms and cancers.}}
\label{fig:shiny2}
\end{figure}

\begin{figure}[hbt!]
\centering
\includegraphics[scale = 0.495]{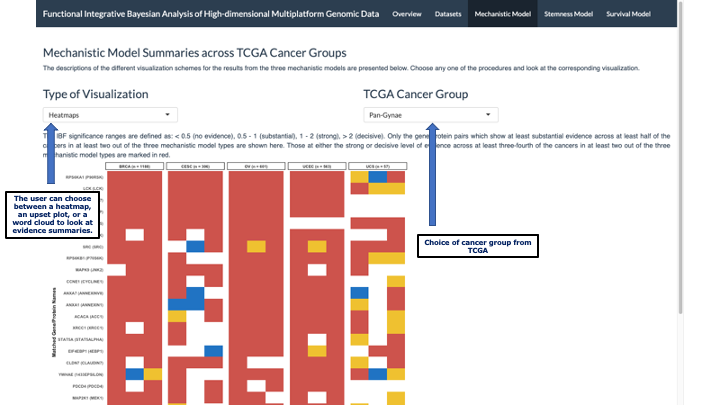}
\caption[Mechanistic model summary figures for the different cancer groups with different visualization schemes.]{\textbf{Mechanistic model summary figures for the different cancer groups with different visualization schemes.}}
\label{fig:shiny3}
\end{figure}

\begin{figure}[hbt!]
\centering
\includegraphics[scale = 0.495]{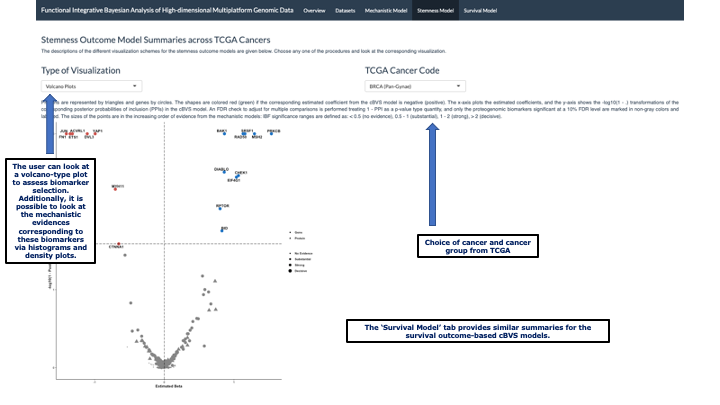}
\caption[Outcome model summaries for the stemness models across different cancers.]{\textbf{Outcome model summaries for the stemness models across different cancers.}}
\label{fig:shiny4}
\end{figure}
        
\end{document}